\documentclass[sigconf]{acmart}
\usepackage{hyperref}
\AtBeginDocument{%
  \providecommand\BibTeX{{%
    \normalfont B\kern-0.5em{\scshape i\kern-0.25em b}\kern-0.8em\TeX}}}

\copyrightyear{2024}
\acmYear{2024}
\setcopyright{rightsretained}
\acmConference[CHI '24]{Proceedings of the CHI Conference on Human Factors in Computing Systems}{May 11--16, 2024}{Honolulu, HI, USA} \acmBooktitle{Proceedings of the CHI Conference on Human Factors in Computing Systems (CHI '24), May 11--16, 2024, Honolulu, HI, USA} \acmDOI{10.1145/3613904.3642013}
\acmISBN{979-8-4007-0330-0/24/05}

\acmPrice{15.00}
\acmISBN{978-1-4503-XXXX-X/18/06}

\begin{document}

\title{Multimodal Healthcare AI: Identifying and Designing Clinically Relevant Vision-Language Applications for Radiology}


\settopmatter{authorsperrow=4}

\author{Nur Yildirim}
\authornote{Work done as a research intern at Microsoft Health Futures.}
\authornotemark[0]
\affiliation{%
  \institution{Carnegie Mellon University}
  \city{Pittsburgh}
  \country{USA}}
\email{yildirim@cmu.edu}

\author{Hannah Richardson}
\affiliation{%
 \institution{Microsoft Health Futures}
 \city{Cambridge}
 \country{UK}}

\author{Maria T. Wetscherek}
\affiliation{%
 \institution{Cambridge University Hospitals}
 \city{Cambridge}
 \country{UK}}

\author{Junaid Bajwa}
\affiliation{%
 \institution{Microsoft Health Futures}
 \city{Cambridge}
 \country{UK}}

\author{Joseph Jacob}
\affiliation{%
 \institution{University College London and UCL Hospitals}
 \city{London}
 \country{UK}}

\author{Mark A. Pinnock}
\affiliation{%
 \institution{University College London}
 \city{London}
 \country{UK}}

\author{Stephen Harris}
\affiliation{%
 \institution{University College London and UCL Hospitals}
 \city{London}
 \country{UK}}

\author{Daniel Coelho de Castro}
\affiliation{%
 \institution{Microsoft Health Futures}
 \city{Cambridge}
 \country{UK}}

\author{Shruthi Bannur}
\affiliation{%
 \institution{Microsoft Health Futures}
 \city{Cambridge}
 \country{UK}}

\author{Stephanie L. Hyland}
\affiliation{%
 \institution{Microsoft Health Futures}
 \city{Cambridge}
 \country{UK}}

\author{Pratik Ghosh}
\affiliation{%
 \institution{Microsoft Health Futures}
 \city{Cambridge}
 \country{UK}}

\author{Mercy Ranjit}
\affiliation{%
 \institution{Microsoft Research India}
 \city{Bengaluru}
 \country{India}}

\author{Kenza Bouzid}
\affiliation{%
 \institution{Microsoft Health Futures}
 \city{Cambridge}
 \country{UK}}

\author{Anton Schwaighofer}
\affiliation{%
 \institution{Microsoft Health Futures}
 \city{Cambridge}
 \country{UK}}

\author{Fernando Pérez-García}
\affiliation{%
 \institution{Microsoft Health Futures}
 \city{Cambridge}
 \country{UK}}

\author{Harshita Sharma}
\affiliation{%
 \institution{Microsoft Health Futures}
 \city{Cambridge}
 \country{UK}}

\author{Ozan Oktay}
\affiliation{%
 \institution{Microsoft Health Futures}
 \city{Cambridge}
 \country{UK}}

\author{Matthew Lungren}
\affiliation{%
 \institution{Microsoft Health \& Life Sciences}
 \city{Palo Alto}
 \country{USA}}

\author{Javier Alvarez-Valle}
\affiliation{%
 \institution{Microsoft Health Futures}
 \city{Cambridge}
 \country{UK}}

\author{Aditya Nori}
\affiliation{%
 \institution{Microsoft Health Futures}
 \city{Cambridge}
 \country{UK}}

\author{Anja Thieme}
\affiliation{%
 \institution{Microsoft Health Futures}
 \city{Cambridge}
 \country{UK}}
\email{anthie@microsoft.com}

\renewcommand{\shortauthors}{Yildirim, et al.}

\begin{abstract}
Recent advances in AI combine large language models (LLMs) with vision encoders that bring forward unprecedented technical capabilities to leverage for a wide range of healthcare applications. Focusing on the domain of radiology, vision-language models (VLMs) achieve good performance results for tasks such as generating radiology findings based on a patient's medical image, or answering visual questions (e.g., ``Where are the nodules in this chest X-ray?''). However, the clinical utility of potential applications of these capabilities is currently underexplored. We engaged in an iterative, multidisciplinary design process to envision clinically relevant VLM interactions, and co-designed four VLM use concepts: Draft Report Generation, Augmented Report Review, Visual Search and Querying, and Patient Imaging History Highlights. We studied these concepts with 13 radiologists and clinicians who assessed the VLM concepts as valuable, yet articulated many design considerations. Reflecting on our findings, we discuss implications for integrating VLM capabilities in radiology, and for healthcare AI more generally.
\end{abstract}



\maketitle

\section{Introduction}
Artificial Intelligence (AI) is increasingly recognized as an important application in radiology~\cite{huang2023generative, liu2023exploring, patel2019human, strohm2020implementation}. In particular, the latest advancements in the creation and adaptation of multimodal foundation models (e.g., BioViL(-T)~\cite{bannur2023learning, boecking2022making}, ELIXR~\cite{xu2023elixr}, MAIRA~\cite{Hyland2023maira}, Med-PaLM M~\cite{tu2023towards}) invite high expectations of how the use of AI may transform clinical practice through efficiency and quality gains~\cite{strohm2020implementation}; and improved overall patient care. By leveraging rich, multimodal data that particularly characterizes the healthcare domain, advanced AI models can achieve impressive new and improved capabilities. In this work, we focus particularly on the combination of large language models (LLMs) with vision capabilities -- as so called vision-language models (VLMs). In the context of radiology imaging, this modality combination enables tasks such as: automatically generating a radiology report from a medical image (e.g., ~\cite{huang2023generative, Hyland2023maira, yu2023evaluating}); using text queries to answer questions about a radiology image (cf. ~\cite{ xu2023elixr}); or detecting errors in a radiology report text through its comparison with the image.

Despite great AI advances both in natural language processing and image-based analysis, translating recent research and development successes into clinical practice remains challenging~\cite{coiera2019last, galsgaard2022artificial, ontika2023pairads, osman2021realizing, verma2021improving, yu2018artificial, tulk2022inclusion, strohm2020implementation, yildirim2021technical, zajkac2023clinician}. Factors hindering successful AI implementation in radiology are wide ranging and include: skepticism due to inconsistent AI performance; lack of trust and overreliance in AI-generated outputs; 
and the need for clinical effectiveness trials (cf. ~\cite{galsgaard2022artificial}). 
A key underlying factor is uncertainty about the value that AI applications bring to clinical practice. In what has been described as ``a race for getting the technology right before exposing human-end users to new promising AI tools''~\cite{osman2021realizing}, the field of AI has been criticized for its development ``in a vacuum''~\cite{miller2017explainable}, disconnected from well-defined needs of intended users or use contexts~\cite{liao2022connecting, thieme2023foundation}. Seeking to close the gap between technical proof-of-concepts and lab experiments towards the successful integration and deployment of AI-enabled systems within routine care requires the adoption of human-centered, participatory approaches~\cite{ontika2022exploring, thieme2023designing}. This involves engagement with relevant stakeholders throughout AI system development, starting as early as the ideation and problem formulation stages \cite{ismail2018bridging, kross2021orienting, morrison2021social, yildirim2023investigating, wilcox2023ai, cai2021onboarding}.

Within this broader context, we set out to better understand the design space of VLMs in healthcare, specifically in the context of radiology. Radiology imaging workflows involve \textit{referring clinicians} who request an imaging test for a patient; and \textit{radiologists} who examine the image and describe their \textit{findings} and \textit{clinical impression}. The resulting report goes back to referring clinicians to inform patient care and treatment~\cite{langlotz2015radiology}. Building on the recent advances in AI research, we focused on \textit{designing the right thing} \cite{buxton2010sketching}: What might be clinically relevant use cases for VLMs to enhance radiology imaging workflows for radiologists and clinicians? Would radiologists want to engage with a draft report generated by AI? Would clinicians find it useful to have report findings visually annotated on an image? What questions might radiologists and clinicians ask if they could query a patient X-ray or CT scan? 

As a team of human-computer interaction (HCI) researchers, AI researchers, radiologists and clinicians, we engaged in an iterative design process to explore these questions. We conducted a three-phase study. The first phase involved in-depth discussions and \textcolor{black}{brainstorming sessions} within our team to elicit our clinical team members' domain expertise, and ideate use cases with VLM capabilities. We discussed how radiologists interpret images and write reports, and how clinicians review these to make patient care decisions. We brainstormed VLM-based interactions using sketches, scenarios and wireflows to identify what would be useful and acceptable. In the second phase, we selected four specific use cases to further detail as design concepts: \textit{Draft Report Generation, Augmented Report Review, Visual Search and Querying}, and \textit{Patient Imaging History Highlights}. In the third phase, we recruited 13 radiologists and clinicians to conduct user feedback sessions probing whether and how these concepts might be useful for clinical practice, and potential concerns. 

Overall, participants perceived the VLM concepts as valuable, but articulated many design requirements for these to be usable and acceptable. Particularly, they shared expectations of AI performance, workflow integration (e.g., well-defined, tool-based interactions rather than open-ended queries), and a desire for context-specificity.

\textcolor{black}{This paper makes two main contributions. 
First, we identify and design VLM use cases to support radiology workflows, and offer initial insights into the perceived value of these concepts. Second, we present a reflective account of our design process as a case study of early phase AI innovation with clinical stakeholders, from brainstorming to prioritization, concept generation and initial assessment.}
We discuss the design implications and future research directions for integrating VLM capabilities into radiology, and healthcare more generally.

\section{Related Work}

\subsection{VLMs: Multimodal Foundation Models}
Recently, considerable excitement has developed around a new class of AI models that have been termed foundation models (FMs)~\cite{bommasani2021opportunities}. These models are trained on broad data at \textit{immense scale}, which results in powerful general-purpose models that can be adapted and more flexibly (re)used for a wide range of tasks and domains – including healthcare ~\cite{brown2020language, singhal2023large, thieme2023foundation, wojcik2022foundation}. 
While FMs can in principle be developed for any data modality (text, image, audio, video, etc.) or their combination, we have seen most advances with large language models (LLMs) that demonstrate impressive capabilities to generate coherent human-like text. Prominent examples include OpenAI's GPT-4~\cite{openai2023gpt4} and Google's LaMDA~\cite{Collins_Ghahramani_2021} models that power conversation-based AI innovations such as ChatGPT \cite{chatGPT}, Microsoft's Copilot \cite{copilot}, and Google Bard~\cite{GoogleBard}. 

In its basic function, an LLM generates statistically likely continuations of word sequences~\cite{shanahan2022talking}. For example, given a specific text fragment such as ``Pneumonia is…'' , the LLM may complete this fragment with ``an infection that inflames the air sacs in one or both lungs'', because it is statistically a likely continuation given the words' distribution in the vast collective corpus of human-generated (English) texts that the model was trained on. As a result, latest generations of LLMs (e.g., 
PaLM 2~\cite{anil2023palm}, 
FLAN-T5~\cite{chung2022scaling}, LLaMA 2~\cite{touvron2023llama}
), and especially those trained on additional medical data (e.g., Med-PALM 2~\cite{singhal2023towards}), do not only interpret and respond in plain language. Having clinical representations encoded~\cite{singhal2023large}, they also exhibit a certain `comprehension' in the medical domain~\cite{Corrado_Matias_2023} as illustrated in their ability to correctly answer medical exam questions~\cite{oh2023chatgpt, nori2023capabilities, singhal2023large, singhal2023towards}. As a result, LLMs are being explored for healthcare tasks such as: medical knowledge extraction~\cite{preston2023toward}, literature search and medical article writing~\cite{gu2023distilling, lecler2023revolutionizing, yang2023harnessing}; medical text simplification~\cite{jeblick2022chatgpt} and clinical notes summarization~\cite{krishna2020generating, NablaCopilot_2023, nuanceNuanceMicrosoft}; and as medical question-answering~\cite{singhal2023towards} and chatbot applications~\cite{lee2023benefits}.

Most recently, LLMs are combined with other modalities~\cite{ Corrado_Matias_2023}, such as medical images (e.g., BioViL(-T)~\cite{bannur2023learning, boecking2022making}, MAIRA~\cite{Hyland2023maira}
) to better leverage rich, \textit{multimodal} data that particularly characterizes healthcare; seeking to achieve new, improved or more efficient AI architectures and capabilities~\cite{tu2023towards, xu2023elixr}. Most relevant to our work are AI models that leverage both \textit{radiology images} and their associated \textit{report text} (e.g., ~\cite{Hyland2023maira, xu2023elixr}). Combining an LLM with an image encoder, a vision-language model (VLM) permits tasks such as: automatic generation of report text from a radiology image; 
text-image retrieval (e.g., \textit{Show me examples of left lower lobe pneumonia}); visual question-answering (e.g., \textit{Does the patient have lung nodules or an infection?}); or error detection in reports (e.g., detecting clinical findings in the image that are not reported in the text). 

While prior work suggests the applicability of new AI capabilities as ``AI Mentor''~\cite{xu2023elixr} or ``Autopilot for Radiologists''~\cite{langlotz2019will}, how these could be usefully configured to enrich human-AI radiology workflows warrants further study. 
Furthermore, many critical challenges need to be addressed to ensure  safe and responsible VLM system design for clinical practice: The growing complexity of the underlying AI models makes it difficult, if not impossible, to understand their workings, or recognize when the AI might fail~\cite{bommasani2021opportunities, moor2023foundation,thieme2023foundation}. Other issues include questions around domain specificity and quality of model input data~\cite{gilbert2023large, lecler2023revolutionizing} as well as societal biases that are inherent in that data~\cite{bernhardt2022bias},  which increase with model scale and multimodality~\cite{bommasani2021opportunities}, risking harms by exacerbating health disparities and social inequalities~\cite{bender2021dangers, singhal2023large, thieme2023foundation, verma2021improving}. And while LLMs tend to show robust performance to out-of-distribution cases, they are sensitive to the phrasing of prompts; generate hallucinations~\cite{jeblick2022chatgpt, lee2023benefits}); or give high confidence indications even for wrong results~\cite{sasha2023mind}; posing significant challenges for AI trust and adoption~\cite{gilbert2023large}.

Our work seeks to bring a human-centered approach to the VLM-assisted radiology workflows by engaging radiologists and clinicians in early phase brainstorming and concept development.

\subsection{Human-Centered Medical AI}
Developing AI systems for healthcare is a complex space with many, wide-ranging sociotechnical challenges~\cite{beede2020human, jacobs2021designing, ghosh2023framing, andersen2023introduction, zajkac2023clinician}, spanning: (i) concerns about patient autonomy and ability to explicitly consent or withdraw from healthcare data uses, and its privacy protection in AI development or use~\cite{thieme2020machine, wilcox2023ai}; (ii) investigations into AI workflow integration~\cite{burgess2023healthcare, beede2020human, calisto2021introduction} and how best to configure clinician-AI relationships to effectively empower care providers~\cite{gu2023augmenting, thieme2023designing, hirsch2018s, yang2019unremarkable, yildirim2024investigating}; as well as (iii) challenges around acceptance, trust and adoption of AI insights into clinical practice~\cite{yang2023harnessing, jacobs2021designing, matthiesen2021clinician, sendak2020human, henry2022human}. This is mostly addressed in the field of eXplainable AI (XAI) through research into AI transparency via explanations and other mechanisms to help clinicians contest~\cite{hirsch2017designing} or learn about AI outputs~\cite{cai2019hello} to be able to develop an appropriate mental model of AI capabilities and their limitations. Where uses of AI are especially proposed to support health screening, triage or treatment recommendations, research explores (iv) risks of inequality and unfair discrimination, which extends to clinical trial design~\cite{chien2022multi}
. All this further requires (v) robust evaluation frameworks and carefully defined AI model or system performance metrics~\cite{liang2022holistic}; and is interlinked with (vi) broader organizational challenges and regulatory approval requirements that pose additional questions about clinical accountability and taking medical-legal responsibility for any AI-assisted decision by individual users, healthcare institutions, or insurance providers (e.g., ~\cite{gilbert2023large, petersen2022responsible, procter2023holding, zajkac2023clinician}). 

Within this vast, growing space, our research and design exploration within medical AI imaging (e.g., ophthalmology~\cite{bach2023if, beede2020human}, pathology~\cite{cai2019human, gu2023augmenting, gu2023improving, lindvall2021rapid}), specifically in radiology~\cite{atad2022chexplaining, bernstein2023can, calisto2021introduction, calisto2022breastscreening, calisto2023assertiveness, ontika2023pairads, verma2021improving, xie2020chexplain}, seeks to better understand -- early within AI development processes -- if and how specific, anticipated VLM capabilities could be beneficial in assisting clinical workflows.

\subsection{Designing AI with Domain Stakeholders}

\textcolor{black}{HCI research highlights the difficulty of eliciting input from domain stakeholders in AI design and development, especially in early ideation and problem formulation phases to inform \textit{what is the right thing to design} \cite{piorkowski2021ai, kross2021orienting, delgado2021stakeholder, buxton2010sketching}. Prior work noted that stakeholders with little to no background in data science or AI (e.g., domain experts, UX designers, policymakers, etc) might be involved in the design of an AI system’s user interface, but rarely in conversations around the objective of the underlying model or the overall problem formulation \cite{delgado2021stakeholder, yildirim2023investigating, wang2023designing, robertson2020if, deng2023investigating}. Recently, a growing body of work in HCI and AI has called for human-centered approaches for broadening participation in AI design to meaningfully engage domain stakeholders to brainstorm and reflect on whether an envisioned future technology is in fact addressing the right problem in the first place \cite{delgado2023participatory, kuo2023understanding, corbett2023power, bell2023think, zhang2023deliberating, cooper2022systematic, liao2023ai}.}

However, designing AI-based systems presents unique challenges even for experienced practitioners \cite{yang2020re, subramonyam2022solving, liao2023designerly}.
Recent work investigating best practices for designing AI products highlights that effective innovation teams work with AI capabilities to scaffold cross-disciplinary ideation~\cite{morrison2021social, yildirim2023investigating, yang2019sketching, liu2023human, yildirim2023creating, yildirim2022experienced}. Inspired by matchmaking \cite{bly1999design}, this approach proposes considering user needs \textit{and} AI capabilities simultaneously to explore matches in a problem-solution space.
Several researchers note that these capability abstractions, sketches, and prototypes serve as boundary objects to help bridge the knowledge gap between AI experts and domain stakeholders, allowing domain stakeholders to gain an understanding of what AI can do to articulate their desired futures \cite{yang2023harnessing, cai2021onboarding, ayobi2023computational, yildirim2024sketching}.
Researchers also point out that innovation teams often focus on use cases that require high task expertise (e.g., clinical decision making), where near-perfect AI performance is needed for a concept to be useful \cite{yildirim2023creating, yang2020re, feng2023ux}. Instead, researchers suggest focusing on where moderately performing, imperfect AI can still create value. We draw on this literature to explore \textit{VLMs as a design material} \cite{yang2020re, morris2023design} for radiology workflows, investigating clinically relevant and valuable use cases for radiologists and clinicians.

\textcolor{black}{\section{Overview of Radiology Workflows}}

\begin{figure*}
\centering
  \includegraphics[width=2.1\columnwidth]{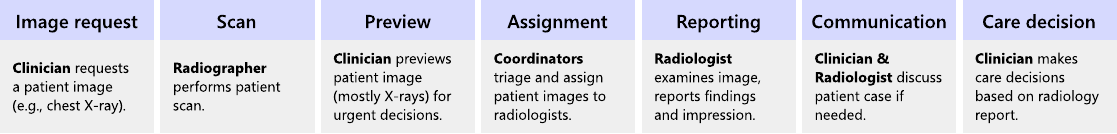}
  \caption{\label{fig:workflow} Overview of the radiology workflow. See Supplementary Material for details on pain points and opportunities.
  \Description{A figure presenting an overview of the radiology workflow and how the work is distributed between clinicians and radiologists. There are seven boxes detailing phases for image request; scan; preview; assignment; reporting; communication; care decision.}
}
\end{figure*}

\textcolor{black}{Radiology workflows unfold across many clinician roles (Figure~\ref{fig:workflow}). First, \textit{referring clinicians} request an imaging study for a patient (e.g., a chest X-ray). Next, \textit{radiographers} perform patient scans, and \textit{radiology coordinators} may prioritize and assign patient images to radiologists. Next, \textit{radiologists} examine patient images, and document their \textit{findings} – descriptions of normal or abnormal observations, such as lesions or nodules – and their \textit{clinical impression} – a summary that synthesizes the findings and suggest possible causes or further tests. Referring clinicians then review the radiology report, and may consult radiologists for further questions or clarifications before making care decisions. In some cases, patient images are brought to multidisciplinary team meetings (MDT) to discuss patient treatment~\cite{langlotz2015radiology}.}

\textcolor{black}{A radiology report (Figure~\ref{fig:report} in the Appendix) typically consists of a \textit{Background} section that describes the patient information and the clinical question that referring clinicians seek to answer, and \textit{Findings} and \textit{Impression} sections that communicates radiologists’ interpretation \cite{kahn2009toward}.
Different imaging modalities have different workflows. For instance, plain (2D) imaging, such as X-rays, are high volume and fast-paced, taking minutes to review \cite{cowan2013measuring}. Complex (3D) imaging on the other hand, such as CTs and MRIs, take more time (10-20 minutes) and cognitive effort \cite{cowan2013measuring}.
Reports are often in the form of prose (sometimes called \textit{narrative report}), while there is also research that calls for structured reporting approaches (e.g., short, bullet-point style sentences) for improved clarity \cite{ganeshan2018structured}.
Reports are usually written using voice dictation, often utilizing templates or draft reports produced by radiology trainees (interns or residents in the US context) in hospital settings.}

\textcolor{black}{Depending on the imaging modality and context, clinicians may review images –especially plain images such as X-rays– before a radiology report becomes available. For example, intensive care physicians immediately review X-rays that are taken to assess if a feeding tube is inserted correctly \cite{law2014radiographers}. Regardless of whether acted upon or not, all images require a radiology report as it serves as a legal document in a patient’s record \cite{clinger1988radiology}. A major challenge within the radiology workflow is the sheer volume of scans, leading to a backlog of unreported images \cite{rimmer2017radiologist}. Wait times might be a few days to a week for radiology reports \cite{naik2001radiology}. In recent years, the majority of radiology services in the UK and the US have been outsourced to private vendors to reduce costs and wait times \cite{rimmer2017radiologist, berry2021high}.}

The majority of human-centered AI research on radiology imaging has focused on mechanisms to explain AI outputs to domain experts ~\cite{atad2022chexplaining, calisto2021introduction, calisto2022breastscreening, ontika2023pairads}, such as explaining the diagnostic outputs for specific chest X-ray findings (e.g., cardiomegaly) by highlighting what feature changes in the medical image would lead the AI system to give a different diagnosis~\cite{atad2022chexplaining}. Other work explored AI acceptance or the impact of using AI systems on radiologist diagnostic performance ~\cite{bernstein2023can, calisto2022breastscreening, calisto2023assertiveness}.
Relatively little work investigated current radiology workflows or asked radiologists where they needed support~\cite{verma2021improving, xie2020chexplain, ontika2023pairads}. Xie et al.'s work presents a rare example of an early phase needfinding and design study, where they conducted a three-phase design process 
to explore opportunities for AI-assisted radiology in the context of X-rays~\cite{xie2020chexplain}.
We build on this existing body of work by investigating radiologists' and clinicians' current needs and desired futures for VLM-assisted radiology workflows.

\hfill 
\section{Method}
As a multidisciplinary team, we engaged in an iterative, reflective design process \cite{zimmerman2007research} to explore \textit{VLMs as a design material} for radiology (cf.~\cite{yang2020re, yildirim2023creating}). \textcolor{black}{We had two high-level research questions: \textbf{(RQ1)} \textit{What might be the clinically relevant use cases for vision-language model capabilities in radiology?} \textbf{(RQ2)} \textit{Whether, how, and in what situations these use cases might provide value for radiologists and/or clinicians?}}
Our three-phase study first included formative work to better understand current radiology workflows and brainstorm VLM use cases. In the second phase, we sketched design concepts for four specific use cases we identified. In the third phase, we sought feedback from 13 radiologists and clinicians outside of our team to investigate if and how these concepts might be useful for clinical practice. Below, we detail the study method and design activities for each phase.

\subsection{Phase 1: Brainstorming VLM Use Cases} 

Phase one included: in-depth discussions to establish an understanding of current radiology workflows within our team, \textcolor{black}{and brainstorming sessions to ideate clinically relevant VLM use cases.}


\subsubsection{In-depth Discussions}
We conducted 7 in-depth discussions with our clinical team members (4 sessions with a cardiothoracic radiologist (R1F); 3 sessions with a general practitioner clinician (C1F)) to form a collective understanding of radiology workflows. Each session lasted 30 mins and was led by an HCI researcher in the form of one-to-one remote, semi-structured interviews. 
Our discussions probed current workflows and pain points through targeted questions, such as: \textit{How do radiologists read a medical image (e.g., an X-ray or CT scan)? How do they describe their findings and impressions in radiology reports?
How do clinicians interact with radiologists to discuss radiology images and reports?} Where possible, clinical team members shared their screen to walk through their process of prioritizing, selecting, and interpreting images, and performing online searches for their information needs. The sessions also involved reviews of VLM capabilities from recent literature (e.g., \cite{singhal2022large, boecking2022making}) to discuss opportunity areas. 

\subsubsection{Brainstorming Sessions}
\textcolor{black}{Following the formative in-depth discussions, we conducted brainstorming sessions to ideate clinically relevant use cases that leverage VLM capabilities. We conducted two one-hour sessions with two groups (four hours in total) involving different team members to previous engagements. Each group consisted of three team members that brought: clinical domain expertise, AI expertise, and HCI expertise. The first group included an intensive care clinician (C2F), an AI researcher, and an HCI researcher. The second group included a cardiothoracic radiologist (R2F), an AI researcher, and an HCI/RAI researcher. Sessions were hybrid (in person + remote) and were facilitated by the same HCI researcher as the initial in-depth discussions.}

Building on the insights from the formative discussions, \textcolor{black}{brainstorming sessions} probed specific VLM capabilities and use cases, such as working with an AI-generated draft list of findings or visually selecting and querying a region on an image. We created sketches, scenarios, and wireframes using a Figma board \cite{figma} to scaffold discussion around each use case. Drawing on each team member's respective expertise, we elaborated on design ideas, discussing the clinical relevance, feasibility, data requirements, data availability, and desired system behavior. At the end of each session, we prioritized and ranked ideas for further development into design concepts.

\subsubsection{Data Collection and Analysis}
\textcolor{black}{We audio and video recorded and transcribed all sessions using video conferencing software. We analyzed the data using a combination of affinity diagramming \cite{martin2012universal}, interpretation sessions \cite{holtzblatt2014field}, and service blueprinting \cite{bitner2008service}. We chose to use affinity diagrams and interpretation sessions --contextual design \cite{karen2017contextual} methods that are commonly used in practice-based HCI research \cite{yang2019unremarkable, yang2023harnessing, holstein2019improving}-- over other data analysis methods such as grounded theory, as our focus was on discovering opportunities for future uses of technology rather than building a detailed theory of current practices and workflows. We reviewed interview and brainstorming session transcripts in interpretation sessions, where the lead researcher retold each session, and the team members built on the insights and pulled out design implications. Using affinity diagrams, we documented key insights, questions, and vignettes capturing our process for exploring this problem-solution space. Through service blueprinting, we traced current workflows capturing how patient images are taken, processed, reported, and reviewed across several clinical roles to inform our understanding.}

\subsection{Phase 2: Sketching VLM Concepts}

\textcolor{black}{Following our formative work that broadly explored VLM opportunities to support radiology workflows, we narrowed our focus to four specific use cases: \textit{Draft Report Generation, Augmented Report Review, Visual Search and Querying}, and \textit{Patient Imaging History Highlights} (Section \ref{phase2} details their rationale).
We translated each use case into design concepts by sketching simple, click-through Figma \cite{figma} prototypes. 
We detailed the use cases based on the scenarios and examples from our brainstorming sessions.
We then populated the prototypes with relevant images and reports from the open source MIMIC-CXR X-ray dataset~\cite{johnson2019mimic}, and placeholders to suggest different image modalities (e.g., CT).
We reviewed and validated the plausibility of each design concept with a radiologist team member.}

\textcolor{black}{Our goal was not to generate fully fleshed out design proposals.
Instead, we wanted the concepts to serve as probes to help clinicians envision possible futures. Therefore, we produced high-fidelity prototypes with only enough detail to probe context-specific questions. While we considered using actual, VLM-generated details in the prototypes (e.g., generated text findings from a radiology image), we concluded that it was irrelevant as the focus of the study was not to evaluate model performance.
Instead, we sought to probe perceived usefulness and clinician acceptance to inform overall system design whilst vision-language models become more capable.}

\subsection{Phase 3: User Feedback Sessions}

\subsubsection{Participants}
\textcolor{black}{We recruited 13 clinical stakeholders across eight hospitals in the UK and the US (5 radiologists, 8 clinicians, 12 male, 1 female) who had not been involved in our design process.}
We contacted an initial set of participants through our collaborating hospitals and our clinician team members' professional connections. We then expanded this set through snowball sampling~\cite{Simkus_2023}, asking each participant to share any contacts with the relevant clinical experience.
Participants represented a range of clinical specialties including: intensive care, emergency care, pediatrics, family medicine, and other domains.
The majority of participants described themselves as `somewhat' or `very familiar' with AI in healthcare.  Table \ref{tab:participants} provides an overview of our participants' clinical roles and experience. 

\subsubsection{Procedure and Data Analysis}
Following the capture of demographic information, our feedback session protocol
was matched to the participant's role, either showing \textit{radiologist} or \textit{clinician} use cases. 
\textcolor{black}{We then investigated the four design concepts, probing the perceived usefulness along with context-specific questions for each concept (detailed in Section~\ref{designconcepts}).} 
Each feedback session lasted 1 hour and was conducted remotely via video conferencing software. We audio and video recorded the sessions. Audio recordings were transcribed using automated transcription and corrected manually by the lead researcher. The data was analyzed using affinity diagramming~\cite{Dam_Siang_2022} to iteratively generate codes for participant utterances, which were then synthesized into high-level themes related to specific use cases; including concerns and desires for additional support.

The study was approved by our institutional ethics review board (IRB: R\&CT 6532, ERP 10690). Informed consent was sought in writing prior to the feedback session. All participants received a £50 (or equivalent) Amazon gift voucher to compensate for their time spent in contributing to the research. Each participant has been given a unique identification number to protect their anonymity, reported as R1-R5 for radiologists and C1-C8 for clinicians.

\begin{table}
  \caption{Participants in user feedback sessions. `Consultant’ denotes a senior doctor with specialist training (the equivalent title in the US is ‘physician’.) (*) denotes clinical trainees (interns or residents in the US context).}
  \label{tab:participants}
  \begin{tabular}{llll}
    \toprule
    ID&Professional Role&Exp.&AI Familiarity\\
    \midrule
    R1 & Emergency Care Radiologist & 12yr & Very familiar\\
    R2 & Pediatric Radiologist & 15yr & Very familiar\\
    R3 & Uroradiologist & 10yr & Somewhat fam.\\
    R4* & Gastrointestinal Radiologist & 4yr & Somewhat fam.\\
    R5 & Cardiothoracic Radiologist& 10yr& Very familiar\\
    C1 & Intensive Care Consultant & 10yr & Very familiar\\
    C2* & Intensive Care Fellow & 1.5yr & Somewhat fam.\\
    C3 & Intensive Care Consultant & 8yr & Very familiar\\
    C4 & Public Health Physician& 11yr & Somewhat fam.\\
    C5 & Internal Medicine Consultant& 7yr& Somewhat fam.\\
    C6 & Cardiothoracic Consultant& 20+yr& Not familiar\\
    C7 & Consultant Oncologist& 20+yr& Very familiar\\
    C8 & Pediatrician& 19yr& Somewhat fam.\\
  \bottomrule
\end{tabular}
\end{table}

\subsection{Study Limitations}
\textcolor{black}{Our study has three major limitations: the study sample, design instantiations, and the scope on clinician acceptance and desirability.}

As mentioned in other HCI healthcare work, recruiting healthcare experts, who are extremely busy professionals presents a challenge~\cite{verma2021improving}. As such, our participants present a convenience sample of UK or US-based individuals, who we either collaborated with previously or who were suggested to us through our clinician contacts. The sample is also biased towards more senior clinicians, higher levels of familiarity with AI, and included only one female. Predominantly, interviewees also had a dual clinical care and academic role, which suggests likely differences in AI expectations as well as experiences to those working in private practice care.

The examples used in the designs evolved from our formative brainstorming work and were reviewed by a radiologist collaborator. Nonetheless, the feedback given on these non-functional prototypes remains speculative; suggesting a need for further interaction design and in-situ workflow integration to substantiate, test and challenge the insights and assumptions that are presented in this work.

Our work is also limited in its particular focus on identifying clinically relevant uses and potential benefits of VLMs for radiology with little insight into their concrete risks and limitations. While we surfaced preferences and initial requirements for those designs and probed into the potential acceptance of, or readiness to correct AI errors in different scenarios, more work is needed into AI risks and strategies for their mitigation to ensure their responsible use in healthcare.

\section{Phase 1: Brainstorming VLM Use Cases} \label{phase1-findings}
\textcolor{black}{Our discussions and brainstorming sessions surfaced many challenges, ranging from requesting a patient scan to prioritization, reporting, and assessment. Our team generated many ideas for improvement (some of which are discussed in prior literature \cite{sectra2013}), such as detecting redundant scan orders; detecting poor quality images at the time of scan to reduce rescans; and optimizing image triage and assignment based on patient urgency and provider subspeciality. We provide a broad overview of these challenges and opportunities using a customer journey map of the radiology workflow (see Supplementary Material).}

\textcolor{black}{In this section, we detail our insights into VLM-specific use cases, mainly around radiology reporting and report review, as our focus was on probing the potential utility of VLM capabilities to support radiologists and clinicians. Where relevant, we provide direct quotes from our clinical team members that were involved in in-depth discussions (R1F, C1F) and brainstorming sessions (R2F, C2F) – denoted with \textit{F (formative study)} to distinguish clinical team members from the user feedback study participants.}

\subsection{Use Cases for Draft Report Generation} \label{reportgeneration}
In considering how VLM capabilities can support radiology image review and reporting, we discussed whether an AI-generated draft report might provide any value. Interestingly, our radiology team members likened these to reports they receive from their trainees: \textit{``I would treat it as a draft report coming from my trainee.'' (R2F)}R2F touched on the difference between draft and preliminary reports, noting that only senior radiology trainees were allowed to make a report `prelim' – which would be available to the clinical team, and would later get `amended' by senior radiologists for any changes. 

This insight led to a detailed discussion on how radiologists currently review, edit, and sign draft or preliminary reports. R1F shared that he looked at the indication (why the request was made) and the image first to form their opinion before looking at the impression, whereas R2F preferred to immediately review the indication and the impression to decide whether she agrees or disagrees. As to how much effort was involved in reviewing and editing these reports, R2F shared: \textit{``Junior trainees' reports will require more work. Depending on how good it is, I might dictate from scratch … Senior trainees, I usually look at [their reports] and sign. I'll just say `I agree'. I'm not going to correct a typo. I might do small edits to say `there is also this' … If I disagree, I will say ``My interpretation is this…'' I will dictate if it's a few sentences or type a few words here and there.''}

Throughout our discussions, we repeatedly asked: What makes `a good AI experience’ in radiology? Elaborating on what makes a radiology report `good’, we teased out three aspects: the report is (1) accurate (i.e. findings are correct); (2) complete (i.e., there are no missing findings); and (3) error-free (i.e. report does not have typos).
\textcolor{black}{This led us to further probe the value proposition AI might bring into radiology in the form of improved report quality and reduced reporting time. Radiology team members pointed out that they often prioritize speed over quality; they had to work really quickly due to the large number of images waiting to be reported. A team member asked whether AI-generated findings in the form of bullet points would provide any value if radiologists still had to dictate the report by themselves (to reduce the risk of errors). Radiology team members pushed back, noting that the system would not save them time in reporting, thus it would provide little value.
They recalled instances where the voice recognition system introduced transcription errors, and stressed that they do not want to spend additional time correcting an AI system’s errors:} \textit{`[recounting an incorrect transcription of `abdominal viscera' as `animal viscera'] It was embarrassing. It should be able to correct these, so that I can sign without having to read what I dictated.'' (R2F)} These discussions hinted at time savings as a key design requirement for clinician acceptance.

Finally, our conversations brought up the questions: Should a draft report be shown to clinicians?
R2F reflected that this may lead to tensions in terms of responsibility and radiologist acceptance: \textit{``There is an issue of responsibility. Radiologists might think they're out of the loop'' (R2F)}. Both clinicians and radiologists proposed that AI-generated findings could be used for triage and early flagging of critical findings without presenting too much detail. This became one of the central themes of exploration in our later study.

\subsection{Use Cases for Visual Search and Querying} \label{VQA}
When reviewing visual question-answering capabilities, both clinicians and radiologists brought up that \textcolor{black}{they regularly perform web searches to look for similar images or clinical information relevant to the patient case. These included medical databases and clinical guidelines (e.g., nice.org.uk – The National Institute for Health and Care Excellence guidelines), as well as websites that provide peer-reviewed patient cases (e.g., gpnotebook.com, radiopaedia.org, radiologyassistant.nl, uptodate.com). R2F described two scenarios where searching similar images was helpful. The first case included situations where she would suspect that there is a pattern in the patient image, but cannot be sure what anomaly it might be: \textit{``I know there is a pattern but I don't know what it is.''} She would use search queries that described the pattern (e.g., glass opacities CT lung) to find similar images to help with diagnostic assessment. The second case was having diagnostic uncertainty about the suspected pattern: \textit{``I think this is crazy paving, but I haven't seen crazy paving in a while.''} She would search for a certain pattern in trusted websites (e.g., \textit{``crazy paving chest ct radiopaedia''}) to see examples of that particular pattern to help disambiguate possible interpretations.}

Both radiologist and clinician team members indicated forming search queries with the abnormality and imaging modality to find similar cases with an overview of pathologies listing common causes: \textit{``I'll look at the differential diagnoses [listed] … [which makes me think] I haven't considered that, but knowing what I know about the patient, yeah that makes sense.'' (R2F)} \textcolor{black}{We discussed how radiologists might perform visual searches if they had the ability to query a region in a patient image, for instance, drawing a bounding box and typing `is this normal or abnormal'} (image query, text query, or image and text query). R1F shared that text query might be preferable: \textit{``I would prefer text, because if I'm selecting a lump, anything might look like a lump.''} R2F however preferred the following search query type: \textit{``If I could snip a region ... so that I don't have to translate that to a text query.''}\textcolor{black}{; suggesting variations in search preferences}.

Our discussions also touched on clinician-radiologist interactions, and the types of questions asked. Clinicians shared that they might ask clarifying questions for less visible findings: \textit{``You said in the image [there is this] ... Where is it? Is this normal?'' (C2F)} Both radiologists and clinicians noted that image annotation tools were part of the reporting software, yet were rarely used. Clinicians also sought information on next steps: \textit{``Do you think we need to act on this? What [additional] imaging should we order? Who should we call about this?'' (C2F)} Radiologist team members shared that such clarification interactions can be overwhelming: \textit{``Sometimes clinicians want to hear from their favorite radiologists that they've built a trust relationship over the years, which can be overwhelming for the radiologist.'' (R2F)} We discussed that visual annotations and image search capabilities might reduce some of the back and forth.

\subsection{Use Cases for Longitudinal Imaging} VLM capabilities enable the comparison of a patient's prior images for longitudinal assessment, a core practice in radiology reporting \cite{aideyan1995influence, sherry2022acr}. Reflecting on situations where this capability could be useful, R2F spoke of the challenge of tracking the size of nodules over time: \textit{``It might look like the size hasn't changed much [compared to the most recent image], but actually it's grown 5 millimeters compared to two years ago.''} We envisioned that a system could summarize past images and reports to provide key highlights, such as chronic events, operations, and the trajectory of abnormalities.

\section{Phase 2: Sketching VLM Concepts} \label{phase2}

We identified four VLM use cases to further design and investigate:

\begin{enumerate}
    \item Draft Report Generation \textit{(radiologist only)}
    \item Augmented Report Review \textit{(clinician only)}
    \item Visual Search and Querying \textit{(clinician \& radiologist)}
    \item Patient Imaging History Highlights \textit{(clinician \& radiologist)}
\end{enumerate}

This section details our design goals and strategies in selecting each of these VLM use cases, and elaborates their design.

\subsection{Concept Prioritization} 
To select the use cases, we focused on concepts that (i) \textit{leverage VLM capabilities that combine radiology image-text pairs} as recently exemplified by~\cite{bannur2023learning, boecking2022making, xu2023elixr}. These technical works commonly propose capabilities for tasks such as: visual question-answering, text-image retrieval and report generation. Furthermore, we sought to (ii) \textit{cover a breadth of required task expertise-AI performance} within the design space. Recent research highlighted that innovators mainly focus on use cases that require high-expertise and near-perfect AI performance (e.g., clinical decision making), yet should investigate beyond~\cite{yildirim2023creating}. In those use cases, requirements for high AI performance are bound-up with greater risks if AI comes to fail – both for patient care and clinician acceptance. We therefore deliberately included use cases that help us explore where lower, yet `good enough' AI performance may still provide utility (e.g., visual search, summarizing prior patient reports) in addition to more common higher-risk, higher-value proposals (e.g., report generation).  

\begin{figure*}
\centering
  \includegraphics[width=2\columnwidth]{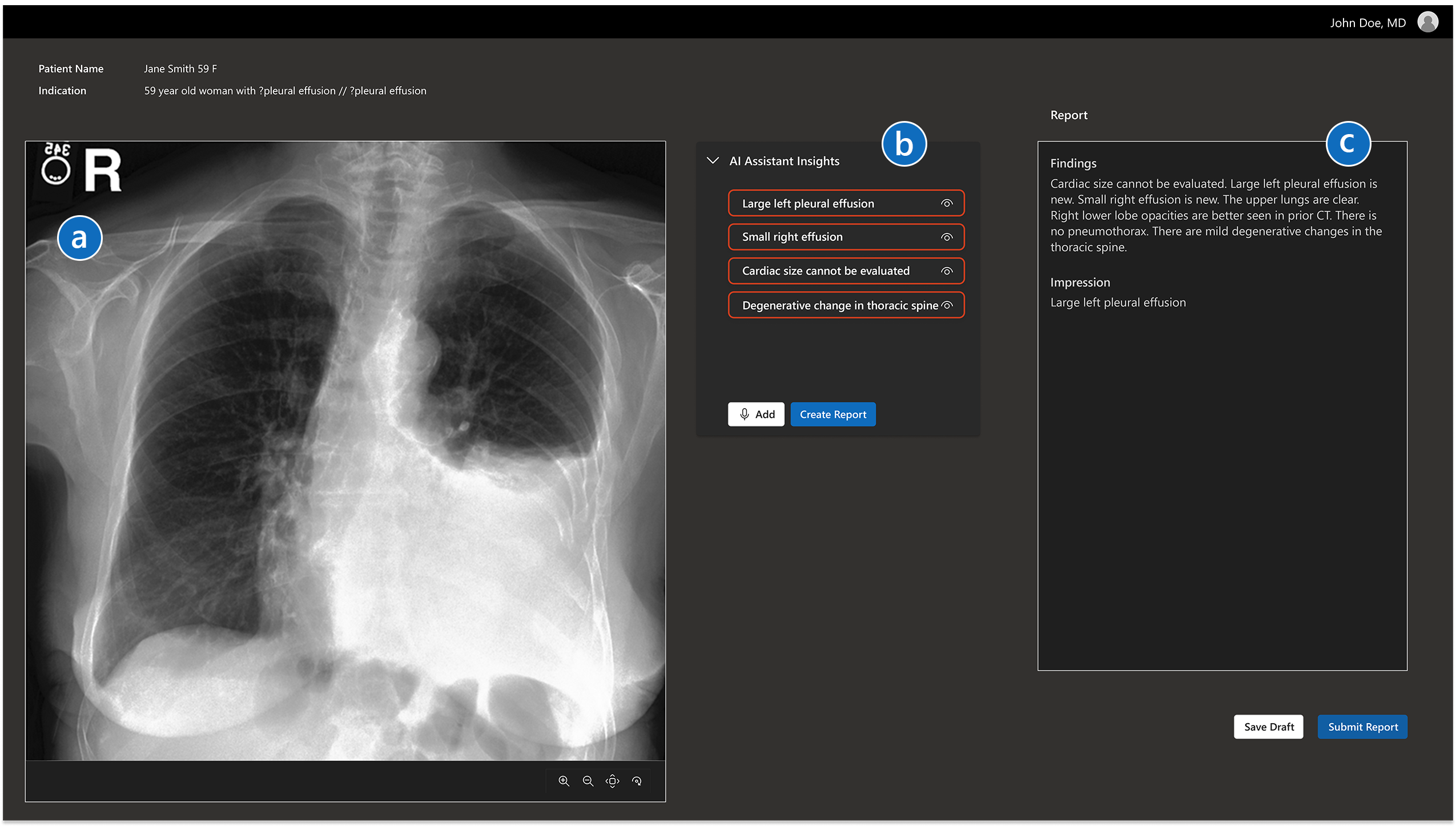}
  \caption{\label{fig:concept1} The Draft Report Generation \textcolor{black}{(radiologist only)} concept displayed (a) a chest X-ray image with patient information and clinical indication, (b) an AI-generated report in bullet point form, and (c) a narrative report created using the bullet points.
  \Description{A prototype displaying the Draft Report Generation concept with a chest x-ray image on the left, bullet point report findings in the middle, and prose report on the right.}
}
\end{figure*}

\begin{figure*}
\centering
  \includegraphics[width=2\columnwidth]{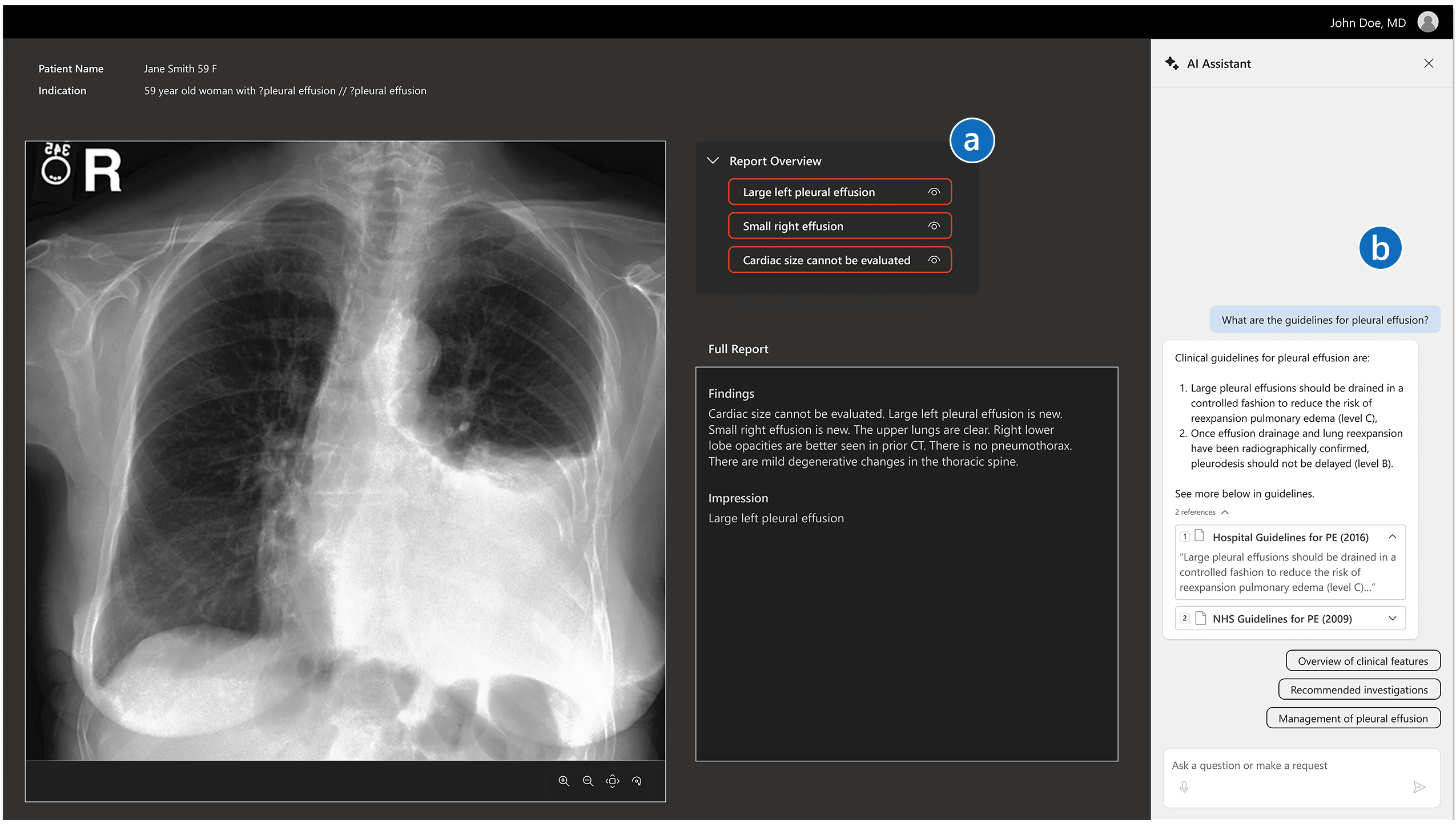}
  \caption{\label{fig:concept2} The Augmented Report Review \textcolor{black}{(clinician only)} concept displayed (a) a report overview feature above the full report, and (b) an AI assistant feature.
  \Description{A prototype displaying the Augmented Report Review concept with a chest x-ray image on the left, report overview section with bullet point findings in the middle, and an AI assistant section with several prompts on the left. A selected prompt displays a result for the query ``What are guidelines for pleural effusion?''}
}
\end{figure*}

\begin{figure*}
\centering
  \includegraphics[width=2\columnwidth]{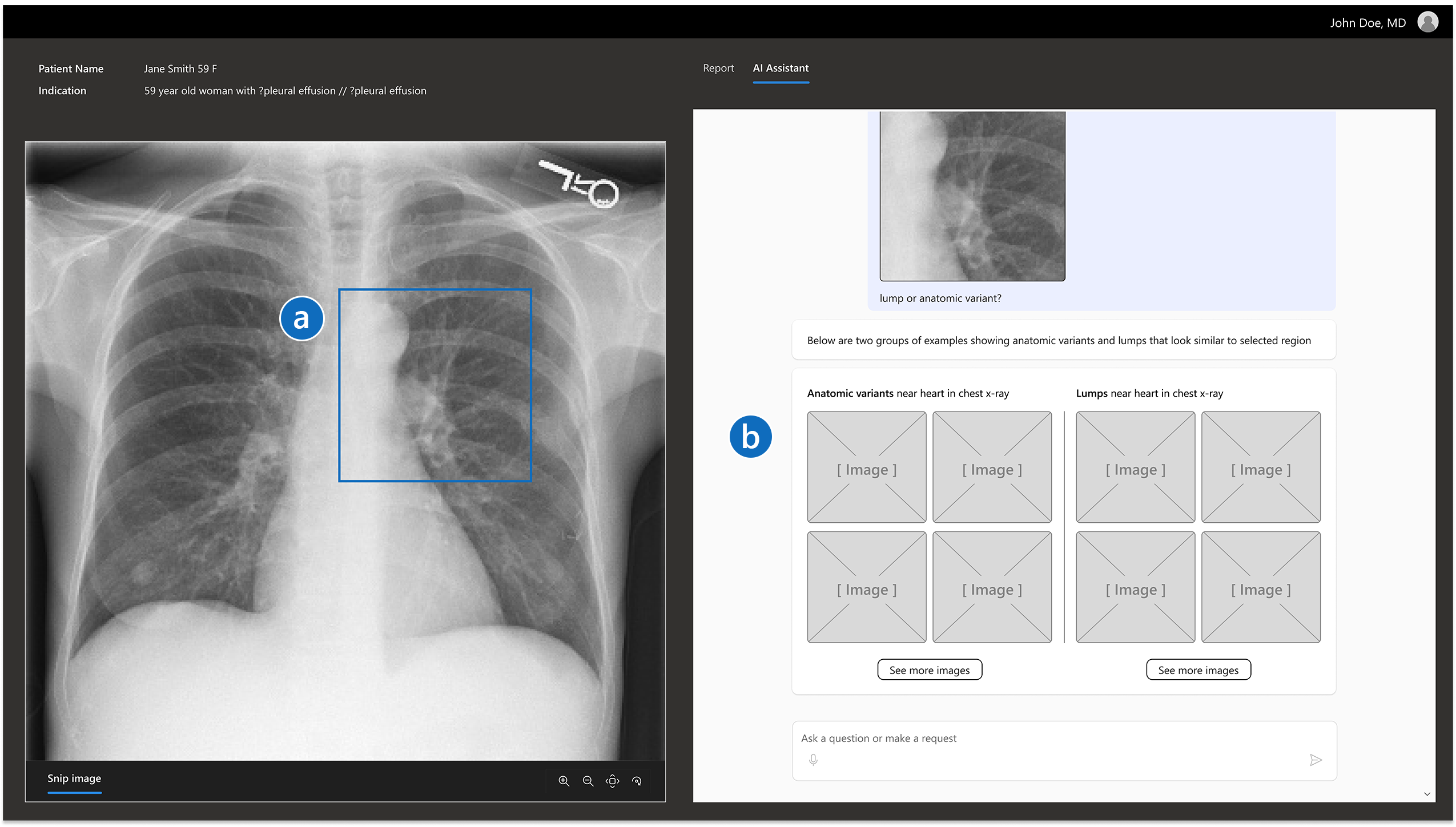}
  \caption{\label{fig:concept3} The Visual Search and Querying concept displayed (a) a visual selection tool that enabled image search or image and text queries, (b) an AI assistant that returned query results \textcolor{black}{without providing an interpretative answer}.
  \Description{A prototype displaying the Visual Search and Querying concept with a chest x-ray image on the right that has a selected region, and an AI assistant on the right that returned query results for the selected region showing similar images diagnosed as anatomic variants versus similar images diagnosed as lump.}
}
\end{figure*}

\begin{figure*}
\centering
  \includegraphics[width=2\columnwidth]{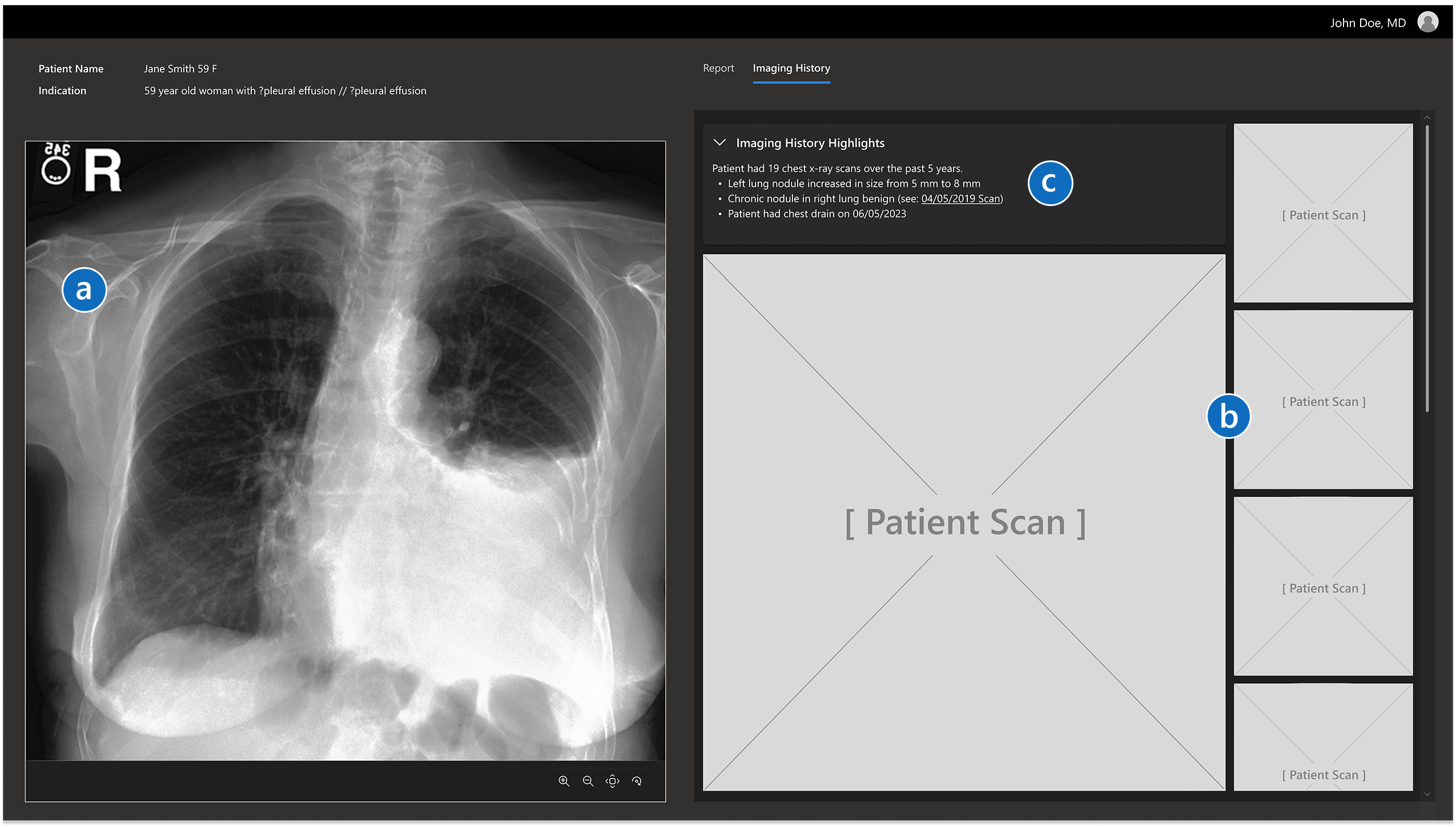}
  \caption{\label{fig:concept4} The Patient Imaging History Highlights concept displayed (a) a new X-ray scan, (b) prior patient images, and (c) an AI-generated summary of prior images and/or reports.
  \Description{A prototype displaying the Patient Imaging History Highlights concept with a chest x-ray on the right, several past patient images as thumbnails on the left, and an AI-generated short summary at the top.}
}
\end{figure*}

\subsection{Design Concepts} \label{designconcepts}

Below we provide an overview and the rationale behind each concept, and enlist the research questions they sought to explore. Click-through prototypes are further illustrated in the Appendix.

\subsubsection{Draft Report Generation}
Motivated by the insight that radiologists are accustomed to working with draft reports from their trainees, the first concept explored the idea of an AI-generated radiology report as a `draft'. The Draft Report Generation concept (Figure~\ref{fig:concept1}) displayed (a) a chest X-ray image with patient information and clinical information, (b) an AI-generated report in short sentence form, and (c) a narrative report created using the short form report. It demonstrated a scenario where the radiologist could review the findings to see annotations in the image, and edit the draft in short form (e.g., crossing out, editing, or adding bullet-point style sentences). The short form text -- illustrated as bullet-points -- was sought to assist in spotting mistakes and enables linking the outputs to source materials (e.g., referencing to previous scans or reports, localizing text findings in the image). Our goal was to explore the balance between \textit{introducing friction} and slowing down radiologists by having them verify the report, and yet still have them achieve time savings overall.

Some concept details were intentionally left open to interpretation. For instance, the prototype did not detail whether only certain parts (e.g., findings, impression) or the entire report should be generated as draft. We also did not list the whole report in short form; instead we floated the idea of listing abnormalities to report what remains as `normal' (a common practice when using templates). 

The concept aimed to explore the following questions: (1) When and how would radiologists want to interact with an AI-generated draft report, if at all? (2) Could there be utility to having a short form report (e.g., bullet points)? (3) Should the draft report be available to clinicians? If so, in what level of detail? (4) What is considered as `good enough' AI performance for draft reports to be useful?






\subsubsection{Augmented Report Review}
Based on our insights on clinicians' information needs, we explored how their review of radiology reports could be augmented with VLM capabilities. The Augmented Report Review concept (Figure~\ref{fig:concept2}) had two main features: (a) a report overview feature shown above the full report, and (b) an AI assistant feature. The report overview displayed a list of abnormal findings extracted from the report that can be visually highlighted in the patient image to facilitate its localization (e.g., \textit{large right pleural effusion}). The AI assistant showcased numerous prompts inspired by clinician questions (e.g., \textit{Given the image-based findings, what are the clinical guidelines for pleural effusion?}). For this concept, a critical design consideration was around latency: vision-language models are currently slow and costly. 
We speculated whether contextual queries can be pre-run prompts, where answers could be displayed immediately (see Appendix). 
As an alternative, we also sketched the AI assistant feature as a chatbot with a text input field to provide contrasting options. The prototype displayed example prompts (e.g., guidelines, suggested investigations) as conversation starters to help clinicians envision what might be useful. 

The concept aimed to explore: (1) Would clinicians want to review AI-generated annotations? If so, which findings are helpful to highlight for different image modalities (e.g., CT)? (2) Could there be any utility to having contextual information when reviewing images? (3) What would clinicians query? What would they never query? (4) Would there be a need for follow up queries (e.g., a chatbot style interaction that can maintain context)?


\subsubsection{Visual Search and Querying}
Building on the insight that radiologists and clinicians perform image searches online, the Visual Search and Querying concept (Figure~\ref{fig:concept3}) explored potential utility by displaying: (a) a visual selection tool that enabled image search (e.g., \textit{Find similar images that look like this region}) or image and text queries (e.g., \textit{``Is this lump or anatomical variant?''}). In line with recent literature showing clinicians look for evidence rather than explanations~\cite{yang2023harnessing}, we envisioned this concept to return groups of similar images instead of providing an interpretative answer (e.g., \textit{``Below are two groups of examples showing anatomic variants and lumps that look similar to the selected region.''}) (Figure~\ref{fig:concept3}b). 

The concept aimed to explore: (1) What would clinicians and radiologists visually query? (2) Could there be utility in performing image \textit{and} text queries? (3) Would clinicians prefer to have an answer along with image examples (e.g., \textit{``Region likely normal''})? (4) What might be the data requirements for finding similar images (e.g., past images and reports from a hospital database)?

\subsubsection{Patient Imaging History Highlights}
Given that clinicians and radiologists commonly review patients' prior images, the Patient Imaging History Highlights concept explored extracting and highlighting key insights across a patient's image history. The prototype (Figure~\ref{fig:concept4}) displayed: (a) a new X-ray scan, (b) prior images, and (c) an AI-generated summary of prior images and/or reports. Example highlights included changes in abnormalities (e.g., \textit{Left lung nodule increased in size from 5 to 8 millimeters}); chronic conditions (e.g., \textit{Chronic nodule in right lung benign, see image reference}); and operations (e.g., \textit{Patient had chest drain on this date}).

The concept aimed to explore: (1) What is relevant to highlight in a prior imaging summary? (2) Would a summary based only on reports be still useful; what is the least AI can do? (3) Would clinicians query prior images? If so, how (e.g., \textit{``Show me only abdomen CTs''})? (4) How would clinicians envision prior imaging summary to best be presented?

\section{Phase 3: Eliciting User Feedback}\label{phase3}
In the third phase, we sought feedback from a broader set of clinicians to understand whether, how and when the VLM-assisted radiology imaging concepts might be useful for clinical practice. This section reports participants' feedback on each design concept, capturing perceived benefits and suggestions for improvement.

\subsection{Draft Report Generation} 
\textbf{Expectation of near-perfect AI performance: }All radiologists expressed that having an AI-generated draft report would be valuable as long as the model performed really well; with high sensitivity and specificity\textcolor{black}{. Describing how AI reporting errors could add burden, one radiologist explained: }\textit{``If it misses something, I've got to say that. If it's false positive, I either have to click to remove it from the report entirely, or I have to change something.''} (R2) \textcolor{black}{To better understand what would be considered as good enough AI performance for this use case, we} asked \textit{``Out of 10 reports, how many are you willing to correct?''}. Almost all replied \textit{``1 out of 10''} (R1, R2, R3) or \textit{``5 to 10 out of 100''} (R5); suggesting the need for near-perfect performance for \textcolor{black}{AI-generated draft reports} to provide real utility. Only one radiologist, a trainee, responded \textit{``3 out of 10''}, noting that the system could make them more confident even if it did not reduce their workload: \textit{``It [would be] getting stuff right enough for me to feel comfortable just to edit the 30\% of cases where it's going to be wrong.'' (R4)}\textcolor{black}{; suggesting potentially added benefits for trainee learning.} 

\textbf{Accounting for fast-paced practice \& high workload:} Echoing our initial findings, radiologists noted that their practice is fast-paced and high volume: \textit{``It is literally going as fast as humanly possible. Scrolling through things, looking at image, saying whatever I can, go over the spellchecks. Make sure I didn't say anything really wrong and then sign and get on the next one. ...  I just need to get my job done fast. I don't get paid more for quality.''} (R2). 
Consequently, participants mainly spoke of value as time savings, especially when reading multi-slice images \textcolor{black}{such as those captured by CT that take significantly longer to review and report than i.e. X-rays, and images} that are outside of their subspecialty (R1, R2, R3, R5): \textit{``I might be a seasoned reporter for lung or cardiac, but as every week it happens, we'll get a neck CT ... when you're not doing it day in day out, it's extremely difficult. You would love an AI which is at least giving you the salient findings.'' (R5)} \textcolor{black}{This suggests a draft report may reduce risks of key clinical observations being missed and could assist with image interpretation confidence. Apart from time savings, participants also mentioned potential benefits in reduced cognitive burden. For simpler X-ray images, R2 for example} mentioned: \textit{``I can do [X-rays] in 10 seconds... [but] there's the cognitive burden. Having to say the words and go through it all is painful.''} R4, who was a trainee, \textcolor{black}{reflected that the main benefit of the system would be reducing reporting time rather than the time spent for image interpretation}: \textit{``Regardless of what the system says, I'm still going to go through my same search patterns for the findings and interpreting those ... the only area where it's going to be saving time is in creating that draft [prose] report because then I don't have to worry about the wording and if I've missed something''}. 

\textbf{Preference for short, standardized reporting:} Interestingly, when probed whether short form sentences could be useful, all radiologists shared that they prefer to work with bullet point style findings instead of prose text. Several participants highlighted the literature on \textit{structured reporting}, which is proposed as a solution for improving report quality and consistency \cite{ganeshan2018structured}:

\begin{quote}

    \textit{``The idea of a narrative report happened in 1898 and we've not moved on from it. It's full of hedging, it's full of weird language that only radiologists use: `likely to be', `cannot exclude'. [This is] what we should be moving away from rather than using the technology to reverse engineer the future into what we got.'' (R3)}
    
\end{quote}

Commenting on how the bullet list findings in the prototype were presented, R1 reflected \textit{``My reporting style is much more telegraphic. So I'll say `large right pleural effusion', that's exactly how I'd phrase. I wouldn't say `there is' or `is seen' or all those kinds of phrases. I don't think [they] are helpful, especially for findings.''} Similarly, R3 advocated for structured findings for consistency and objectivity: \textit{``Rather than saying `suspected mild cardiomegaly', you say `heart is enlarged' or `heart enlarged', which is a statement. It may be right or wrong, but it's objective.''} \textcolor{black}{All these suggest a preference for concise, accurate and consistent reporting over the historic use of more ambiguous prose text, something that AI reporting could assist in standardizing}.

\textbf{Favoring prioritized findings \& confidence indications \textcolor{black}{to assist image interpretation}:} Additionally, radiologists described the benefits of having findings structured by their clinical relevance and the systems' confidence in the generated outputs. For example, a systems capability to compare a current study to a patient's prior image enables ordering report findings by: what is new, what has changed or is unchanged\textcolor{black}{, which gives important context} to aid image interpretation \textcolor{black}{and subsequent clinical action. For example, the sudden `new' appearance of a pneumothorax would require urgent clinical attention whilst a reduction in consolidation in the patients chest upon pneumonia diagnosis may suggest that antibiotic treatment is working}. Furthermore, all participants (R1, R2, R3, R5) suggested having confidence intervals to communicate \textcolor{black}{AI} uncertainty: \textit{``Rather than using `likely to be', `unlikely to be', `possibly' ... `Likely prostate cancer 4 out of 5', [which is] more robust and easier to interpret.'' (R3)} One radiologist suggested displaying the model confidence and ranking findings on this basis: \textit{``[Say for a finding] I don't totally agree, I don't disagree. But if it's confidence is only like 56\%, I'm just going to knock that out.'' (R2)}

\textbf{Impressions present key interpretative work:} While short form, structured reporting was preferred for findings, some radiologists (R1, R3) shared that having unstructured, prose text is more appropriate for the impression section which is the \textit{``non-objective, doctor bit'' (R3)}: \textit{``The main focus of communication between us and the team taking care of the patient is that impression part of the report. So it's really important to me to have that correctly crafted.'' (R1)} R5 reflected that findings could be useful, yet the impression will be more difficult to get right: \textit{``We get a lot of [outsourced] reports from teleradiology, which just tell you what the findings are. A clinician will want to know the clinical impression. ... Is a report better than no report? I think it is fine if it gets the findings right, even if it doesn't do all the synthesis clinically.''} \textcolor{black}{Given the importance of the impression section and its broader interpretative work that may include additional contextual information, the feedback from our participants suggests that clinicians may want to remain in charge of this task; positioning AI's role closer to the extraction of relevant findings from an image rather than its overall clinical interpretation}.

\textbf{Broading uses of (prose) draft reports:} When asked how an AI-generated draft report should be presented, all radiologists suggested having both bullet points and prose report presented together \textcolor{black}{whereby} bullet points \textcolor{black}{serve to assist the} review, and prose for clinical communication: \textit{``I could just get rid of [a bullet point] and it takes it out of the report, that's great. Because editing at that level is so much easier than editing on the report.'' (R2)} A few radiologists noted that a patient-facing report could also be generated based on the list of findings (R1, R3); suggesting additional use cases and user groups. 

In response to making an AI-generated draft report available to clinicians, all radiologists thought the AI-generated report could be useful for triage purposes, especially in situations where clinicians could escalate cases -- as long as it did not look \textit{too final}: \textit{``The subtlety there is that a draft report sounds too final in the health culture. But a `prelim' or a `wet read', that's a very rough, not final thing. The clinicians would take that information and use their judgement to call the radiologist or wait for the report.'' (R2)} Alongside legal,  regulatory and other organizational requirements to approve any such AI \textcolor{black}{use, this requires} a system design that appropriately communicates and \textcolor{black}{clearly} discloses the nature of preliminary AI-generated contents.

\subsection{Augmented Report Review}

\textbf{\textcolor{black}{Locating~image findings \& their prioritization by clinical relevance}: }Exploring how VLM capabilities could be utilized to augment the experiences of clinicians when reviewing the radiology report, all described finding image annotations helpful, especially for complex images like CTs. Most clinicians shared that they do not receive training to read CTs: \textit{``I look at CT scans, but I'm not trained to look at CT scans. I'm trained to look at X-rays.'' (C5)} Some (C3, C6, C7) noted that they are comfortable reading CTs mainly within their subspeciality: \textit{``[In a brain scan] I would 100\% be able to localize where things are. But if it was a report of a liver I would struggle.'' (C7)} They pointed out that for \textcolor{black}{such} multi-slice images, current systems require them to manually navigate to the image slice indicated in the report to view abnormalities. Having ``clickable'' findings, either on the report itself or in an overview section, that would direct them to the image location of relevance, was perceived valuable to save time and make it easier to differentiate what is in the image: \textit{``[Looking at a CT scan that had multiple areas of edema infarction] As a clinician, you're like, well, this must be the bit that's bleeding, but this must be the inflamed bit. But they look similar to me.'' (C1)} Clinicians \textcolor{black}{additionally} described several abnormalities that can be difficult to interpret: \textit{``Lymph nodes are the thing that people often miss on chest X-rays. Small pneumothoraces are difficult to see. The difference between a pneumothorax and a bullae [is] a common problem with the misreading of chest X-rays.'' (C6)} \textcolor{black}{As such, they ascribed value to AI image annotations in aiding their understanding of the reported findings. Furthermore, s}imilar to radiologists' feedback, clinicians  reflected that an overview section could highlight the most important and actionable findings: \textit{``Report overview would work best if you constrain it to show the top 6 salient features. We can get a lot of information overload if there are 25 of them.'' (C7)}

\textbf{Building an appropriate mental model of the AI:} 
\textcolor{black}{When discussing more broadly how AI assistance could feature within workflows, o}ne clinician differentiated for example a radiology assistant from a clinical assistant\textcolor{black}{, whereby the former is embedded within the image viewer for radiology-specific tasks, whereas the latter --which is conceived as answering broader clinical questions-- would be expected to sit within the EHR system}: \textit{``If I've got a radiologist at my fingertips, I'd restrict to asking it the kind of questions I might be asking the radiologist. Therefore it belongs in  \textcolor{black}{[the radiology]} screen, whereas some of the other things like, how should I treat this patient? I think that belongs in the main body of EHR rather than in this radiology reporting system.'' (C4)} \textcolor{black}{ This commentary highlights} the importance of workflow integration for building an appropriate mental model of the AI's likely purpose and capabilities. 

\textbf{Cautioning about chat format \& too complex queries:} In response to the AI assistant embodied as chatbot, several clinicians (C1, C3, C5, C7) commented that they were unlikely to use an assistant in chat form due to time-demands and lack of trust in generated, potentially high-risk responses: \textit{``I don't need a chatbot function where I'm talking and stuff. I haven't got the time for it.'' (C5)} Some clinicians raised concerns about responsibility in clinical decision making: \textit{``I'm not all of a sudden going to ask ChatGPT `What am I going to do with the brain tumor?' I'm going to ask my friend who's a specialist of this. There's a question of responsibility.
'' (C1)} Similarly, in answers to questions what clinicians would not want to use an AI assistant for (whether in chat or any other form), C7 -- an oncologist -- emphasized that he would not use it as a prognostic tool: \textit{``The radiology assistant shouldn't be used to make predictions. It's not a radiomic analysis in that sense.''} Similarly, a cardiothoracic physician indicated that she would not ask what's unknowable: \textit{``You wouldn't ask things that are impossible to know. Things that are too complicated, like [the patient is] on six other drugs, how are they going to interact in combination? I wouldn't bother asking, I wouldn't trust the answer cause it's too individualized.'' (C6)} Another concern was around the reinforcement of radiology observations that present negative findings. Here, clinicians stressed that they weigh positive findings more than negatives: \textit{``[If someone asks] `Can you confirm there really isn't a small pneumothorax on this?' Then the answer from the assistant should be `No, you can't'.'' (C7)} \textcolor{black}{In other words, clinicians cautioned the uses of AI for more ambitious, high-risk VLM use cases involving prognosis, more complex patient cases, or a definite negation of abnormalities -- given more likely chances of errors and their negative implications on patient care.}

\textbf{Focusing on task- and patient-specific, functional queries:} However, clinicians described an array of rather functional, task-specific queries where they could imagine AI to assist by either connecting them to, or extracting information on their behalf. 
For example, clinicians envisioned the AI assistant to perform image-based \textcolor{black}{quantifications} such as calculations of the cardiothoracic ratio (calculated by measuring the maximum diameter of the heart and thoracic cavity); Mirels' score (indicating the risk of bone fracture); sarcopenia index (muscle-fat ratio to track weight loss in cancer patients); and waist-to-hip ratio in CT scans. All of these are currently calculated manually, often using phone apps: \textit{``It would be perceived added value if it could be quickly extracted from [an image] read, as you wouldn't calculate it unless you needed.'' (C7)} \textcolor{black}{In keeping with these more functional tasks, p}articipants often envisioned AI assisting interactions in familiar forms, such as tool buttons, alerts or reminders for specific conditions and workflows\textcolor{black}{; thereby describing expectations of the AI being designed as a workflow tool. One clinician expressed:} \textit{``I almost would want the prompt `Have you thought about this?''' (C5)} \textcolor{black}{whilst simultaneously cautioning that such} prompts could easily become annoying: \textit{``[For guidelines] I want to be able to click [on a finding], guidance, then it searches and brings it up for me. I don't want pop-up fatigue.'' (C5)}

Furthermore, clinicians described how such \textcolor{black}{practical, patient-specific AI functionality could be achieved even more effectively if VLM capabilities were combined with patient EHR data}:

\begin{quote}
    \textit{``You want it to give you, here's their allergies, here's their weight, here's their renal function, here's their swallow plan. Do they have a cannula in place? And here's their other medications that could interact with that medication. If it can pull from the system that type of information, excellent, you're saving me a huge amount of time.'' (C5)}
\end{quote}

Criticizing many of the more generic information that were probed in our concept sketch (e.g., clinical features, differential diagnoses), clinicians emphasized the benefits of including additional \textcolor{black}{EHR} data to provide patient-context relevant information: \textit{``I don't need [it to remind me] the 10 common causes of pleural effusion. What will be really helpful is for it to know that actually in this context, hypothyroidism becomes not the 29th thing, but actually upping [that to] your top five you should be considering ... because this patient's got some other clues or signs.'' (C3)} Similarly, surfacing a patient's eligibility for clinical trials or surfacing \textcolor{black}{specific} hospital or NHS level guidelines were described useful (C1, C2, C5, C6, C7)\textcolor{black}{; re-emphasizing the need for AI information provision to be adapted to each patient's specific context.}  

\subsection{Visual Search and Querying}

\textbf{Aiding interpretation via comparison with relevant patient cases: }All clinicians and radiologists shared that they perform web searches to find similar images, though not too frequently (e.g., 1/week). \textcolor{black}{For this concept, being able} to visually search radiology images and reports within the context of their hospital and patient population \textcolor{black}{was valued the most}: \textit{``Often you look at a CT scan on [internet] and you go `my CT scans don't look anything like that' [because it was a different generation CT scanner]. So it's very important to visualize the abnormality in the context of the type of imaging you would see in your center.'' (C7)} Most clinicians and radiologists wanted to query what is normal, 
\textcolor{black}{or} queries with age and sex: \textit{``Recently we had a big debate: What does a 16 year old thymus look like normally?'' (C6)} \textcolor{black}{An intensive care unit (ICU) clinician also described the difficulty of assessing rare conditions where they overlap with other abnormalities, because such cases are too infrequent and unfamiliar:} 

\begin{quote}
\textit{``Nasogastric (NG) tubes in the wrong place on a chest X-ray on someone in ICU with pneumonia is even less common [than misplaced NG tubes alone]. So people have to simulate abnormalities in their head and compare the X-ray with their simulation. Showing [cases] similar to your patient would be useful.'' (C1)} 
\end{quote}

\textcolor{black}{All this suggests potential benefits of VLM use in retrieving or simulating other patient cases that enable comparative image assessments for either rare and complex (e.g., querying `NG tube' + `pneumonia'), or normal cases to assist interpretation. For such uses, participants again positioned the AI system as a tool for extracting, searching or filtering information rather than as a conversational interface: \textit{``I'd have it as a tool that I can work with, and not conversation.'' (R1)} Describing how they would use queries to refine image search, one clinician added}: \textit{``To then be able to type in pneumonia for example, and then the other [search results] go away. `Just female patients' or `I'm only interested in people over 75'.'' (C7)}

\textbf{\textcolor{black}{AI insights to provide reassurance to `human' interpretation}:} Reflecting on \textit{when} in their workflow \textcolor{black}{visual search and query capabilities} could be useful, some clinicians suggested their use for \textcolor{black}{follow-up questions about the} radiology report: \textit{``Radiologist might have looked at it, but just not commented on it. I just want the reassurance, is that normal or not? Is it a nodule? Is it a mass? Is it a piece of consolidation? Same goes with head scans. Does this look like quite a full brain? Does the patient have hydrocephalus or not?'' (C5)}. \textcolor{black}{Yet, other clinicians reflected that even with AI functionality to retrieve i.e., similar images,} they might still want to ask a radiologist to be assured: \textit{``Would I be reassured if it flashed up a whole load of other people's chest X-rays and said, this was reported as normal and this was reported as normal, for yours is probably normal. I'm not sure that I would, but maybe.'' (C6)} Interestingly, none of the participants expected the system to provide an answer\textcolor{black}{, and preferred example patient cases} to inform their decisions: \textit{``Here's a bunch of pictures, you decide. And that's reasonable, right? I'm not asking some kind of segmentation to then take responsibility for the decisions.'' (C1)}. \textcolor{black}{This suggests preferences for AI use to reassure and aid human image interpretation rather than its use as an interpretative agent in itself.} 

\subsection{Patient Imaging History Highlights}
\textbf{Reducing laborious information gathering:} All radiologists and clinicians highly valued having a summary of a patient's prior images highlighting key events and chronic conditions. Searching through a patient's history was a major part of the clinical workflow. Recognizing the potential for time savings: \textit{``Half of my life is kind of spent chasing notes and pre-existing conditions. A sentence or two, just about the radiology, would save me a lot of time.'' (C1)} Some clinicians (C3, C7) spoke of a time-reward trade-off: \textit{``The problem with image interpretation is, how far back do you look when interpreting for change?'' (C7)} They expressed feelings of guilt as they mostly look through recent reports, but not images, due to lack of time. Radiologists, on the other hand, shared they take a thorough look at past images, yet expressed desires for an automated summary: \textit{``That is a pretty standard practice already for radiologists, but certainly being able to more easily get at that imaging history is going to be a help.'' (R1)}

\textbf{Facilitating relevant patient information access:} Probing what would be useful to highlight, participants mainly described the historical status of the patient, such as the baseline lung architecture before a patient had pneumonia. Examples included past operations (e.g., \textit{Do they have a collapsed lung?}), key events (e.g., \textit{When their pacemaker first appeared or their sternotomy wires first went in?}) and changes in abnormalities (e.g., \textit{New masses, fluid consolidation, rib fractures, are they old or new?}). When asked whether a text summary would be still useful in comparison to more multimodal, VML capabilities (e.g., text summary of key events along with image annotations), most participants commented that linked reports and visual highlights could aid verification: \textit{``If you clicked on it [for it to show you annotated images], then you can corroborate.'' (C6)} However, several participants emphasized that even a text summary would provide an improvement to the current state: \textit{``We would willingly ingest that information even if it was a little bit more clunky.'' (C7)} Finally, a few clinicians (C3, C6) pointed out that unlike radiologists, the interface they use to review prior reports only presents a list view without images. As such, they thought of AI to still be useful if it could point them, at least, to important reports to guide their navigation to the relevant image: \textit{``I have to click on each one individually, wait for it to load. ... Even if I had a little red flag next to it saying `open this one, this has got money in it'.'' (C3)}. This again highlights the prospective utility of AI in surfacing the clinically most relevant insights; and suggests that utility may already be achieved with simpler AI capabilities. 

\section{Discussion}
Aiming to narrow the gap between \textcolor{black}{(multimodal) frontier AI  model} advances and their successful translation into clinical practice~\cite{coiera2019last, galsgaard2022artificial, ontika2023pairads, osman2021realizing, verma2021improving, yu2018artificial, tulk2022inclusion}, our work engaged in early phase design and user research to identify and co-create clinically relevant use cases of VLM capabilities for radiology. Below, we first discuss the findings from our user feedback sessions, detailing design requirements for each use case. We then share our broader reflections on these insights \textcolor{black}{for human-AI interaction design in radiology and healthcare more generally}. We conclude with our thoughts on the design process for ideating, prioritizing, and sketching VLM concepts.

\subsection{Implications for VLMs in Radiology}
Below, we expand on key learnings and critical design considerations that surfaced in our study to inform if, and how, the identified VLM use cases can assist radiologists and clinicians in their work.

\subsubsection{Chat and Question-Answering}
Our findings indicate that clinicians are unlikely to use \textit{conversational} systems within their radiology work, especially to seek clinical decision support (e.g., making prognosis on complex patient cases) \textcolor{black}{due to a lack of time to engage in dialogue and lack of trust in generated responses. Instead, p}articipants in our study expressed a preference for \textit{tool-based} interactions for well-defined tasks. \textcolor{black}{This suggests the need for a broadened focus beyond `chatbots' and `AI agents' as concepts for how we envision AI to benefit clinical workflows; and requires the development of a richer design vocabulary for AI tooling in healthcare}.  
This is not to say that user acceptance \textcolor{black}{of chat interactions} could not evolve as use cases become more concrete and familiar. Furthermore, unlike clinicians who work under time pressure, \textcolor{black}{conversational UIs} might be more useful for patients in interacting with their medical imaging findings. Exploring if, and how, VLM capabilities could support other user groups marks a clear direction for future research. 

\subsubsection{Image-Text Search}
Our study revealed that searching for similar images using credible web sources is a common practice in radiology image interpretation when reviewing cases with uncertainty. Both clinicians and radiologists expressed a desire to be able to search patient images and reports within their internal hospital databases, as images found online were often poor quality and did not reflect the local patient population. Echoing recent literature findings, participants in our study preferred the AI system to provide evidence rather than answers \cite{yang2023harnessing}. That being said, VLMs are unlikely to replace web search: participants emphasized that providing generic information would not be helpful, as \textcolor{black}{such can} easily and reliably \textcolor{black}{be retrieved} through web lookups. Instead, our design exploration focused on \textit{comparative} queries (e.g., \textit{Show me similar images that are pneumothorax versus bullae}\textcolor{black}{; \textit{What is considered normal in the context of a particular age group or rare condition}}), which might allow for more flexible queries than web search. Future work should explore how VLM capabilities lend themselves to unique search, filter and retrieval experiences that go beyond what web search can provide.

\subsubsection{Report Generation}
In line with research on practice guidelines for radiology reporting ~\cite{sherry2022acr}, our findings surfaced the need for more effective, precise articulation of imaging findings. All expressed a preference for short form findings (e.g., bullet points) over prose, calling for more structured representations clearly indicating findings that are new, changed, unchanged, and normal. Interestingly, participants described the impression section as the ``doctor bit''; the more carefully crafted, interpretative piece that may include other contextual information (cf. findings by~\cite{xie2020chexplain}), and that acts as the main communication within the care team. Radiologists regarded the impression section as `harder to get right' compared to the findings, which are based on observations alone.
Interestingly, radiologists seemed comfortable with the idea of making draft reports available to clinicians, as long as the report did not look too final, but was a `wet read’.
These insights \textcolor{black}{suggest that} AI research efforts for generating draft reports could find more acceptance for auto-generated findings, as opposed to an auto-generated impression (cf. ~\cite{ma2023impressiongpt}). Should the entire report be generated as a draft, or only certain parts? How best to juxtapose AI and human generated contents for easy review and editing/correcting? All these remain open questions for future research.

\textcolor{black}{For the draft report generation concept, we took inspiration from current practices whereby draft or preliminary reports by a junior or senior trainee are reviewed and amended if needed by a senior radiologists. The use of AI-generated draft reports brings up many questions in terms of the impact on trainees’ education and learning experience. While AI-generated draft reports may result in time savings, it could take away important opportunities for trainees developing their clinical reporting skills. On the other hand, an AI draft report could also be seen to provide a useful resource to assist trainees' learning journey and confidence as they acquire image interpretation competencies. This design and research space, as well as questions on how to develop an appropriate AI reliance --upskilling trainees without over-dependency on AI-- are complex, and present open challenges for future work.} 

\subsection{Implications for Designing AI in Healthcare}
\textcolor{black}{In this section, we take the learnings from our VLM design explorations to reflect on the question “What makes a good AI experience in healthcare?”} 

\subsubsection{Focusing on clinically useful AI applications}
\textcolor{black}{Above and beyond the vast technical possibilities that new AI foundation models provide, this section describes the necessity of balancing these with clinical utility, risks of AI fallibility, and AI acceptance.}

\textcolor{black}{\textbf{Prioritizing lower risk-medium reward use cases}:} Recent research highlighted that AI innovators mainly focus on use cases that require high task expertise and near-perfect AI performance (e.g., clinical decision making) \cite{yildirim2023creating}. Instead, researchers point to use cases where moderate-performance AI still could be useful in clinical tasks and workflows, such as in triage, workload management, and resource optimization. Echoing this \cite{yildirim2023creating, feng2023ux}, our case study found that clinicians expect a near-perfect performance for use cases that require high clinician expertise (e.g., draft report generation). However, we also identified use cases that required medium-expertise, moderate-performance, yet were perceived as high value. For example, summarizing prior patient reports requires relatively lower expertise –a medical student level task– where having a `good enough' summary could be still more useful than no summary. As argued elsewhere~\cite{jha2016adapting, singhal2023large, bossen2023batman}, this suggests a focus in AI development on simpler, more standardized use cases as a potentially lower risk and more responsible approach when starting to introduce AI innovations into clinical practice.

\textcolor{black}{\textbf{AI as information resource, not clinical decision agent:}} \textcolor{black}{Amongst the many new capabilities afforded by VLMs (and other multimodal FMs) such as uses of natural language conversation; ability to generate or translate content; or predict specific outcomes; we found most uptake for proposals that involved relevant information retrieval or summarization. Notably, participants expressed low acceptance around ambitious, higher-risk VLM concepts, such as generalist medical AI uses for prognosis or to offer counterfactual reasoning due to a lack of trust in generated outputs; and n}one of them characterized our design proposals as clinical decision support that provides clinician interpretations or treatment recommendations. Instead, they recognized the potential value as reduced time and cognitive effort. For instance, draft report generation was perceived as \textcolor{black}{valuable for radiologist time savings (if correct), and for nudging} clinicians to escalate cases to radiologists to seek assistance. \textcolor{black}{Radiologists also described wanting to remain in charge of the `doctor bit' and more complicated image interpretations that are harder for AI to get right.}

\textcolor{black}{\textbf{Empowering humans in their expert work}: As indicated above, instead of applying VLM capabilities as an interpretative  agent, participants proposed their use for more functional,  tedious} tasks such as calculating medical ratios and scores; assessing organ sizes, volumes or density (cf. ~\cite{langlotz2019will, ontika2023pairads}); retrieving EHR patient data; or for administrative support (cf.~\cite{sahniadministrative}). 
\textcolor{black}{Outside these \textit{practical tasks}, AI was positioned as \textit{an information resource for assisting human interpretation} by extracting, highlighting or summarizing clinically relevant observations. Examples include AI use to provide evidence (e.g. via comparative queries and similar image retrieval, or locating image findings in complex CT scans); for ordering report findings by clinical relevance (or urgency); or deriving imaging history highlights. Conceiving of AI’s role as assisting, and ideally empowering, healthcare professionals in their expert work likely also plays a crucial role for its acceptance and clinical adoption.}

\subsubsection{Integrating AI seamlessly into clinical workflows} \textcolor{black}{Intertwined with the above AI use cases are considerations of how best to integrate AI into fast-paced, high caseload workflows to provide utility whilst inviting appropriate frictions to check or correct AI outputs.} 

\textcolor{black}{\textbf{Designing context-specific “workflow tools”}:} \textcolor{black}{Especially in responses to proposals for an `ask me anything' type AI assistant feature, we found that such openness did not help clinicians in forming mental models; it was not clear to them what they could ask, or how the AI assistant would know the answers. Instead, they wanted the AI system to be more specialized and focus on specific tasks within their workflow (e.g., filtering search results, querying past patient reports). This} aligns with user expectations around agent expertise that are well documented in the human-robot interaction literature~\cite{luria2019re, reig2020not}. As a consequence, we suspect that current research aspirations for creating a `generalist medical AI'~\cite{moor2023foundation, tu2023generalist} are unlikely to correctly capture clinicians' mental model, which presents a key consideration for the usability of such AI systems in healthcare. \textcolor{black}{All this suggests that while (multimodal) AI can technically be leveraged for vastly different tasks, it may be beneficial to design them as context-specific workflow tools with clearly defined purposes to increase their understanding and practical utilization.}

\textcolor{black}{\textbf{Demonstrating clear benefits for disruption or change}:} \textcolor{black}{Given the growing complexity of VLMs and other advanced AI models, it is questionable how well, if at all, their workings can be explained, or their likelihood to fail be detected~\cite{thieme2023foundation}. While it is common to think of human-in-the-loop approaches to verify and correct potentially fallible AI outcomes (e.g., review AI-generated draft reports), where such requirements add extra time burden to clinicians, or present tasks that they are less interested to perform, it risks reducing its utility and uptake in practice (cf.~\cite{jacobs2021designing, sekhon2017acceptability, sendak2020human, thieme2023designing}). If} we were to accept that no AI model is always correct, designers need to give more consideration to strategies that enable clinicians not only to \textit{easily verify \textcolor{black}{or quickly correct} VLM outputs} – beyond technical solutions (e.g., self-consistency prompting~\cite{singhal2023large}, LLM-generated explanations~\cite{nori2023capabilities}, correctness predictions ~\cite{kadavath2022language}); but also \textcolor{black}{develop better strategies} to \textit{assist AI output justifications}. Re-visiting notions of AI uses as provision of evidence rather than clinical interpretations, connecting AI outputs to other criteria ‘external’ to the model~\cite{shanahan2022talking} may aid clinicians to triangulate those outputs across different, trustworthy information sources \textcolor{black}{such that they can more effectively be accepted or rejected as part of their work (cf.~\cite{thieme2020interpretability, thieme2023designing}).}

\textcolor{black}{Further, we have to take into account that adoption of new AI-assisted practices can be greeted with reluctance where clinicians are asked to change established practices (e.g., move away from prose sentence dictation). Readiness to adapt work styles to accommodate AI likely requires sufficient benefits of a new approach.}

\subsection{Sketching VLM Experiences}

Research investigating the best practices around designing AI products and services noted emergent approaches that blend human-centered and tech-centered processes, and the use of AI capabilities for sensitizing domain stakeholders \cite{yildirim2023creating, yildirim2023investigating}. Our case study demonstrates that a capability-based approach –starting with both user needs and AI capabilities to find matches in the problem-solution space– proved effective for multidisciplinary brainstorming to identify clinically relevant AI use cases. Moreover, using multiple sketches that are framed as \textit{instantiations of capabilities} rather than concrete design proposals to probe clinicians and radiologists (e.g., \textit{Knowing that AI can do this, can you think of situations where this capability would be useful?}) seemed to work well.

While \textit{sketching} \cite{buxton2010sketching} with VLM capabilities scaffolded ideation, separating the underlying capability from the form was a challenge. For example, the AI literature often uses the term `Visual Question-Answering' to refer to AI tasks around image-to-text or text-to-image capabilities, yet these capabilities do not necessarily require a conversational form. Similarly, we struggled to envision novel VLM interactions that go beyond chatbots, alerts, and recommenders, a well-known challenge in AI design literature \cite{yang2020re}. We approached this challenge by framing VLM capabilities as \textit{queries} that can be formed in different ways (e.g., conversational questions, pre-run prompts, alerts, visual annotations, etc). Interestingly, the way participants described VLM interactions resembled \textit{robotic process automation}: AI that fetches data in the background and presents it in an unremarkable manner \cite{yang2019unremarkable} that can either be included or easily ignored. These findings point to a need for new design patterns beyond current paradigms of LLM or VLM uses as chat or conversational queries -- especially in workflow-oriented contexts.

\textcolor{black}{Additionally, VLMs as a design material centers considerations on balancing a seamless user experience with the time latency and financial costs (e.g., pre-running complex queries on large volumes of data to surface what prompts might be relevant; applying self-verification strategies to reduce risks of AI errors). This may determine choices to prioritize smaller, more efficient VLM models. Furthermore, designers should evaluate whether VLM (or other multimodal AI) capabilities are truly needed and appropriate for a task and explore alternatives. During our user feedback sessions, we probed into VLM boundaries by asking `Can there be simpler, dumber versions of these concepts?' The Patient Imaging History Highlights concept demonstrates this approach well: while a multimodal model can summarize rich patient image-report data; text-only models may already create value by summarizing previous report texts or pointing to important reports without text extraction and summarization.}

Finally, we see opportunities for design research to investigate how to effectively sketch and prototype VLM interactions. In this paper, we utilized click-through sketches to scaffold clinicians' thinking around what AI can do and how the system might behave in specific use cases. While the concepts were used to probe into clinician expectations and AI acceptance, further research is needed to substantiate, test, and challenge our insights. Recent literature highlights \textit{prompt prototyping} as a potential research direction for experience prototyping with LLMs and generative AI capabilities~\cite{valencia2023less, petridis2023promptinfuser, jiang2022promptmaker}. Future work should detail and refine interactions for the identified use cases; and extend insights into challenges of workflow integration, task completion time, as well as error types and their likely implications and mitigation.

\section{Conclusion}
Intersecting the fields of HCI, AI and Healthcare, the work in this paper presents a first investigation into the potential utility and design requirements for leveraging vision-language model (VLM) capabilities in the context of radiology. To this end, we conducted a three phase study that involved brainstorming with clinical experts and sketching four specific VLM use cases.
Our findings from feedback sessions with 13 clinicians and radiologists
provide initial insights into clinician acceptance and desirability for \textcolor{black}{ various identified VLM use cases, and advance this research by capturing nuanced design considerations.} 
\textcolor{black}{Against this backdrop, our work surfaced broader insights and challenges for human centered-AI systems in healthcare. While much emphasis is currently placed on developing more general-purpose AI models that can flexibly be adapted and scaled across different healthcare contexts, our research highlights the importance of bringing new AI capabilities into the focus of specific, practical tasks and use contexts to achieve effective workflow integration and the formation of useful mental models for AI and its intended uses. Notably, we found lesser interest in more ambitious VLM concepts that may offer value in terms of predictive diagnosis or counterfactual reasoning on medical outcomes. Instead, participants positioned AI as being assistive to health experts’ work to help with mundane information extraction and processing tasks; thereby serving as a resource for human interpretation (e.g., by offering access to clinical evidence in a patient-specific manner). We further highlight the various trade-offs that are needed to ensure that AI's utility is balanced with the cost of AI risks; human effort in checking or correcting AI outputs; changes in work practices; as well as AI output latency and compute requirements.}
We encourage HCI researchers to further explore the benefits, risks, and limitations of VLMs in radiology workflows, and healthcare in general.

\begin{acks}
We thank our participants for their time and valuable input, and the reviewers of this paper for their thoughtful feedback. Joseph Jacob was supported by the Wellcome Trust [209553/Z/17/Z] and the NIHR UCLH Biomedical Research Centre, UK.
\end{acks}

\bibliographystyle{ACM-Reference-Format}
\bibliography{radiology}


\begin{thebibliography}{152}


\ifx \showCODEN    \undefined \def \showCODEN     #1{\unskip}     \fi
\ifx \showDOI      \undefined \def \showDOI       #1{#1}\fi
\ifx \showISBNx    \undefined \def \showISBNx     #1{\unskip}     \fi
\ifx \showISBNxiii \undefined \def \showISBNxiii  #1{\unskip}     \fi
\ifx \showISSN     \undefined \def \showISSN      #1{\unskip}     \fi
\ifx \showLCCN     \undefined \def \showLCCN      #1{\unskip}     \fi
\ifx \shownote     \undefined \def \shownote      #1{#1}          \fi
\ifx \showarticletitle \undefined \def \showarticletitle #1{#1}   \fi
\ifx \showURL      \undefined \def \showURL       {\relax}        \fi
\providecommand\bibfield[2]{#2}
\providecommand\bibinfo[2]{#2}
\providecommand\natexlab[1]{#1}
\providecommand\showeprint[2][]{arXiv:#2}

\bibitem[AI(2022)]%
        {chatGPT}
\bibfield{author}{\bibinfo{person}{Open AI}.} \bibinfo{year}{2022}\natexlab{}.
\newblock \bibinfo{title}{chatGPT}.
\newblock
\newblock
\urldef\tempurl%
\url{https://chat.openai.com}
\showURL{%
\tempurl}


\bibitem[Aideyan et~al\mbox{.}(1995)]%
        {aideyan1995influence}
\bibfield{author}{\bibinfo{person}{Uwa~O Aideyan}, \bibinfo{person}{Kevin
  Berbaum}, {and} \bibinfo{person}{Wilbur~L Smith}.}
  \bibinfo{year}{1995}\natexlab{}.
\newblock \showarticletitle{Influence of prior radiologic information on the
  interpretation of radiographic examinations}.
\newblock \bibinfo{journal}{\emph{Academic Radiology}} \bibinfo{volume}{2},
  \bibinfo{number}{3} (\bibinfo{year}{1995}), \bibinfo{pages}{205--208}.
\newblock


\bibitem[Andersen et~al\mbox{.}(2023)]%
        {andersen2023introduction}
\bibfield{author}{\bibinfo{person}{Tariq~Osman Andersen},
  \bibinfo{person}{Francisco Nunes}, \bibinfo{person}{Lauren Wilcox},
  \bibinfo{person}{Enrico Coiera}, {and} \bibinfo{person}{Yvonne Rogers}.}
  \bibinfo{year}{2023}\natexlab{}.
\newblock \bibinfo{title}{Introduction to the Special Issue on Human-Centred AI
  in Healthcare: Challenges Appearing in the Wild}.
\newblock , \bibinfo{numpages}{11}~pages.
\newblock


\bibitem[Anil et~al\mbox{.}(2023)]%
        {anil2023palm}
\bibfield{author}{\bibinfo{person}{Rohan Anil}, \bibinfo{person}{Andrew~M Dai},
  \bibinfo{person}{Orhan Firat}, \bibinfo{person}{Melvin Johnson},
  \bibinfo{person}{Dmitry Lepikhin}, \bibinfo{person}{Alexandre Passos},
  \bibinfo{person}{Siamak Shakeri}, \bibinfo{person}{Emanuel Taropa},
  \bibinfo{person}{Paige Bailey}, \bibinfo{person}{Zhifeng Chen},
  {et~al\mbox{.}}} \bibinfo{year}{2023}\natexlab{}.
\newblock \showarticletitle{Palm 2 technical report}.
\newblock \bibinfo{journal}{\emph{arXiv preprint arXiv:2305.10403}}
  (\bibinfo{year}{2023}).
\newblock


\bibitem[Atad et~al\mbox{.}(2022)]%
        {atad2022chexplaining}
\bibfield{author}{\bibinfo{person}{Matan Atad}, \bibinfo{person}{Vitalii
  Dmytrenko}, \bibinfo{person}{Yitong Li}, \bibinfo{person}{Xinyue Zhang},
  \bibinfo{person}{Matthias Keicher}, \bibinfo{person}{Jan Kirschke},
  \bibinfo{person}{Bene Wiestler}, \bibinfo{person}{Ashkan Khakzar}, {and}
  \bibinfo{person}{Nassir Navab}.} \bibinfo{year}{2022}\natexlab{}.
\newblock \showarticletitle{Chexplaining in style: Counterfactual explanations
  for chest x-rays using stylegan}.
\newblock \bibinfo{journal}{\emph{arXiv preprint arXiv:2207.07553}}
  (\bibinfo{year}{2022}).
\newblock


\bibitem[Ayobi et~al\mbox{.}(2023)]%
        {ayobi2023computational}
\bibfield{author}{\bibinfo{person}{Amid Ayobi}, \bibinfo{person}{Jacob Hughes},
  \bibinfo{person}{Christopher~J Duckworth}, \bibinfo{person}{Jakub~J Dylag},
  \bibinfo{person}{Sam James}, \bibinfo{person}{Paul Marshall},
  \bibinfo{person}{Matthew Guy}, \bibinfo{person}{Anitha Kumaran},
  \bibinfo{person}{Adriane Chapman}, \bibinfo{person}{Michael Boniface},
  {et~al\mbox{.}}} \bibinfo{year}{2023}\natexlab{}.
\newblock \showarticletitle{Computational Notebooks as Co-Design Tools:
  Engaging Young Adults Living with Diabetes, Family Carers, and Clinicians
  with Machine Learning Models}. In \bibinfo{booktitle}{\emph{Proceedings of
  the 2023 CHI Conference on Human Factors in Computing Systems}}.
  \bibinfo{pages}{1--20}.
\newblock


\bibitem[Bach et~al\mbox{.}(2023)]%
        {bach2023if}
\bibfield{author}{\bibinfo{person}{Anne Kathrine~Petersen Bach},
  \bibinfo{person}{Trine~Munch N{\o}rgaard}, \bibinfo{person}{Jens~Christian
  Brok}, {and} \bibinfo{person}{Niels van Berkel}.}
  \bibinfo{year}{2023}\natexlab{}.
\newblock \showarticletitle{“If I Had All the Time in the World”:
  Ophthalmologists’ Perceptions of Anchoring Bias Mitigation in Clinical AI
  Support}. In \bibinfo{booktitle}{\emph{Proceedings of the 2023 CHI Conference
  on Human Factors in Computing Systems}}. \bibinfo{pages}{1--14}.
\newblock


\bibitem[Bannur et~al\mbox{.}(2023)]%
        {bannur2023learning}
\bibfield{author}{\bibinfo{person}{Shruthi Bannur}, \bibinfo{person}{Stephanie
  Hyland}, \bibinfo{person}{Qianchu Liu}, \bibinfo{person}{Fernando
  Perez-Garcia}, \bibinfo{person}{Maximilian Ilse}, \bibinfo{person}{Daniel~C
  Castro}, \bibinfo{person}{Benedikt Boecking}, \bibinfo{person}{Harshita
  Sharma}, \bibinfo{person}{Kenza Bouzid}, \bibinfo{person}{Anja Thieme},
  {et~al\mbox{.}}} \bibinfo{year}{2023}\natexlab{}.
\newblock \showarticletitle{Learning to exploit temporal structure for
  biomedical vision-language processing}. In
  \bibinfo{booktitle}{\emph{Proceedings of the IEEE/CVF Conference on Computer
  Vision and Pattern Recognition}}. \bibinfo{pages}{15016--15027}.
\newblock


\bibitem[Beede et~al\mbox{.}(2020)]%
        {beede2020human}
\bibfield{author}{\bibinfo{person}{Emma Beede}, \bibinfo{person}{Elizabeth
  Baylor}, \bibinfo{person}{Fred Hersch}, \bibinfo{person}{Anna Iurchenko},
  \bibinfo{person}{Lauren Wilcox}, \bibinfo{person}{Paisan Ruamviboonsuk},
  {and} \bibinfo{person}{Laura~M Vardoulakis}.}
  \bibinfo{year}{2020}\natexlab{}.
\newblock \showarticletitle{A human-centered evaluation of a deep learning
  system deployed in clinics for the detection of diabetic retinopathy}. In
  \bibinfo{booktitle}{\emph{Proceedings of the 2020 CHI conference on human
  factors in computing systems}}. \bibinfo{pages}{1--12}.
\newblock


\bibitem[Bell et~al\mbox{.}(2023)]%
        {bell2023think}
\bibfield{author}{\bibinfo{person}{Andrew Bell}, \bibinfo{person}{Oded Nov},
  {and} \bibinfo{person}{Julia Stoyanovich}.} \bibinfo{year}{2023}\natexlab{}.
\newblock \showarticletitle{Think about the stakeholders first! Toward an
  algorithmic transparency playbook for regulatory compliance}.
\newblock \bibinfo{journal}{\emph{Data \& Policy}}  \bibinfo{volume}{5}
  (\bibinfo{year}{2023}), \bibinfo{pages}{e12}.
\newblock


\bibitem[Bender et~al\mbox{.}(2021)]%
        {bender2021dangers}
\bibfield{author}{\bibinfo{person}{Emily~M Bender}, \bibinfo{person}{Timnit
  Gebru}, \bibinfo{person}{Angelina McMillan-Major}, {and}
  \bibinfo{person}{Shmargaret Shmitchell}.} \bibinfo{year}{2021}\natexlab{}.
\newblock \showarticletitle{On the Dangers of Stochastic Parrots: Can Language
  Models Be Too Big?}. In \bibinfo{booktitle}{\emph{Proceedings of the 2021 ACM
  Conference on Fairness, Accountability, and Transparency}}.
  \bibinfo{pages}{610--623}.
\newblock


\bibitem[Bernhardt et~al\mbox{.}(2022)]%
        {bernhardt2022bias}
\bibfield{author}{\bibinfo{person}{M\'{e}lanie Bernhardt},
  \bibinfo{person}{Charles Jones}, {and} \bibinfo{person}{Ben Glocker}.}
  \bibinfo{year}{2022}\natexlab{}.
\newblock \showarticletitle{Potential sources of dataset bias complicate
  investigation of underdiagnosis by machine learning algorithms}.
\newblock \bibinfo{journal}{\emph{Nature Medicine}} \bibinfo{volume}{28},
  \bibinfo{number}{6} (\bibinfo{date}{June} \bibinfo{year}{2022}),
  \bibinfo{pages}{1157--1158}.
\newblock
\urldef\tempurl%
\url{https://doi.org/10.1038/s41591-022-01846-8}
\showDOI{\tempurl}


\bibitem[Bernstein et~al\mbox{.}(2023)]%
        {bernstein2023can}
\bibfield{author}{\bibinfo{person}{Michael~H Bernstein},
  \bibinfo{person}{Michael~K Atalay}, \bibinfo{person}{Elizabeth~H Dibble},
  \bibinfo{person}{Aaron~WP Maxwell}, \bibinfo{person}{Adib~R Karam},
  \bibinfo{person}{Saurabh Agarwal}, \bibinfo{person}{Robert~C Ward},
  \bibinfo{person}{Terrance~T Healey}, {and} \bibinfo{person}{Grayson~L
  Baird}.} \bibinfo{year}{2023}\natexlab{}.
\newblock \showarticletitle{Can incorrect artificial intelligence (AI) results
  impact radiologists, and if so, what can we do about it? A multi-reader pilot
  study of lung cancer detection with chest radiography}.
\newblock \bibinfo{journal}{\emph{European Radiology}} (\bibinfo{year}{2023}),
  \bibinfo{pages}{1--7}.
\newblock


\bibitem[Berry et~al\mbox{.}(2021)]%
        {berry2021high}
\bibfield{author}{\bibinfo{person}{Leonard~L Berry}, \bibinfo{person}{Sunjay
  Letchuman}, \bibinfo{person}{Nandini Ramani}, {and} \bibinfo{person}{Paul
  Barach}.} \bibinfo{year}{2021}\natexlab{}.
\newblock \showarticletitle{The high stakes of outsourcing in health care}. In
  \bibinfo{booktitle}{\emph{Mayo Clinic Proceedings}},
  Vol.~\bibinfo{volume}{96}. Elsevier, \bibinfo{pages}{2879--2890}.
\newblock


\bibitem[Bitner et~al\mbox{.}(2008)]%
        {bitner2008service}
\bibfield{author}{\bibinfo{person}{Mary~Jo Bitner}, \bibinfo{person}{Amy~L
  Ostrom}, {and} \bibinfo{person}{Felicia~N Morgan}.}
  \bibinfo{year}{2008}\natexlab{}.
\newblock \showarticletitle{Service blueprinting: a practical technique for
  service innovation}.
\newblock \bibinfo{journal}{\emph{California management review}}
  \bibinfo{volume}{50}, \bibinfo{number}{3} (\bibinfo{year}{2008}),
  \bibinfo{pages}{66--94}.
\newblock


\bibitem[Bly and Churchill(1999)]%
        {bly1999design}
\bibfield{author}{\bibinfo{person}{Sara Bly} {and} \bibinfo{person}{Elizabeth~F
  Churchill}.} \bibinfo{year}{1999}\natexlab{}.
\newblock \showarticletitle{Design through matchmaking: technology in search of
  users}.
\newblock \bibinfo{journal}{\emph{interactions}} \bibinfo{volume}{6},
  \bibinfo{number}{2} (\bibinfo{year}{1999}), \bibinfo{pages}{23--31}.
\newblock


\bibitem[Boecking et~al\mbox{.}(2022)]%
        {boecking2022making}
\bibfield{author}{\bibinfo{person}{Benedikt Boecking}, \bibinfo{person}{Naoto
  Usuyama}, \bibinfo{person}{Shruthi Bannur}, \bibinfo{person}{Daniel~C
  Castro}, \bibinfo{person}{Anton Schwaighofer}, \bibinfo{person}{Stephanie
  Hyland}, \bibinfo{person}{Maria Wetscherek}, \bibinfo{person}{Tristan
  Naumann}, \bibinfo{person}{Aditya Nori}, \bibinfo{person}{Javier
  Alvarez-Valle}, {et~al\mbox{.}}} \bibinfo{year}{2022}\natexlab{}.
\newblock \showarticletitle{Making the most of text semantics to improve
  biomedical vision--language processing}. In
  \bibinfo{booktitle}{\emph{European conference on computer vision}}. Springer,
  \bibinfo{pages}{1--21}.
\newblock


\bibitem[Bommasani et~al\mbox{.}(2021)]%
        {bommasani2021opportunities}
\bibfield{author}{\bibinfo{person}{Rishi Bommasani}, \bibinfo{person}{Drew~A
  Hudson}, \bibinfo{person}{Ehsan Adeli}, \bibinfo{person}{Russ Altman},
  \bibinfo{person}{Simran Arora}, \bibinfo{person}{Sydney von Arx},
  \bibinfo{person}{Michael~S Bernstein}, \bibinfo{person}{Jeannette Bohg},
  \bibinfo{person}{Antoine Bosselut}, \bibinfo{person}{Emma Brunskill},
  {et~al\mbox{.}}} \bibinfo{year}{2021}\natexlab{}.
\newblock \showarticletitle{On the opportunities and risks of foundation
  models}.
\newblock \bibinfo{journal}{\emph{arXiv preprint arXiv:2108.07258}}
  (\bibinfo{year}{2021}).
\newblock


\bibitem[Bossen and Pine(2023)]%
        {bossen2023batman}
\bibfield{author}{\bibinfo{person}{Claus Bossen} {and}
  \bibinfo{person}{Kathleen~H Pine}.} \bibinfo{year}{2023}\natexlab{}.
\newblock \showarticletitle{Batman and Robin in Healthcare Knowledge Work:
  Human-AI Collaboration by Clinical Documentation Integrity Specialists}.
\newblock \bibinfo{journal}{\emph{ACM Transactions on Computer-Human
  Interaction}} \bibinfo{volume}{30}, \bibinfo{number}{2}
  (\bibinfo{year}{2023}), \bibinfo{pages}{1--29}.
\newblock


\bibitem[Brown et~al\mbox{.}(2020)]%
        {brown2020language}
\bibfield{author}{\bibinfo{person}{Tom Brown}, \bibinfo{person}{Benjamin Mann},
  \bibinfo{person}{Nick Ryder}, \bibinfo{person}{Melanie Subbiah},
  \bibinfo{person}{Jared~D Kaplan}, \bibinfo{person}{Prafulla Dhariwal},
  \bibinfo{person}{Arvind Neelakantan}, \bibinfo{person}{Pranav Shyam},
  \bibinfo{person}{Girish Sastry}, \bibinfo{person}{Amanda Askell},
  {et~al\mbox{.}}} \bibinfo{year}{2020}\natexlab{}.
\newblock \showarticletitle{Language models are few-shot learners}.
\newblock \bibinfo{journal}{\emph{Advances in neural information processing
  systems}}  \bibinfo{volume}{33} (\bibinfo{year}{2020}),
  \bibinfo{pages}{1877--1901}.
\newblock


\bibitem[Burgess et~al\mbox{.}(2023)]%
        {burgess2023healthcare}
\bibfield{author}{\bibinfo{person}{Eleanor~R Burgess}, \bibinfo{person}{Ivana
  Jankovic}, \bibinfo{person}{Melissa Austin}, \bibinfo{person}{Nancy Cai},
  \bibinfo{person}{Adela Kapu{\'s}ci{\'n}ska}, \bibinfo{person}{Suzanne
  Currie}, \bibinfo{person}{J~Marc Overhage}, \bibinfo{person}{Erika~S Poole},
  {and} \bibinfo{person}{Jofish Kaye}.} \bibinfo{year}{2023}\natexlab{}.
\newblock \showarticletitle{Healthcare AI Treatment Decision Support: Design
  Principles to Enhance Clinician Adoption and Trust}. In
  \bibinfo{booktitle}{\emph{Proceedings of the 2023 CHI Conference on Human
  Factors in Computing Systems}}. \bibinfo{pages}{1--19}.
\newblock


\bibitem[Buxton(2010)]%
        {buxton2010sketching}
\bibfield{author}{\bibinfo{person}{Bill Buxton}.}
  \bibinfo{year}{2010}\natexlab{}.
\newblock \bibinfo{booktitle}{\emph{Sketching user experiences: getting the
  design right and the right design}}.
\newblock \bibinfo{publisher}{Morgan kaufmann}.
\newblock


\bibitem[Cai et~al\mbox{.}(2019a)]%
        {cai2019human}
\bibfield{author}{\bibinfo{person}{Carrie~J Cai}, \bibinfo{person}{Emily Reif},
  \bibinfo{person}{Narayan Hegde}, \bibinfo{person}{Jason Hipp},
  \bibinfo{person}{Been Kim}, \bibinfo{person}{Daniel Smilkov},
  \bibinfo{person}{Martin Wattenberg}, \bibinfo{person}{Fernanda Viegas},
  \bibinfo{person}{Greg~S Corrado}, \bibinfo{person}{Martin~C Stumpe},
  {et~al\mbox{.}}} \bibinfo{year}{2019}\natexlab{a}.
\newblock \showarticletitle{Human-centered tools for coping with imperfect
  algorithms during medical decision-making}. In
  \bibinfo{booktitle}{\emph{Proceedings of the 2019 chi conference on human
  factors in computing systems}}. \bibinfo{pages}{1--14}.
\newblock


\bibitem[Cai et~al\mbox{.}(2019b)]%
        {cai2019hello}
\bibfield{author}{\bibinfo{person}{Carrie~J Cai}, \bibinfo{person}{Samantha
  Winter}, \bibinfo{person}{David Steiner}, \bibinfo{person}{Lauren Wilcox},
  {and} \bibinfo{person}{Michael Terry}.} \bibinfo{year}{2019}\natexlab{b}.
\newblock \showarticletitle{" Hello AI": uncovering the onboarding needs of
  medical practitioners for human-AI collaborative decision-making}.
\newblock \bibinfo{journal}{\emph{Proceedings of the ACM on Human-computer
  Interaction}} \bibinfo{volume}{3}, \bibinfo{number}{CSCW}
  (\bibinfo{year}{2019}), \bibinfo{pages}{1--24}.
\newblock


\bibitem[Cai et~al\mbox{.}(2021)]%
        {cai2021onboarding}
\bibfield{author}{\bibinfo{person}{Carrie~J Cai}, \bibinfo{person}{Samantha
  Winter}, \bibinfo{person}{David Steiner}, \bibinfo{person}{Lauren Wilcox},
  {and} \bibinfo{person}{Michael Terry}.} \bibinfo{year}{2021}\natexlab{}.
\newblock \showarticletitle{Onboarding Materials as Cross-functional Boundary
  Objects for Developing AI Assistants}. In \bibinfo{booktitle}{\emph{Extended
  Abstracts of the 2021 CHI Conference on Human Factors in Computing Systems}}.
  \bibinfo{pages}{1--7}.
\newblock


\bibitem[Calisto et~al\mbox{.}(2023)]%
        {calisto2023assertiveness}
\bibfield{author}{\bibinfo{person}{Francisco~Maria Calisto},
  \bibinfo{person}{Jo{\~a}o Fernandes}, \bibinfo{person}{Margarida Morais},
  \bibinfo{person}{Carlos Santiago}, \bibinfo{person}{Jo{\~a}o~Maria Abrantes},
  \bibinfo{person}{Nuno Nunes}, {and} \bibinfo{person}{Jacinto~C Nascimento}.}
  \bibinfo{year}{2023}\natexlab{}.
\newblock \showarticletitle{Assertiveness-based Agent Communication for a
  Personalized Medicine on Medical Imaging Diagnosis}. In
  \bibinfo{booktitle}{\emph{Proceedings of the 2023 CHI Conference on Human
  Factors in Computing Systems}}. \bibinfo{pages}{1--20}.
\newblock


\bibitem[Calisto et~al\mbox{.}(2021)]%
        {calisto2021introduction}
\bibfield{author}{\bibinfo{person}{Francisco~Maria Calisto},
  \bibinfo{person}{Carlos Santiago}, \bibinfo{person}{Nuno Nunes}, {and}
  \bibinfo{person}{Jacinto~C Nascimento}.} \bibinfo{year}{2021}\natexlab{}.
\newblock \showarticletitle{Introduction of human-centric AI assistant to aid
  radiologists for multimodal breast image classification}.
\newblock \bibinfo{journal}{\emph{International Journal of Human-Computer
  Studies}}  \bibinfo{volume}{150} (\bibinfo{year}{2021}),
  \bibinfo{pages}{102607}.
\newblock


\bibitem[Calisto et~al\mbox{.}(2022)]%
        {calisto2022breastscreening}
\bibfield{author}{\bibinfo{person}{Francisco~Maria Calisto},
  \bibinfo{person}{Carlos Santiago}, \bibinfo{person}{Nuno Nunes}, {and}
  \bibinfo{person}{Jacinto~C Nascimento}.} \bibinfo{year}{2022}\natexlab{}.
\newblock \showarticletitle{BreastScreening-AI: Evaluating medical intelligent
  agents for human-AI interactions}.
\newblock \bibinfo{journal}{\emph{Artificial Intelligence in Medicine}}
  \bibinfo{volume}{127} (\bibinfo{year}{2022}), \bibinfo{pages}{102285}.
\newblock


\bibitem[Chien et~al\mbox{.}(2022)]%
        {chien2022multi}
\bibfield{author}{\bibinfo{person}{Isabel Chien}, \bibinfo{person}{Nina Deliu},
  \bibinfo{person}{Richard Turner}, \bibinfo{person}{Adrian Weller},
  \bibinfo{person}{Sofia Villar}, {and} \bibinfo{person}{Niki Kilbertus}.}
  \bibinfo{year}{2022}\natexlab{}.
\newblock \showarticletitle{Multi-disciplinary fairness considerations in
  machine learning for clinical trials}. In
  \bibinfo{booktitle}{\emph{Proceedings of the 2022 ACM Conference on Fairness,
  Accountability, and Transparency}}. \bibinfo{pages}{906--924}.
\newblock


\bibitem[Chung et~al\mbox{.}(2022)]%
        {chung2022scaling}
\bibfield{author}{\bibinfo{person}{Hyung~Won Chung}, \bibinfo{person}{Le Hou},
  \bibinfo{person}{Shayne Longpre}, \bibinfo{person}{Barret Zoph},
  \bibinfo{person}{Yi Tay}, \bibinfo{person}{William Fedus},
  \bibinfo{person}{Eric Li}, \bibinfo{person}{Xuezhi Wang},
  \bibinfo{person}{Mostafa Dehghani}, \bibinfo{person}{Siddhartha Brahma},
  {et~al\mbox{.}}} \bibinfo{year}{2022}\natexlab{}.
\newblock \showarticletitle{Scaling instruction-finetuned language models}.
\newblock \bibinfo{journal}{\emph{arXiv preprint arXiv:2210.11416}}
  (\bibinfo{year}{2022}).
\newblock


\bibitem[Clinger et~al\mbox{.}(1988)]%
        {clinger1988radiology}
\bibfield{author}{\bibinfo{person}{Neal~J Clinger}, \bibinfo{person}{Tim~B
  Hunter}, {and} \bibinfo{person}{Bruce~J Hillman}.}
  \bibinfo{year}{1988}\natexlab{}.
\newblock \showarticletitle{Radiology reporting: attitudes of referring
  physicians.}
\newblock \bibinfo{journal}{\emph{Radiology}} \bibinfo{volume}{169},
  \bibinfo{number}{3} (\bibinfo{year}{1988}), \bibinfo{pages}{825--826}.
\newblock


\bibitem[Coiera(2019)]%
        {coiera2019last}
\bibfield{author}{\bibinfo{person}{Enrico Coiera}.}
  \bibinfo{year}{2019}\natexlab{}.
\newblock \showarticletitle{The last mile: where artificial intelligence meets
  reality}.
\newblock \bibinfo{journal}{\emph{Journal of medical Internet research}}
  \bibinfo{volume}{21}, \bibinfo{number}{11} (\bibinfo{year}{2019}),
  \bibinfo{pages}{e16323}.
\newblock


\bibitem[Collins and Ghahramani(2021)]%
        {Collins_Ghahramani_2021}
\bibfield{author}{\bibinfo{person}{Eli Collins} {and} \bibinfo{person}{Zoubin
  Ghahramani}.} \bibinfo{year}{2021}\natexlab{}.
\newblock \bibinfo{title}{LAMDA: Our breakthrough conversation technology}.
\newblock
\newblock
\urldef\tempurl%
\url{https://blog.google/technology/ai/lamda/}
\showURL{%
\tempurl}


\bibitem[Cooper et~al\mbox{.}(2022)]%
        {cooper2022systematic}
\bibfield{author}{\bibinfo{person}{Ned Cooper}, \bibinfo{person}{Tiffanie
  Horne}, \bibinfo{person}{Gillian~R Hayes}, \bibinfo{person}{Courtney
  Heldreth}, \bibinfo{person}{Michal Lahav}, \bibinfo{person}{Jess Holbrook},
  {and} \bibinfo{person}{Lauren Wilcox}.} \bibinfo{year}{2022}\natexlab{}.
\newblock \showarticletitle{A systematic review and thematic analysis of
  community-collaborative approaches to computing research}. In
  \bibinfo{booktitle}{\emph{Proceedings of the 2022 CHI Conference on Human
  Factors in Computing Systems}}. \bibinfo{pages}{1--18}.
\newblock


\bibitem[Corbett et~al\mbox{.}(2023)]%
        {corbett2023power}
\bibfield{author}{\bibinfo{person}{Eric Corbett}, \bibinfo{person}{Emily
  Denton}, {and} \bibinfo{person}{Sheena Erete}.}
  \bibinfo{year}{2023}\natexlab{}.
\newblock \showarticletitle{Power and Public Participation in AI}.
\newblock In \bibinfo{booktitle}{\emph{Equity and Access in Algorithms,
  Mechanisms, and Optimization}}. \bibinfo{pages}{1--13}.
\newblock


\bibitem[Corrado and Matias(2023)]%
        {Corrado_Matias_2023}
\bibfield{author}{\bibinfo{person}{Greg Corrado} {and} \bibinfo{person}{Yossi
  Matias}.} \bibinfo{year}{2023}\natexlab{}.
\newblock \bibinfo{title}{Multimodal Medical Ai}.
\newblock
\newblock
\urldef\tempurl%
\url{https://ai.googleblog.com/2023/08/multimodal-medical-ai.html}
\showURL{%
\tempurl}


\bibitem[Cowan et~al\mbox{.}(2013)]%
        {cowan2013measuring}
\bibfield{author}{\bibinfo{person}{Ian~A Cowan}, \bibinfo{person}{Sharyn~LS
  MacDonald}, {and} \bibinfo{person}{Richard~A Floyd}.}
  \bibinfo{year}{2013}\natexlab{}.
\newblock \showarticletitle{Measuring and managing radiologist workload:
  Measuring radiologist reporting times using data from a R adiology I
  nformation S ystem}.
\newblock \bibinfo{journal}{\emph{Journal of medical imaging and radiation
  oncology}} \bibinfo{volume}{57}, \bibinfo{number}{5} (\bibinfo{year}{2013}),
  \bibinfo{pages}{558--566}.
\newblock


\bibitem[Dam and Siang(2022)]%
        {Dam_Siang_2022}
\bibfield{author}{\bibinfo{person}{Rikke~Friis Dam} {and}
  \bibinfo{person}{Teo~Yu Siang}.} \bibinfo{year}{2022}\natexlab{}.
\newblock \bibinfo{title}{Affinity diagrams: How to cluster your ideas and
  reveal insights}.
\newblock
\newblock
\urldef\tempurl%
\url{https://www.interaction-design.org/literature/article/affinity-diagrams-learn-how-to-cluster-and-bundle-ideas-and-facts}
\showURL{%
\tempurl}


\bibitem[Delgado et~al\mbox{.}(2021)]%
        {delgado2021stakeholder}
\bibfield{author}{\bibinfo{person}{Fernando Delgado}, \bibinfo{person}{Stephen
  Yang}, \bibinfo{person}{Michael Madaio}, {and} \bibinfo{person}{Qian Yang}.}
  \bibinfo{year}{2021}\natexlab{}.
\newblock \showarticletitle{Stakeholder Participation in AI: Beyond" Add
  Diverse Stakeholders and Stir"}.
\newblock \bibinfo{journal}{\emph{arXiv preprint arXiv:2111.01122}}
  (\bibinfo{year}{2021}).
\newblock


\bibitem[Delgado et~al\mbox{.}(2023)]%
        {delgado2023participatory}
\bibfield{author}{\bibinfo{person}{Fernando Delgado}, \bibinfo{person}{Stephen
  Yang}, \bibinfo{person}{Michael Madaio}, {and} \bibinfo{person}{Qian Yang}.}
  \bibinfo{year}{2023}\natexlab{}.
\newblock \showarticletitle{The Participatory Turn in AI Design: Theoretical
  Foundations and the Current State of Practice}. In
  \bibinfo{booktitle}{\emph{Proceedings of the 3rd ACM Conference on Equity and
  Access in Algorithms, Mechanisms, and Optimization}}. \bibinfo{pages}{1--23}.
\newblock


\bibitem[Deng et~al\mbox{.}(2023)]%
        {deng2023investigating}
\bibfield{author}{\bibinfo{person}{Wesley~Hanwen Deng}, \bibinfo{person}{Nur
  Yildirim}, \bibinfo{person}{Monica Chang}, \bibinfo{person}{Motahhare
  Eslami}, \bibinfo{person}{Kenneth Holstein}, {and} \bibinfo{person}{Michael
  Madaio}.} \bibinfo{year}{2023}\natexlab{}.
\newblock \showarticletitle{Investigating Practices and Opportunities for
  Cross-functional Collaboration around AI Fairness in Industry Practice}. In
  \bibinfo{booktitle}{\emph{Proceedings of the 2023 ACM Conference on Fairness,
  Accountability, and Transparency}}. \bibinfo{pages}{705--716}.
\newblock


\bibitem[Feng et~al\mbox{.}(2023)]%
        {feng2023ux}
\bibfield{author}{\bibinfo{person}{KJ~Kevin Feng},
  \bibinfo{person}{Maxwell~James Coppock}, {and} \bibinfo{person}{David~W
  McDonald}.} \bibinfo{year}{2023}\natexlab{}.
\newblock \showarticletitle{How Do UX Practitioners Communicate AI as a Design
  Material? Artifacts, Conceptions, and Propositions}. In
  \bibinfo{booktitle}{\emph{Proceedings of the 2023 ACM Designing Interactive
  Systems Conference}}. \bibinfo{pages}{2263--2280}.
\newblock


\bibitem[Figma(2023)]%
        {figma}
\bibfield{author}{\bibinfo{person}{Figma}.} \bibinfo{year}{2023}\natexlab{}.
\newblock \bibinfo{title}{Figma: the collaborative interface design tool.}
\newblock
\newblock
\urldef\tempurl%
\url{https://www.figma.com/}
\showURL{%
\tempurl}


\bibitem[Galsgaard et~al\mbox{.}(2022)]%
        {galsgaard2022artificial}
\bibfield{author}{\bibinfo{person}{Astrid Galsgaard}, \bibinfo{person}{Tom
  Doorschodt}, \bibinfo{person}{Ann-Louise Holten},
  \bibinfo{person}{Felix~Christoph M{\"u}ller}, \bibinfo{person}{Mikael~Ploug
  Boesen}, {and} \bibinfo{person}{Mario Maas}.}
  \bibinfo{year}{2022}\natexlab{}.
\newblock \showarticletitle{Artificial intelligence and multidisciplinary team
  meetings; a communication challenge for radiologists' sense of agency and
  position as spider in a web?}
\newblock \bibinfo{journal}{\emph{European Journal of Radiology}}
  \bibinfo{volume}{155} (\bibinfo{year}{2022}), \bibinfo{pages}{110231}.
\newblock


\bibitem[Ganeshan et~al\mbox{.}(2018)]%
        {ganeshan2018structured}
\bibfield{author}{\bibinfo{person}{Dhakshinamoorthy Ganeshan},
  \bibinfo{person}{Phuong-Anh~Thi Duong}, \bibinfo{person}{Linda Probyn},
  \bibinfo{person}{Leon Lenchik}, \bibinfo{person}{Tatum~A McArthur},
  \bibinfo{person}{Michele Retrouvey}, \bibinfo{person}{Emily~H Ghobadi},
  \bibinfo{person}{Stephane~L Desouches}, \bibinfo{person}{David Pastel}, {and}
  \bibinfo{person}{Isaac~R Francis}.} \bibinfo{year}{2018}\natexlab{}.
\newblock \showarticletitle{Structured reporting in radiology}.
\newblock \bibinfo{journal}{\emph{Academic radiology}} \bibinfo{volume}{25},
  \bibinfo{number}{1} (\bibinfo{year}{2018}), \bibinfo{pages}{66--73}.
\newblock


\bibitem[Ghosh et~al\mbox{.}(2023)]%
        {ghosh2023framing}
\bibfield{author}{\bibinfo{person}{Pratik Ghosh}, \bibinfo{person}{Karen~L
  Posner}, \bibinfo{person}{Stephanie~L Hyland}, \bibinfo{person}{Wil
  Van~Cleve}, \bibinfo{person}{Melissa Bristow}, \bibinfo{person}{Dustin~R
  Long}, \bibinfo{person}{Konstantina Palla}, \bibinfo{person}{Bala Nair},
  \bibinfo{person}{Christine Fong}, \bibinfo{person}{Ronald Pauldine},
  {et~al\mbox{.}}} \bibinfo{year}{2023}\natexlab{}.
\newblock \showarticletitle{Framing Machine Learning Opportunities for
  Hypotension Prediction in Perioperative Care: A Socio-Technical Perspective}.
\newblock \bibinfo{journal}{\emph{ACM Transactions on Computer-Human
  Interaction}} (\bibinfo{year}{2023}).
\newblock


\bibitem[Gilbert et~al\mbox{.}(2023)]%
        {gilbert2023large}
\bibfield{author}{\bibinfo{person}{Stephen Gilbert}, \bibinfo{person}{Hugh
  Harvey}, \bibinfo{person}{Tom Melvin}, \bibinfo{person}{Erik Vollebregt},
  {and} \bibinfo{person}{Paul Wicks}.} \bibinfo{year}{2023}\natexlab{}.
\newblock \showarticletitle{Large language model AI chatbots require approval
  as medical devices}.
\newblock \bibinfo{journal}{\emph{Nature Medicine}} (\bibinfo{year}{2023}),
  \bibinfo{pages}{1--3}.
\newblock


\bibitem[Google(2023)]%
        {GoogleBard}
\bibfield{author}{\bibinfo{person}{Google}.} \bibinfo{year}{2023}\natexlab{}.
\newblock \bibinfo{title}{Bard - Chat Based AI Tool from Google, Powered by
  PaLM 2}.
\newblock
\newblock
\urldef\tempurl%
\url{https://bard.google.com/}
\showURL{%
\tempurl}


\bibitem[Gu et~al\mbox{.}(2023a)]%
        {gu2023improving}
\bibfield{author}{\bibinfo{person}{Hongyan Gu}, \bibinfo{person}{Yuan Liang},
  \bibinfo{person}{Yifan Xu}, \bibinfo{person}{Christopher~Kazu Williams},
  \bibinfo{person}{Shino Magaki}, \bibinfo{person}{Negar Khanlou},
  \bibinfo{person}{Harry Vinters}, \bibinfo{person}{Zesheng Chen},
  \bibinfo{person}{Shuo Ni}, \bibinfo{person}{Chunxu Yang}, {et~al\mbox{.}}}
  \bibinfo{year}{2023}\natexlab{a}.
\newblock \showarticletitle{Improving workflow integration with XPath: Design
  and evaluation of a human-AI diagnosis system in pathology}.
\newblock \bibinfo{journal}{\emph{ACM Transactions on Computer-Human
  Interaction}} \bibinfo{volume}{30}, \bibinfo{number}{2}
  (\bibinfo{year}{2023}), \bibinfo{pages}{1--37}.
\newblock


\bibitem[Gu et~al\mbox{.}(2023b)]%
        {gu2023augmenting}
\bibfield{author}{\bibinfo{person}{Hongyan Gu}, \bibinfo{person}{Chunxu Yang},
  \bibinfo{person}{Mohammad Haeri}, \bibinfo{person}{Jing Wang},
  \bibinfo{person}{Shirley Tang}, \bibinfo{person}{Wenzhong Yan},
  \bibinfo{person}{Shujin He}, \bibinfo{person}{Christopher~Kazu Williams},
  \bibinfo{person}{Shino Magaki}, {and} \bibinfo{person}{Xiang'Anthony' Chen}.}
  \bibinfo{year}{2023}\natexlab{b}.
\newblock \showarticletitle{Augmenting Pathologists with NaviPath: Design and
  Evaluation of a Human-AI Collaborative Navigation System}. In
  \bibinfo{booktitle}{\emph{Proceedings of the 2023 CHI Conference on Human
  Factors in Computing Systems}}. \bibinfo{pages}{1--19}.
\newblock


\bibitem[Gu et~al\mbox{.}(2023c)]%
        {gu2023distilling}
\bibfield{author}{\bibinfo{person}{Yu Gu}, \bibinfo{person}{Sheng Zhang},
  \bibinfo{person}{Naoto Usuyama}, \bibinfo{person}{Yonas Woldesenbet},
  \bibinfo{person}{Cliff Wong}, \bibinfo{person}{Praneeth Sanapathi},
  \bibinfo{person}{Mu Wei}, \bibinfo{person}{Naveen Valluri},
  \bibinfo{person}{Erika Strandberg}, \bibinfo{person}{Tristan Naumann},
  {et~al\mbox{.}}} \bibinfo{year}{2023}\natexlab{c}.
\newblock \showarticletitle{Distilling large language models for biomedical
  knowledge extraction: A case study on adverse drug events}.
\newblock \bibinfo{journal}{\emph{arXiv preprint arXiv:2307.06439}}
  (\bibinfo{year}{2023}).
\newblock


\bibitem[Henry et~al\mbox{.}(2022)]%
        {henry2022human}
\bibfield{author}{\bibinfo{person}{Katharine~E Henry}, \bibinfo{person}{Rachel
  Kornfield}, \bibinfo{person}{Anirudh Sridharan}, \bibinfo{person}{Robert~C
  Linton}, \bibinfo{person}{Catherine Groh}, \bibinfo{person}{Tony Wang},
  \bibinfo{person}{Albert Wu}, \bibinfo{person}{Bilge Mutlu}, {and}
  \bibinfo{person}{Suchi Saria}.} \bibinfo{year}{2022}\natexlab{}.
\newblock \showarticletitle{Human--machine teaming is key to AI adoption:
  clinicians’ experiences with a deployed machine learning system}.
\newblock \bibinfo{journal}{\emph{NPJ digital medicine}} \bibinfo{volume}{5},
  \bibinfo{number}{1} (\bibinfo{year}{2022}), \bibinfo{pages}{97}.
\newblock


\bibitem[Hirsch et~al\mbox{.}(2017)]%
        {hirsch2017designing}
\bibfield{author}{\bibinfo{person}{Tad Hirsch}, \bibinfo{person}{Kritzia
  Merced}, \bibinfo{person}{Shrikanth Narayanan}, \bibinfo{person}{Zac~E Imel},
  {and} \bibinfo{person}{David~C Atkins}.} \bibinfo{year}{2017}\natexlab{}.
\newblock \showarticletitle{Designing contestability: Interaction design,
  machine learning, and mental health}. In
  \bibinfo{booktitle}{\emph{Proceedings of the 2017 Conference on Designing
  Interactive Systems}}. \bibinfo{pages}{95--99}.
\newblock


\bibitem[Hirsch et~al\mbox{.}(2018)]%
        {hirsch2018s}
\bibfield{author}{\bibinfo{person}{Tad Hirsch}, \bibinfo{person}{Christina
  Soma}, \bibinfo{person}{Kritzia Merced}, \bibinfo{person}{Patty Kuo},
  \bibinfo{person}{Aaron Dembe}, \bibinfo{person}{Derek~D Caperton},
  \bibinfo{person}{David~C Atkins}, {and} \bibinfo{person}{Zac~E Imel}.}
  \bibinfo{year}{2018}\natexlab{}.
\newblock \showarticletitle{" It's hard to argue with a computer" Investigating
  Psychotherapists' Attitudes towards Automated Evaluation}. In
  \bibinfo{booktitle}{\emph{Proceedings of the 2018 Designing Interactive
  Systems Conference}}. \bibinfo{pages}{559--571}.
\newblock


\bibitem[Holstein et~al\mbox{.}(2019)]%
        {holstein2019improving}
\bibfield{author}{\bibinfo{person}{Kenneth Holstein}, \bibinfo{person}{Jennifer
  Wortman~Vaughan}, \bibinfo{person}{Hal Daum{\'e}~III}, \bibinfo{person}{Miro
  Dudik}, {and} \bibinfo{person}{Hanna Wallach}.}
  \bibinfo{year}{2019}\natexlab{}.
\newblock \showarticletitle{Improving fairness in machine learning systems:
  What do industry practitioners need?}. In
  \bibinfo{booktitle}{\emph{Proceedings of the 2019 CHI conference on human
  factors in computing systems}}. \bibinfo{pages}{1--16}.
\newblock


\bibitem[Holtzblatt and Beyer(2014)]%
        {holtzblatt2014field}
\bibfield{author}{\bibinfo{person}{Karen Holtzblatt} {and}
  \bibinfo{person}{Hugh Beyer}.} \bibinfo{year}{2014}\natexlab{}.
\newblock \showarticletitle{Field research: data collection and
  interpretation}.
\newblock In \bibinfo{booktitle}{\emph{Contextual Design: Evolved}}.
  \bibinfo{publisher}{Springer}, \bibinfo{pages}{11--20}.
\newblock


\bibitem[Huang et~al\mbox{.}(2023)]%
        {huang2023generative}
\bibfield{author}{\bibinfo{person}{Jonathan Huang}, \bibinfo{person}{Luke
  Neill}, \bibinfo{person}{Matthew Wittbrodt}, \bibinfo{person}{David Melnick},
  \bibinfo{person}{Matthew Klug}, \bibinfo{person}{Michael Thompson},
  \bibinfo{person}{John Bailitz}, \bibinfo{person}{Timothy Loftus},
  \bibinfo{person}{Sanjeev Malik}, \bibinfo{person}{Amit Phull},
  {et~al\mbox{.}}} \bibinfo{year}{2023}\natexlab{}.
\newblock \showarticletitle{Generative Artificial Intelligence for Chest
  Radiograph Interpretation in the Emergency Department}.
\newblock \bibinfo{journal}{\emph{JAMA network open}} \bibinfo{volume}{6},
  \bibinfo{number}{10} (\bibinfo{year}{2023}),
  \bibinfo{pages}{e2336100--e2336100}.
\newblock


\bibitem[Hyland et~al\mbox{.}(2023)]%
        {Hyland2023maira}
\bibfield{author}{\bibinfo{person}{Stephanie Hyland}, \bibinfo{person}{Shruthi
  Bannur}, \bibinfo{person}{Kenza Bouzid}, \bibinfo{person}{Daniel~C Castro},
  \bibinfo{person}{Mercy Ranjit}, \bibinfo{person}{Anton Schwaighofer},
  \bibinfo{person}{Fernando P{\'e}rez-Garc{\'\i}a}, {et~al\mbox{.}}}
  \bibinfo{year}{2023}\natexlab{}.
\newblock \showarticletitle{MAIRA-1: A specialised large multimodal model for
  radiology report generation}.
\newblock \bibinfo{journal}{\emph{arXiv preprint arXiv: 2311.13668}}
  (\bibinfo{year}{2023}).
\newblock


\bibitem[Ismail et~al\mbox{.}(2018)]%
        {ismail2018bridging}
\bibfield{author}{\bibinfo{person}{Azra Ismail}, \bibinfo{person}{Naveena
  Karusala}, {and} \bibinfo{person}{Neha Kumar}.}
  \bibinfo{year}{2018}\natexlab{}.
\newblock \showarticletitle{Bridging disconnected knowledges for community
  health}.
\newblock \bibinfo{journal}{\emph{Proceedings of the ACM on Human-Computer
  Interaction}} \bibinfo{volume}{2}, \bibinfo{number}{CSCW}
  (\bibinfo{year}{2018}), \bibinfo{pages}{1--27}.
\newblock


\bibitem[Jacobs et~al\mbox{.}(2021)]%
        {jacobs2021designing}
\bibfield{author}{\bibinfo{person}{Maia Jacobs}, \bibinfo{person}{Jeffrey He},
  \bibinfo{person}{Melanie F.~Pradier}, \bibinfo{person}{Barbara Lam},
  \bibinfo{person}{Andrew~C Ahn}, \bibinfo{person}{Thomas~H McCoy},
  \bibinfo{person}{Roy~H Perlis}, \bibinfo{person}{Finale Doshi-Velez}, {and}
  \bibinfo{person}{Krzysztof~Z Gajos}.} \bibinfo{year}{2021}\natexlab{}.
\newblock \showarticletitle{Designing AI for trust and collaboration in
  time-constrained medical decisions: a sociotechnical lens}. In
  \bibinfo{booktitle}{\emph{Proceedings of the 2021 chi conference on human
  factors in computing systems}}. \bibinfo{pages}{1--14}.
\newblock


\bibitem[Jeblick et~al\mbox{.}(2022)]%
        {jeblick2022chatgpt}
\bibfield{author}{\bibinfo{person}{Katharina Jeblick},
  \bibinfo{person}{Balthasar Schachtner}, \bibinfo{person}{Jakob Dexl},
  \bibinfo{person}{Andreas Mittermeier}, \bibinfo{person}{Anna~Theresa
  St{\"u}ber}, \bibinfo{person}{Johanna Topalis}, \bibinfo{person}{Tobias
  Weber}, \bibinfo{person}{Philipp Wesp}, \bibinfo{person}{Bastian Sabel},
  \bibinfo{person}{Jens Ricke}, {et~al\mbox{.}}}
  \bibinfo{year}{2022}\natexlab{}.
\newblock \showarticletitle{Chatgpt makes medicine easy to swallow: An
  exploratory case study on simplified radiology reports}.
\newblock \bibinfo{journal}{\emph{arXiv preprint arXiv:2212.14882}}
  (\bibinfo{year}{2022}).
\newblock


\bibitem[Jha and Topol(2016)]%
        {jha2016adapting}
\bibfield{author}{\bibinfo{person}{Saurabh Jha} {and} \bibinfo{person}{Eric~J
  Topol}.} \bibinfo{year}{2016}\natexlab{}.
\newblock \showarticletitle{Adapting to artificial intelligence: radiologists
  and pathologists as information specialists}.
\newblock \bibinfo{journal}{\emph{Jama}} \bibinfo{volume}{316},
  \bibinfo{number}{22} (\bibinfo{year}{2016}), \bibinfo{pages}{2353--2354}.
\newblock


\bibitem[Jiang et~al\mbox{.}(2022)]%
        {jiang2022promptmaker}
\bibfield{author}{\bibinfo{person}{Ellen Jiang}, \bibinfo{person}{Kristen
  Olson}, \bibinfo{person}{Edwin Toh}, \bibinfo{person}{Alejandra Molina},
  \bibinfo{person}{Aaron Donsbach}, \bibinfo{person}{Michael Terry}, {and}
  \bibinfo{person}{Carrie~J Cai}.} \bibinfo{year}{2022}\natexlab{}.
\newblock \showarticletitle{Promptmaker: Prompt-based prototyping with large
  language models}. In \bibinfo{booktitle}{\emph{CHI Conference on Human
  Factors in Computing Systems Extended Abstracts}}. \bibinfo{pages}{1--8}.
\newblock


\bibitem[Johnson et~al\mbox{.}(2019)]%
        {johnson2019mimic}
\bibfield{author}{\bibinfo{person}{Alistair~EW Johnson}, \bibinfo{person}{Tom~J
  Pollard}, \bibinfo{person}{Seth~J Berkowitz}, \bibinfo{person}{Nathaniel~R
  Greenbaum}, \bibinfo{person}{Matthew~P Lungren}, \bibinfo{person}{Chih-ying
  Deng}, \bibinfo{person}{Roger~G Mark}, {and} \bibinfo{person}{Steven Horng}.}
  \bibinfo{year}{2019}\natexlab{}.
\newblock \showarticletitle{MIMIC-CXR, a de-identified publicly available
  database of chest radiographs with free-text reports}.
\newblock \bibinfo{journal}{\emph{Scientific data}} \bibinfo{volume}{6},
  \bibinfo{number}{1} (\bibinfo{year}{2019}), \bibinfo{pages}{317}.
\newblock


\bibitem[Kadavath et~al\mbox{.}(2022)]%
        {kadavath2022language}
\bibfield{author}{\bibinfo{person}{Saurav Kadavath}, \bibinfo{person}{Tom
  Conerly}, \bibinfo{person}{Amanda Askell}, \bibinfo{person}{Tom Henighan},
  \bibinfo{person}{Dawn Drain}, \bibinfo{person}{Ethan Perez},
  \bibinfo{person}{Nicholas Schiefer}, \bibinfo{person}{Zac Hatfield-Dodds},
  \bibinfo{person}{Nova DasSarma}, \bibinfo{person}{Eli Tran-Johnson},
  {et~al\mbox{.}}} \bibinfo{year}{2022}\natexlab{}.
\newblock \showarticletitle{Language models (mostly) know what they know}.
\newblock \bibinfo{journal}{\emph{arXiv preprint arXiv:2207.05221}}
  (\bibinfo{year}{2022}).
\newblock


\bibitem[Kahn~Jr et~al\mbox{.}(2009)]%
        {kahn2009toward}
\bibfield{author}{\bibinfo{person}{Charles~E Kahn~Jr},
  \bibinfo{person}{Curtis~P Langlotz}, \bibinfo{person}{Elizabeth~S Burnside},
  \bibinfo{person}{John~A Carrino}, \bibinfo{person}{David~S Channin},
  \bibinfo{person}{David~M Hovsepian}, {and} \bibinfo{person}{Daniel~L Rubin}.}
  \bibinfo{year}{2009}\natexlab{}.
\newblock \showarticletitle{Toward best practices in radiology reporting}.
\newblock \bibinfo{journal}{\emph{Radiology}} \bibinfo{volume}{252},
  \bibinfo{number}{3} (\bibinfo{year}{2009}), \bibinfo{pages}{852--856}.
\newblock


\bibitem[Karen and Sandra(2017)]%
        {karen2017contextual}
\bibfield{author}{\bibinfo{person}{Holtzblatt Karen} {and}
  \bibinfo{person}{Jones Sandra}.} \bibinfo{year}{2017}\natexlab{}.
\newblock \showarticletitle{Contextual inquiry: A participatory technique for
  system design}.
\newblock In \bibinfo{booktitle}{\emph{Participatory design}}.
  \bibinfo{publisher}{CRC Press}, \bibinfo{pages}{177--210}.
\newblock


\bibitem[Krishna et~al\mbox{.}(2020)]%
        {krishna2020generating}
\bibfield{author}{\bibinfo{person}{Kundan Krishna}, \bibinfo{person}{Sopan
  Khosla}, \bibinfo{person}{Jeffrey~P Bigham}, {and} \bibinfo{person}{Zachary~C
  Lipton}.} \bibinfo{year}{2020}\natexlab{}.
\newblock \showarticletitle{Generating SOAP notes from doctor-patient
  conversations using modular summarization techniques}.
\newblock \bibinfo{journal}{\emph{arXiv preprint arXiv:2005.01795}}
  (\bibinfo{year}{2020}).
\newblock


\bibitem[Kross and Guo(2021)]%
        {kross2021orienting}
\bibfield{author}{\bibinfo{person}{Sean Kross} {and} \bibinfo{person}{Philip
  Guo}.} \bibinfo{year}{2021}\natexlab{}.
\newblock \showarticletitle{Orienting, framing, bridging, magic, and
  counseling: How data scientists navigate the outer loop of client
  collaborations in industry and academia}.
\newblock \bibinfo{journal}{\emph{Proceedings of the ACM on Human-Computer
  Interaction}} \bibinfo{volume}{5}, \bibinfo{number}{CSCW2}
  (\bibinfo{year}{2021}), \bibinfo{pages}{1--28}.
\newblock


\bibitem[Kuo et~al\mbox{.}(2023)]%
        {kuo2023understanding}
\bibfield{author}{\bibinfo{person}{Tzu-Sheng Kuo}, \bibinfo{person}{Hong Shen},
  \bibinfo{person}{Jisoo Geum}, \bibinfo{person}{Nev Jones},
  \bibinfo{person}{Jason~I Hong}, \bibinfo{person}{Haiyi Zhu}, {and}
  \bibinfo{person}{Kenneth Holstein}.} \bibinfo{year}{2023}\natexlab{}.
\newblock \showarticletitle{Understanding Frontline Workers’ and Unhoused
  Individuals’ Perspectives on AI Used in Homeless Services}. In
  \bibinfo{booktitle}{\emph{Proceedings of the 2023 CHI Conference on Human
  Factors in Computing Systems}}. \bibinfo{pages}{1--17}.
\newblock


\bibitem[Langlots(2015)]%
        {langlotz2015radiology}
\bibfield{author}{\bibinfo{person}{Curtis~P. Langlots}.}
  \bibinfo{year}{2015}\natexlab{}.
\newblock \bibinfo{booktitle}{\emph{The radiology report: a guide to thoughtful
  communication for radiologists and other medical professionals}}.
\newblock \bibinfo{publisher}{Springer}.
\newblock


\bibitem[Langlotz(2019)]%
        {langlotz2019will}
\bibfield{author}{\bibinfo{person}{Curtis~P Langlotz}.}
  \bibinfo{year}{2019}\natexlab{}.
\newblock \bibinfo{title}{Will artificial intelligence replace radiologists?}
\newblock , \bibinfo{numpages}{e190058}~pages.
\newblock


\bibitem[Law(2014)]%
        {law2014radiographers}
\bibfield{author}{\bibinfo{person}{Robert Law}.}
  \bibinfo{year}{2014}\natexlab{}.
\newblock \showarticletitle{Radiographers,‘never events’ and the
  nasogastric tube}.
\newblock \bibinfo{journal}{\emph{Radiography}} \bibinfo{volume}{20},
  \bibinfo{number}{1} (\bibinfo{year}{2014}), \bibinfo{pages}{2--3}.
\newblock


\bibitem[Lecler et~al\mbox{.}(2023)]%
        {lecler2023revolutionizing}
\bibfield{author}{\bibinfo{person}{Augustin Lecler}, \bibinfo{person}{Lo{\"\i}c
  Duron}, {and} \bibinfo{person}{Philippe Soyer}.}
  \bibinfo{year}{2023}\natexlab{}.
\newblock \showarticletitle{Revolutionizing radiology with GPT-based models:
  Current applications, future possibilities and limitations of ChatGPT}.
\newblock \bibinfo{journal}{\emph{Diagnostic and Interventional Imaging}}
  \bibinfo{volume}{104}, \bibinfo{number}{6} (\bibinfo{year}{2023}),
  \bibinfo{pages}{269--274}.
\newblock


\bibitem[Lee et~al\mbox{.}(2023)]%
        {lee2023benefits}
\bibfield{author}{\bibinfo{person}{Peter Lee}, \bibinfo{person}{Sebastien
  Bubeck}, {and} \bibinfo{person}{Joseph Petro}.}
  \bibinfo{year}{2023}\natexlab{}.
\newblock \showarticletitle{Benefits, Limits, and Risks of GPT-4 as an AI
  Chatbot for Medicine}.
\newblock \bibinfo{journal}{\emph{New England Journal of Medicine}}
  \bibinfo{volume}{388}, \bibinfo{number}{13} (\bibinfo{year}{2023}),
  \bibinfo{pages}{1233--1239}.
\newblock


\bibitem[Liang et~al\mbox{.}(2022)]%
        {liang2022holistic}
\bibfield{author}{\bibinfo{person}{Percy Liang}, \bibinfo{person}{Rishi
  Bommasani}, \bibinfo{person}{Tony Lee}, \bibinfo{person}{Dimitris Tsipras},
  \bibinfo{person}{Dilara Soylu}, \bibinfo{person}{Michihiro Yasunaga},
  \bibinfo{person}{Yian Zhang}, \bibinfo{person}{Deepak Narayanan},
  \bibinfo{person}{Yuhuai Wu}, \bibinfo{person}{Ananya Kumar}, {et~al\mbox{.}}}
  \bibinfo{year}{2022}\natexlab{}.
\newblock \showarticletitle{Holistic evaluation of language models}.
\newblock \bibinfo{journal}{\emph{arXiv preprint arXiv:2211.09110}}
  (\bibinfo{year}{2022}).
\newblock


\bibitem[Liao et~al\mbox{.}(2023)]%
        {liao2023designerly}
\bibfield{author}{\bibinfo{person}{Q~Vera Liao}, \bibinfo{person}{Hariharan
  Subramonyam}, \bibinfo{person}{Jennifer Wang}, {and}
  \bibinfo{person}{Jennifer Wortman~Vaughan}.} \bibinfo{year}{2023}\natexlab{}.
\newblock \showarticletitle{Designerly understanding: Information needs for
  model transparency to support design ideation for AI-powered user
  experience}. In \bibinfo{booktitle}{\emph{Proceedings of the 2023 CHI
  conference on human factors in computing systems}}. \bibinfo{pages}{1--21}.
\newblock


\bibitem[Liao and Vaughan(2023)]%
        {liao2023ai}
\bibfield{author}{\bibinfo{person}{Q~Vera Liao} {and}
  \bibinfo{person}{Jennifer~Wortman Vaughan}.} \bibinfo{year}{2023}\natexlab{}.
\newblock \showarticletitle{AI Transparency in the Age of LLMs: A
  Human-Centered Research Roadmap}.
\newblock \bibinfo{journal}{\emph{arXiv preprint arXiv:2306.01941}}
  (\bibinfo{year}{2023}).
\newblock


\bibitem[Liao et~al\mbox{.}(2022)]%
        {liao2022connecting}
\bibfield{author}{\bibinfo{person}{Q~Vera Liao}, \bibinfo{person}{Yunfeng
  Zhang}, \bibinfo{person}{Ronny Luss}, {et~al\mbox{.}}}
  \bibinfo{year}{2022}\natexlab{}.
\newblock \showarticletitle{Connecting Algorithmic Research and Usage Contexts:
  A Perspective of Contextualized Evaluation for Explainable AI}. In
  \bibinfo{booktitle}{\emph{Proceedings of the AAAI Conference on Human
  Computation and Crowdsourcing}}, Vol.~\bibinfo{volume}{10}.
  \bibinfo{pages}{147--159}.
\newblock


\bibitem[Lindvall et~al\mbox{.}(2021)]%
        {lindvall2021rapid}
\bibfield{author}{\bibinfo{person}{Martin Lindvall}, \bibinfo{person}{Claes
  Lundstr{\"o}m}, {and} \bibinfo{person}{Jonas L{\"o}wgren}.}
  \bibinfo{year}{2021}\natexlab{}.
\newblock \showarticletitle{Rapid assisted visual search: Supporting digital
  pathologists with imperfect AI}. In \bibinfo{booktitle}{\emph{26th
  International Conference on Intelligent User Interfaces}}.
  \bibinfo{pages}{504--513}.
\newblock


\bibitem[Liu et~al\mbox{.}(2023a)]%
        {liu2023human}
\bibfield{author}{\bibinfo{person}{Houjiang Liu}, \bibinfo{person}{Anubrata
  Das}, \bibinfo{person}{Alexander Boltz}, \bibinfo{person}{Didi Zhou},
  \bibinfo{person}{Daisy Pinaroc}, \bibinfo{person}{Matthew Lease}, {and}
  \bibinfo{person}{Min~Kyung Lee}.} \bibinfo{year}{2023}\natexlab{a}.
\newblock \showarticletitle{Human-centered NLP Fact-checking: Co-Designing with
  Fact-checkers using Matchmaking for AI}.
\newblock \bibinfo{journal}{\emph{arXiv preprint arXiv:2308.07213}}
  (\bibinfo{year}{2023}).
\newblock


\bibitem[Liu et~al\mbox{.}(2023b)]%
        {liu2023exploring}
\bibfield{author}{\bibinfo{person}{Qianchu Liu}, \bibinfo{person}{Stephanie
  Hyland}, \bibinfo{person}{Shruthi Bannur}, \bibinfo{person}{Kenza Bouzid},
  \bibinfo{person}{Daniel~C Castro}, \bibinfo{person}{Maria~Teodora
  Wetscherek}, \bibinfo{person}{Robert Tinn}, \bibinfo{person}{Harshita
  Sharma}, \bibinfo{person}{Fernando P{\'e}rez-Garc{\'\i}a},
  \bibinfo{person}{Anton Schwaighofer}, {et~al\mbox{.}}}
  \bibinfo{year}{2023}\natexlab{b}.
\newblock \showarticletitle{Exploring the Boundaries of GPT-4 in Radiology}.
\newblock \bibinfo{journal}{\emph{arXiv preprint arXiv:2310.14573}}
  (\bibinfo{year}{2023}).
\newblock


\bibitem[Luria et~al\mbox{.}(2019)]%
        {luria2019re}
\bibfield{author}{\bibinfo{person}{Michal Luria}, \bibinfo{person}{Samantha
  Reig}, \bibinfo{person}{Xiang~Zhi Tan}, \bibinfo{person}{Aaron Steinfeld},
  \bibinfo{person}{Jodi Forlizzi}, {and} \bibinfo{person}{John Zimmerman}.}
  \bibinfo{year}{2019}\natexlab{}.
\newblock \showarticletitle{Re-Embodiment and Co-Embodiment: Exploration of
  social presence for robots and conversational agents}. In
  \bibinfo{booktitle}{\emph{Proceedings of the 2019 on Designing Interactive
  Systems Conference}}. \bibinfo{pages}{633--644}.
\newblock


\bibitem[Ma et~al\mbox{.}(2023)]%
        {ma2023impressiongpt}
\bibfield{author}{\bibinfo{person}{Chong Ma}, \bibinfo{person}{Zihao Wu},
  \bibinfo{person}{Jiaqi Wang}, \bibinfo{person}{Shaochen Xu},
  \bibinfo{person}{Yaonai Wei}, \bibinfo{person}{Zhengliang Liu},
  \bibinfo{person}{Lei Guo}, \bibinfo{person}{Xiaoyan Cai},
  \bibinfo{person}{Shu Zhang}, \bibinfo{person}{Tuo Zhang}, {et~al\mbox{.}}}
  \bibinfo{year}{2023}\natexlab{}.
\newblock \showarticletitle{ImpressionGPT: an iterative optimizing framework
  for radiology report summarization with chatGPT}.
\newblock \bibinfo{journal}{\emph{arXiv preprint arXiv:2304.08448}}
  (\bibinfo{year}{2023}).
\newblock


\bibitem[Martin et~al\mbox{.}(2012)]%
        {martin2012universal}
\bibfield{author}{\bibinfo{person}{Bella Martin}, \bibinfo{person}{Bruce
  Hanington}, {and} \bibinfo{person}{Bruce~M Hanington}.}
  \bibinfo{year}{2012}\natexlab{}.
\newblock \showarticletitle{Universal methods of design: 100 ways to research
  complex problems}.
\newblock \bibinfo{journal}{\emph{Develop Innovative Ideas, and Design
  Effective Solutions}} (\bibinfo{year}{2012}), \bibinfo{pages}{12--13}.
\newblock


\bibitem[Matthiesen et~al\mbox{.}(2021)]%
        {matthiesen2021clinician}
\bibfield{author}{\bibinfo{person}{Stina Matthiesen},
  \bibinfo{person}{S{\o}ren~Z{\"o}ga Diederichsen}, \bibinfo{person}{Mikkel
  Klitzing~Hartmann Hansen}, \bibinfo{person}{Christina Villumsen},
  \bibinfo{person}{Mats Christian~H{\o}jbjerg Lassen},
  \bibinfo{person}{Peter~Karl Jacobsen}, \bibinfo{person}{Niels Risum},
  \bibinfo{person}{Bo~Gregers Winkel}, \bibinfo{person}{Berit~T Philbert},
  \bibinfo{person}{Jesper~Hastrup Svendsen}, {et~al\mbox{.}}}
  \bibinfo{year}{2021}\natexlab{}.
\newblock \showarticletitle{Clinician preimplementation perspectives of a
  decision-support tool for the prediction of cardiac arrhythmia based on
  machine learning: near-live feasibility and qualitative study}.
\newblock \bibinfo{journal}{\emph{JMIR human factors}} \bibinfo{volume}{8},
  \bibinfo{number}{4} (\bibinfo{year}{2021}), \bibinfo{pages}{e26964}.
\newblock


\bibitem[Microsoft(2023)]%
        {copilot}
\bibfield{author}{\bibinfo{person}{Microsoft}.}
  \bibinfo{year}{2023}\natexlab{}.
\newblock \bibinfo{title}{Microsoft Copilot: Your everyday AI companion}.
\newblock
\newblock
\urldef\tempurl%
\url{https://copilot.microsoft.com/}
\showURL{%
\tempurl}


\bibitem[Miller et~al\mbox{.}(2017)]%
        {miller2017explainable}
\bibfield{author}{\bibinfo{person}{Tim Miller}, \bibinfo{person}{Piers Howe},
  {and} \bibinfo{person}{Liz Sonenberg}.} \bibinfo{year}{2017}\natexlab{}.
\newblock \showarticletitle{Explainable AI: Beware of inmates running the
  asylum or: How I learnt to stop worrying and love the social and behavioural
  sciences}.
\newblock \bibinfo{journal}{\emph{arXiv preprint arXiv:1712.00547}}
  (\bibinfo{year}{2017}).
\newblock


\bibitem[Moor et~al\mbox{.}(2023)]%
        {moor2023foundation}
\bibfield{author}{\bibinfo{person}{Michael Moor}, \bibinfo{person}{Oishi
  Banerjee}, \bibinfo{person}{Zahra Shakeri~Hossein Abad},
  \bibinfo{person}{Harlan~M Krumholz}, \bibinfo{person}{Jure Leskovec},
  \bibinfo{person}{Eric~J Topol}, {and} \bibinfo{person}{Pranav Rajpurkar}.}
  \bibinfo{year}{2023}\natexlab{}.
\newblock \showarticletitle{Foundation models for generalist medical artificial
  intelligence}.
\newblock \bibinfo{journal}{\emph{Nature}} \bibinfo{volume}{616},
  \bibinfo{number}{7956} (\bibinfo{year}{2023}), \bibinfo{pages}{259--265}.
\newblock


\bibitem[Morris et~al\mbox{.}(2023)]%
        {morris2023design}
\bibfield{author}{\bibinfo{person}{Meredith~Ringel Morris},
  \bibinfo{person}{Carrie~J Cai}, \bibinfo{person}{Jess Holbrook},
  \bibinfo{person}{Chinmay Kulkarni}, {and} \bibinfo{person}{Michael Terry}.}
  \bibinfo{year}{2023}\natexlab{}.
\newblock \showarticletitle{The design space of generative models}.
\newblock \bibinfo{journal}{\emph{arXiv preprint arXiv:2304.10547}}
  (\bibinfo{year}{2023}).
\newblock


\bibitem[Morrison et~al\mbox{.}(2021)]%
        {morrison2021social}
\bibfield{author}{\bibinfo{person}{Cecily Morrison}, \bibinfo{person}{Edward
  Cutrell}, \bibinfo{person}{Martin Grayson}, \bibinfo{person}{Anja Thieme},
  \bibinfo{person}{Alex Taylor}, \bibinfo{person}{Geert Roumen},
  \bibinfo{person}{Camilla Longden}, \bibinfo{person}{Sebastian Tschiatschek},
  \bibinfo{person}{Rita Faia~Marques}, {and} \bibinfo{person}{Abigail Sellen}.}
  \bibinfo{year}{2021}\natexlab{}.
\newblock \showarticletitle{Social Sensemaking with AI: Designing an Open-ended
  AI experience with a Blind Child}. In \bibinfo{booktitle}{\emph{Proceedings
  of the 2021 CHI Conference on Human Factors in Computing Systems}}.
  \bibinfo{pages}{1--14}.
\newblock


\bibitem[Nabla(2023)]%
        {NablaCopilot_2023}
\bibfield{author}{\bibinfo{person}{Nabla}.} \bibinfo{year}{2023}\natexlab{}.
\newblock \bibinfo{title}{{N}abla {C}opilot · {E}njoy care again}.
\newblock
\newblock
\urldef\tempurl%
\url{https://www.nabla.com/}
\showURL{%
\tempurl}
\newblock
\shownote{[Accessed 11-08-2023]}.


\bibitem[Naik et~al\mbox{.}(2001)]%
        {naik2001radiology}
\bibfield{author}{\bibinfo{person}{Sandeep~S Naik}, \bibinfo{person}{Anthony
  Hanbidge}, {and} \bibinfo{person}{Stephanie~R Wilson}.}
  \bibinfo{year}{2001}\natexlab{}.
\newblock \showarticletitle{Radiology reports: examining radiologist and
  clinician preferences regarding style and content}.
\newblock \bibinfo{journal}{\emph{American Journal of Roentgenology}}
  \bibinfo{volume}{176}, \bibinfo{number}{3} (\bibinfo{year}{2001}),
  \bibinfo{pages}{591--598}.
\newblock


\bibitem[Nori et~al\mbox{.}(2023)]%
        {nori2023capabilities}
\bibfield{author}{\bibinfo{person}{Harsha Nori}, \bibinfo{person}{Nicholas
  King}, \bibinfo{person}{Scott~Mayer McKinney}, \bibinfo{person}{Dean
  Carignan}, {and} \bibinfo{person}{Eric Horvitz}.}
  \bibinfo{year}{2023}\natexlab{}.
\newblock \showarticletitle{Capabilities of gpt-4 on medical challenge
  problems}.
\newblock \bibinfo{journal}{\emph{arXiv preprint arXiv:2303.13375}}
  (\bibinfo{year}{2023}).
\newblock


\bibitem[Nuance-Microsoft(2023)]%
        {nuanceNuanceMicrosoft}
\bibfield{author}{\bibinfo{person}{Nuance-Microsoft}.}
  \bibinfo{year}{2023}\natexlab{}.
\newblock \bibinfo{title}{{N}uance and {M}icrosoft {A}nnounce the {F}irst
  {F}ully {A}{I}-{A}utomated {C}linical {D}ocumentation {A}pplication for
  {H}ealthcare --- news.nuance.com}.
\newblock
  \bibinfo{howpublished}{\url{https://news.nuance.com/2023-03-20-Nuance-and-Microsoft-Announce-the-First-Fully-AI-Automated-Clinical-Documentation-Application-for-Healthcare}}.
\newblock
\newblock
\shownote{[Accessed 11-08-2023]}.


\bibitem[Oh et~al\mbox{.}(2023)]%
        {oh2023chatgpt}
\bibfield{author}{\bibinfo{person}{Namkee Oh}, \bibinfo{person}{Gyu-Seong
  Choi}, {and} \bibinfo{person}{Woo~Yong Lee}.}
  \bibinfo{year}{2023}\natexlab{}.
\newblock \showarticletitle{ChatGPT goes to the operating room: evaluating
  GPT-4 performance and its potential in surgical education and training in the
  era of large language models}.
\newblock \bibinfo{journal}{\emph{Annals of Surgical Treatment and Research}}
  \bibinfo{volume}{104}, \bibinfo{number}{5} (\bibinfo{year}{2023}),
  \bibinfo{pages}{269}.
\newblock


\bibitem[Ontika et~al\mbox{.}(2023)]%
        {ontika2023pairads}
\bibfield{author}{\bibinfo{person}{Nazmun~Nisat Ontika},
  \bibinfo{person}{Sheree~May Sassmannshausen},
  \bibinfo{person}{Aparecido~Fabiano Pinatti De~Carvalho}, {and}
  \bibinfo{person}{Volkmar Pipek}.} \bibinfo{year}{2023}\natexlab{}.
\newblock \showarticletitle{PAIRADS: Hybrid Interaction Between Humans and AI
  in Radiology}.
\newblock In \bibinfo{booktitle}{\emph{HHAI 2023: Augmenting Human Intellect}}.
  \bibinfo{publisher}{IOS Press}, \bibinfo{pages}{395--397}.
\newblock


\bibitem[Ontika et~al\mbox{.}(2022)]%
        {ontika2022exploring}
\bibfield{author}{\bibinfo{person}{Nazmun~Nisat Ontika},
  \bibinfo{person}{Hussain~Abid Syed}, \bibinfo{person}{Sheree~May
  Sa{\ss}mannshausen}, \bibinfo{person}{Richard~HR Harper},
  \bibinfo{person}{Yunan Chen}, \bibinfo{person}{Sun~Young Park},
  \bibinfo{person}{Miria Grisot}, \bibinfo{person}{Astrid Chow},
  \bibinfo{person}{Nils Blaumer}, \bibinfo{person}{Aparecido~Fabiano Pinatti~de
  Carvalho}, {et~al\mbox{.}}} \bibinfo{year}{2022}\natexlab{}.
\newblock \showarticletitle{Exploring human-centered AI in healthcare:
  diagnosis, explainability, and trust}.
\newblock  (\bibinfo{year}{2022}).
\newblock


\bibitem[OpenAI(2023)]%
        {openai2023gpt4}
\bibfield{author}{\bibinfo{person}{OpenAI}.} \bibinfo{year}{2023}\natexlab{}.
\newblock \bibinfo{title}{GPT-4 Technical Report}.
\newblock
\newblock
\showeprint[arxiv]{2303.08774}~[cs.CL]


\bibitem[Osman~Andersen et~al\mbox{.}(2021)]%
        {osman2021realizing}
\bibfield{author}{\bibinfo{person}{Tariq Osman~Andersen},
  \bibinfo{person}{Francisco Nunes}, \bibinfo{person}{Lauren Wilcox},
  \bibinfo{person}{Elizabeth Kaziunas}, \bibinfo{person}{Stina Matthiesen},
  {and} \bibinfo{person}{Farah Magrabi}.} \bibinfo{year}{2021}\natexlab{}.
\newblock \showarticletitle{Realizing AI in healthcare: challenges appearing in
  the wild}. In \bibinfo{booktitle}{\emph{Extended Abstracts of the 2021 CHI
  Conference on Human Factors in Computing Systems}}. \bibinfo{pages}{1--5}.
\newblock


\bibitem[Patel et~al\mbox{.}(2019)]%
        {patel2019human}
\bibfield{author}{\bibinfo{person}{Bhavik~N Patel}, \bibinfo{person}{Louis
  Rosenberg}, \bibinfo{person}{Gregg Willcox}, \bibinfo{person}{David Baltaxe},
  \bibinfo{person}{Mimi Lyons}, \bibinfo{person}{Jeremy Irvin},
  \bibinfo{person}{Pranav Rajpurkar}, \bibinfo{person}{Timothy Amrhein},
  \bibinfo{person}{Rajan Gupta}, \bibinfo{person}{Safwan Halabi},
  {et~al\mbox{.}}} \bibinfo{year}{2019}\natexlab{}.
\newblock \showarticletitle{Human--machine partnership with artificial
  intelligence for chest radiograph diagnosis}.
\newblock \bibinfo{journal}{\emph{NPJ digital medicine}} \bibinfo{volume}{2},
  \bibinfo{number}{1} (\bibinfo{year}{2019}), \bibinfo{pages}{111}.
\newblock


\bibitem[Petersen et~al\mbox{.}(2022)]%
        {petersen2022responsible}
\bibfield{author}{\bibinfo{person}{Eike Petersen}, \bibinfo{person}{Yannik
  Potdevin}, \bibinfo{person}{Esfandiar Mohammadi}, \bibinfo{person}{Stephan
  Zidowitz}, \bibinfo{person}{Sabrina Breyer}, \bibinfo{person}{Dirk Nowotka},
  \bibinfo{person}{Sandra Henn}, \bibinfo{person}{Ludwig Pechmann},
  \bibinfo{person}{Martin Leucker}, \bibinfo{person}{Philipp Rostalski},
  {et~al\mbox{.}}} \bibinfo{year}{2022}\natexlab{}.
\newblock \showarticletitle{Responsible and regulatory conform machine learning
  for medicine: a survey of challenges and solutions}.
\newblock \bibinfo{journal}{\emph{IEEE Access}}  \bibinfo{volume}{10}
  (\bibinfo{year}{2022}), \bibinfo{pages}{58375--58418}.
\newblock


\bibitem[Petridis et~al\mbox{.}(2023)]%
        {petridis2023promptinfuser}
\bibfield{author}{\bibinfo{person}{Savvas Petridis}, \bibinfo{person}{Michael
  Terry}, {and} \bibinfo{person}{Carrie~Jun Cai}.}
  \bibinfo{year}{2023}\natexlab{}.
\newblock \showarticletitle{PromptInfuser: Bringing User Interface Mock-ups to
  Life with Large Language Models}. In \bibinfo{booktitle}{\emph{Extended
  Abstracts of the 2023 CHI Conference on Human Factors in Computing Systems}}.
  \bibinfo{pages}{1--6}.
\newblock


\bibitem[Piorkowski et~al\mbox{.}(2021)]%
        {piorkowski2021ai}
\bibfield{author}{\bibinfo{person}{David Piorkowski}, \bibinfo{person}{Soya
  Park}, \bibinfo{person}{April~Yi Wang}, \bibinfo{person}{Dakuo Wang},
  \bibinfo{person}{Michael Muller}, {and} \bibinfo{person}{Felix Portnoy}.}
  \bibinfo{year}{2021}\natexlab{}.
\newblock \showarticletitle{How ai developers overcome communication challenges
  in a multidisciplinary team: A case study}.
\newblock \bibinfo{journal}{\emph{Proceedings of the ACM on Human-Computer
  Interaction}} \bibinfo{volume}{5}, \bibinfo{number}{CSCW1}
  (\bibinfo{year}{2021}), \bibinfo{pages}{1--25}.
\newblock


\bibitem[Preston et~al\mbox{.}(2023)]%
        {preston2023toward}
\bibfield{author}{\bibinfo{person}{Sam Preston}, \bibinfo{person}{Mu Wei},
  \bibinfo{person}{Rajesh Rao}, \bibinfo{person}{Robert Tinn},
  \bibinfo{person}{Naoto Usuyama}, \bibinfo{person}{Michael Lucas},
  \bibinfo{person}{Yu Gu}, \bibinfo{person}{Roshanthi Weerasinghe},
  \bibinfo{person}{Soohee Lee}, \bibinfo{person}{Brian Piening},
  {et~al\mbox{.}}} \bibinfo{year}{2023}\natexlab{}.
\newblock \showarticletitle{Toward structuring real-world data: Deep learning
  for extracting oncology information from clinical text with patient-level
  supervision}.
\newblock \bibinfo{journal}{\emph{Patterns}} \bibinfo{volume}{4},
  \bibinfo{number}{4} (\bibinfo{year}{2023}).
\newblock


\bibitem[Procter et~al\mbox{.}(2023)]%
        {procter2023holding}
\bibfield{author}{\bibinfo{person}{Rob Procter}, \bibinfo{person}{Peter
  Tolmie}, {and} \bibinfo{person}{Mark Rouncefield}.}
  \bibinfo{year}{2023}\natexlab{}.
\newblock \showarticletitle{Holding AI to account: Challenges for the delivery
  of trustworthy AI in healthcare}.
\newblock \bibinfo{journal}{\emph{ACM Transactions on Computer-Human
  Interaction}} \bibinfo{volume}{30}, \bibinfo{number}{2}
  (\bibinfo{year}{2023}), \bibinfo{pages}{1--34}.
\newblock


\bibitem[Reig et~al\mbox{.}(2020)]%
        {reig2020not}
\bibfield{author}{\bibinfo{person}{Samantha Reig}, \bibinfo{person}{Michal
  Luria}, \bibinfo{person}{Janet~Z Wang}, \bibinfo{person}{Danielle Oltman},
  \bibinfo{person}{Elizabeth~Jeanne Carter}, \bibinfo{person}{Aaron Steinfeld},
  \bibinfo{person}{Jodi Forlizzi}, {and} \bibinfo{person}{John Zimmerman}.}
  \bibinfo{year}{2020}\natexlab{}.
\newblock \showarticletitle{Not Some Random Agent: Multi-person interaction
  with a personalizing service robot}. In \bibinfo{booktitle}{\emph{Proceedings
  of the 2020 ACM/IEEE international conference on human-robot interaction}}.
  \bibinfo{pages}{289--297}.
\newblock


\bibitem[Rimmer(2017)]%
        {rimmer2017radiologist}
\bibfield{author}{\bibinfo{person}{Abi Rimmer}.}
  \bibinfo{year}{2017}\natexlab{}.
\newblock \showarticletitle{Radiologist shortage leaves patient care at risk,
  warns royal college}.
\newblock \bibinfo{journal}{\emph{BMJ: British Medical Journal (Online)}}
  \bibinfo{volume}{359} (\bibinfo{year}{2017}).
\newblock


\bibitem[Robertson and Salehi(2020)]%
        {robertson2020if}
\bibfield{author}{\bibinfo{person}{Samantha Robertson} {and}
  \bibinfo{person}{Niloufar Salehi}.} \bibinfo{year}{2020}\natexlab{}.
\newblock \showarticletitle{What If I Don't Like Any Of The Choices? The Limits
  of Preference Elicitation for Participatory Algorithm Design}.
\newblock \bibinfo{journal}{\emph{arXiv preprint arXiv:2007.06718}}
  (\bibinfo{year}{2020}).
\newblock


\bibitem[Sahni et~al\mbox{.}({[n.\,d.]})]%
        {sahniadministrative}
\bibfield{author}{\bibinfo{person}{NR Sahni}, \bibinfo{person}{P Mishra},
  \bibinfo{person}{B Carrus}, {and} \bibinfo{person}{DM Cutler}.}
  \bibinfo{year}{[n.\,d.]}\natexlab{}.
\newblock \bibinfo{title}{Administrative Simplification: How to Save a
  Quarter-Trillion Dollars in US Healthcare. McKinsey \& Company. October 20,
  2021}.
\newblock
\newblock


\bibitem[Sasha~Luccioni and Rogers(2023)]%
        {sasha2023mind}
\bibfield{author}{\bibinfo{person}{Alexandra Sasha~Luccioni} {and}
  \bibinfo{person}{Anna Rogers}.} \bibinfo{year}{2023}\natexlab{}.
\newblock \showarticletitle{Mind your Language (Model): Fact-Checking LLMs and
  their Role in NLP Research and Practice}.
\newblock \bibinfo{journal}{\emph{arXiv e-prints}} (\bibinfo{year}{2023}),
  \bibinfo{pages}{arXiv--2308}.
\newblock


\bibitem[Sectra(2013)]%
        {sectra2013}
\bibfield{author}{\bibinfo{person}{Sectra}.} \bibinfo{year}{2013}\natexlab{}.
\newblock \bibinfo{title}{How radiology can improve communication with
  referring physicians}.
\newblock
\newblock
\urldef\tempurl%
\url{https://sectraprodstorage01.blob.core.windows.net/medical-uploads/2017/09/report-how-radiology-can-improve-communication-with-referring-physicians.pdf}
\showURL{%
\tempurl}
\newblock
\shownote{[Accessed 11-22-2023]}.


\bibitem[Sekhon et~al\mbox{.}(2017)]%
        {sekhon2017acceptability}
\bibfield{author}{\bibinfo{person}{Mandeep Sekhon}, \bibinfo{person}{Martin
  Cartwright}, {and} \bibinfo{person}{Jill~J Francis}.}
  \bibinfo{year}{2017}\natexlab{}.
\newblock \showarticletitle{Acceptability of healthcare interventions: an
  overview of reviews and development of a theoretical framework}.
\newblock \bibinfo{journal}{\emph{BMC health services research}}
  \bibinfo{volume}{17}, \bibinfo{number}{1} (\bibinfo{year}{2017}),
  \bibinfo{pages}{1--13}.
\newblock


\bibitem[Sendak et~al\mbox{.}(2020)]%
        {sendak2020human}
\bibfield{author}{\bibinfo{person}{Mark Sendak},
  \bibinfo{person}{Madeleine~Clare Elish}, \bibinfo{person}{Michael Gao},
  \bibinfo{person}{Joseph Futoma}, \bibinfo{person}{William Ratliff},
  \bibinfo{person}{Marshall Nichols}, \bibinfo{person}{Armando Bedoya},
  \bibinfo{person}{Suresh Balu}, {and} \bibinfo{person}{Cara O'Brien}.}
  \bibinfo{year}{2020}\natexlab{}.
\newblock \showarticletitle{" The human body is a black box" supporting
  clinical decision-making with deep learning}. In
  \bibinfo{booktitle}{\emph{Proceedings of the 2020 conference on fairness,
  accountability, and transparency}}. \bibinfo{pages}{99--109}.
\newblock


\bibitem[Shanahan(2022)]%
        {shanahan2022talking}
\bibfield{author}{\bibinfo{person}{Murray Shanahan}.}
  \bibinfo{year}{2022}\natexlab{}.
\newblock \showarticletitle{Talking About Large Language Models}.
\newblock \bibinfo{journal}{\emph{arXiv preprint arXiv:2212.03551}}
  (\bibinfo{year}{2022}).
\newblock


\bibitem[Sherry et~al\mbox{.}(2022)]%
        {sherry2022acr}
\bibfield{author}{\bibinfo{person}{C Sherry}, \bibinfo{person}{M Adams},
  \bibinfo{person}{L Berlin}, \bibinfo{person}{L Fajardo}, \bibinfo{person}{G
  Gazelle}, \bibinfo{person}{DB Haseman}, {et~al\mbox{.}}}
  \bibinfo{year}{2022}\natexlab{}.
\newblock \showarticletitle{ACR practice guideline for communication of
  diagnostic imaging findings}.
\newblock \bibinfo{journal}{\emph{American College of Radiology}}
  (\bibinfo{year}{2022}).
\newblock


\bibitem[Simkus(2023)]%
        {Simkus_2023}
\bibfield{author}{\bibinfo{person}{Julia Simkus}.}
  \bibinfo{year}{2023}\natexlab{}.
\newblock \bibinfo{title}{Snowball sampling method: Definition, Techniques \&
  Examples}.
\newblock
\newblock
\urldef\tempurl%
\url{https://www.simplypsychology.org/snowball-sampling.html}
\showURL{%
\tempurl}


\bibitem[Singhal et~al\mbox{.}(2022)]%
        {singhal2022large}
\bibfield{author}{\bibinfo{person}{Karan Singhal}, \bibinfo{person}{Shekoofeh
  Azizi}, \bibinfo{person}{Tao Tu}, \bibinfo{person}{S~Sara Mahdavi},
  \bibinfo{person}{Jason Wei}, \bibinfo{person}{Hyung~Won Chung},
  \bibinfo{person}{Nathan Scales}, \bibinfo{person}{Ajay Tanwani},
  \bibinfo{person}{Heather Cole-Lewis}, \bibinfo{person}{Stephen Pfohl},
  {et~al\mbox{.}}} \bibinfo{year}{2022}\natexlab{}.
\newblock \showarticletitle{Large Language Models Encode Clinical Knowledge}.
\newblock \bibinfo{journal}{\emph{arXiv preprint arXiv:2212.13138}}
  (\bibinfo{year}{2022}).
\newblock


\bibitem[Singhal et~al\mbox{.}(2023a)]%
        {singhal2023large}
\bibfield{author}{\bibinfo{person}{Karan Singhal}, \bibinfo{person}{Shekoofeh
  Azizi}, \bibinfo{person}{Tao Tu}, \bibinfo{person}{S~Sara Mahdavi},
  \bibinfo{person}{Jason Wei}, \bibinfo{person}{Hyung~Won Chung},
  \bibinfo{person}{Nathan Scales}, \bibinfo{person}{Ajay Tanwani},
  \bibinfo{person}{Heather Cole-Lewis}, \bibinfo{person}{Stephen Pfohl},
  {et~al\mbox{.}}} \bibinfo{year}{2023}\natexlab{a}.
\newblock \showarticletitle{Large language models encode clinical knowledge}.
\newblock \bibinfo{journal}{\emph{Nature}} (\bibinfo{year}{2023}),
  \bibinfo{pages}{1--9}.
\newblock


\bibitem[Singhal et~al\mbox{.}(2023b)]%
        {singhal2023towards}
\bibfield{author}{\bibinfo{person}{Karan Singhal}, \bibinfo{person}{Tao Tu},
  \bibinfo{person}{Juraj Gottweis}, \bibinfo{person}{Rory Sayres},
  \bibinfo{person}{Ellery Wulczyn}, \bibinfo{person}{Le Hou},
  \bibinfo{person}{Kevin Clark}, \bibinfo{person}{Stephen Pfohl},
  \bibinfo{person}{Heather Cole-Lewis}, \bibinfo{person}{Darlene Neal},
  {et~al\mbox{.}}} \bibinfo{year}{2023}\natexlab{b}.
\newblock \showarticletitle{Towards expert-level medical question answering
  with large language models}.
\newblock \bibinfo{journal}{\emph{arXiv preprint arXiv:2305.09617}}
  (\bibinfo{year}{2023}).
\newblock


\bibitem[Strohm et~al\mbox{.}(2020)]%
        {strohm2020implementation}
\bibfield{author}{\bibinfo{person}{Lea Strohm}, \bibinfo{person}{Charisma
  Hehakaya}, \bibinfo{person}{Erik~R Ranschaert}, \bibinfo{person}{Wouter~PC
  Boon}, {and} \bibinfo{person}{Ellen~HM Moors}.}
  \bibinfo{year}{2020}\natexlab{}.
\newblock \showarticletitle{Implementation of artificial intelligence (AI)
  applications in radiology: hindering and facilitating factors}.
\newblock \bibinfo{journal}{\emph{European radiology}}  \bibinfo{volume}{30}
  (\bibinfo{year}{2020}), \bibinfo{pages}{5525--5532}.
\newblock


\bibitem[Subramonyam et~al\mbox{.}(2022)]%
        {subramonyam2022solving}
\bibfield{author}{\bibinfo{person}{Hariharan Subramonyam},
  \bibinfo{person}{Jane Im}, \bibinfo{person}{Colleen Seifert}, {and}
  \bibinfo{person}{Eytan Adar}.} \bibinfo{year}{2022}\natexlab{}.
\newblock \showarticletitle{Solving separation-of-concerns problems in
  collaborative design of human-AI systems through leaky abstractions}. In
  \bibinfo{booktitle}{\emph{Proceedings of the 2022 CHI Conference on Human
  Factors in Computing Systems}}. \bibinfo{pages}{1--21}.
\newblock


\bibitem[Thieme et~al\mbox{.}(2020a)]%
        {thieme2020machine}
\bibfield{author}{\bibinfo{person}{Anja Thieme}, \bibinfo{person}{Danielle
  Belgrave}, {and} \bibinfo{person}{Gavin Doherty}.}
  \bibinfo{year}{2020}\natexlab{a}.
\newblock \showarticletitle{Machine learning in mental health: A systematic
  review of the HCI literature to support the development of effective and
  implementable ML systems}.
\newblock \bibinfo{journal}{\emph{ACM Transactions on Computer-Human
  Interaction (TOCHI)}} \bibinfo{volume}{27}, \bibinfo{number}{5}
  (\bibinfo{year}{2020}), \bibinfo{pages}{1--53}.
\newblock


\bibitem[Thieme et~al\mbox{.}(2020b)]%
        {thieme2020interpretability}
\bibfield{author}{\bibinfo{person}{Anja Thieme}, \bibinfo{person}{Ed Cutrell},
  \bibinfo{person}{Cecily Morrison}, \bibinfo{person}{Alex Taylor}, {and}
  \bibinfo{person}{Abigail Sellen}.} \bibinfo{year}{2020}\natexlab{b}.
\newblock \showarticletitle{Interpretability as a dynamic of human-AI
  interaction}.
\newblock \bibinfo{journal}{\emph{Interactions}} \bibinfo{volume}{27},
  \bibinfo{number}{5} (\bibinfo{year}{2020}), \bibinfo{pages}{40--45}.
\newblock


\bibitem[Thieme et~al\mbox{.}(2023a)]%
        {thieme2023designing}
\bibfield{author}{\bibinfo{person}{Anja Thieme}, \bibinfo{person}{Maryann
  Hanratty}, \bibinfo{person}{Maria Lyons}, \bibinfo{person}{Jorge Palacios},
  \bibinfo{person}{Rita~Faia Marques}, \bibinfo{person}{Cecily Morrison}, {and}
  \bibinfo{person}{Gavin Doherty}.} \bibinfo{year}{2023}\natexlab{a}.
\newblock \showarticletitle{Designing human-centered AI for mental health:
  Developing clinically relevant applications for online CBT treatment}.
\newblock \bibinfo{journal}{\emph{ACM Transactions on Computer-Human
  Interaction}} \bibinfo{volume}{30}, \bibinfo{number}{2}
  (\bibinfo{year}{2023}), \bibinfo{pages}{1--50}.
\newblock


\bibitem[Thieme et~al\mbox{.}(2023b)]%
        {thieme2023foundation}
\bibfield{author}{\bibinfo{person}{Anja Thieme}, \bibinfo{person}{Aditya Nori},
  \bibinfo{person}{Marzyeh Ghassemi}, \bibinfo{person}{Rishi Bommasani},
  \bibinfo{person}{Tariq~Osman Andersen}, {and} \bibinfo{person}{Ewa Luger}.}
  \bibinfo{year}{2023}\natexlab{b}.
\newblock \showarticletitle{Foundation Models in Healthcare: Opportunities,
  Risks \& Strategies Forward}. In \bibinfo{booktitle}{\emph{Extended Abstracts
  of the 2023 CHI Conference on Human Factors in Computing Systems}}.
  \bibinfo{pages}{1--4}.
\newblock


\bibitem[Touvron et~al\mbox{.}(2023)]%
        {touvron2023llama}
\bibfield{author}{\bibinfo{person}{Hugo Touvron}, \bibinfo{person}{Louis
  Martin}, \bibinfo{person}{Kevin Stone}, \bibinfo{person}{Peter Albert},
  \bibinfo{person}{Amjad Almahairi}, \bibinfo{person}{Yasmine Babaei},
  \bibinfo{person}{Nikolay Bashlykov}, \bibinfo{person}{Soumya Batra},
  \bibinfo{person}{Prajjwal Bhargava}, \bibinfo{person}{Shruti Bhosale},
  {et~al\mbox{.}}} \bibinfo{year}{2023}\natexlab{}.
\newblock \showarticletitle{Llama 2: Open foundation and fine-tuned chat
  models}.
\newblock \bibinfo{journal}{\emph{arXiv preprint arXiv:2307.09288}}
  (\bibinfo{year}{2023}).
\newblock


\bibitem[Tu et~al\mbox{.}(2023a)]%
        {tu2023towards}
\bibfield{author}{\bibinfo{person}{Tao Tu}, \bibinfo{person}{Shekoofeh Azizi},
  \bibinfo{person}{Danny Driess}, \bibinfo{person}{Mike Schaekermann},
  \bibinfo{person}{Mohamed Amin}, \bibinfo{person}{Pi-Chuan Chang},
  \bibinfo{person}{Andrew Carroll}, \bibinfo{person}{Chuck Lau},
  \bibinfo{person}{Ryutaro Tanno}, \bibinfo{person}{Ira Ktena},
  {et~al\mbox{.}}} \bibinfo{year}{2023}\natexlab{a}.
\newblock \showarticletitle{Towards generalist biomedical ai}.
\newblock \bibinfo{journal}{\emph{arXiv preprint arXiv:2307.14334}}
  (\bibinfo{year}{2023}).
\newblock


\bibitem[Tu et~al\mbox{.}(2023b)]%
        {tu2023generalist}
\bibfield{author}{\bibinfo{person}{Tao Tu}, \bibinfo{person}{Shekoofeh Azizi},
  \bibinfo{person}{Danny Driess}, \bibinfo{person}{Mike Schaekermann},
  \bibinfo{person}{Mohamed Amin}, \bibinfo{person}{Pi-Chuan Chang},
  \bibinfo{person}{Andrew Carroll}, \bibinfo{person}{Chuck Lau},
  \bibinfo{person}{Ryutaro Tanno}, \bibinfo{person}{Ira Ktena},
  \bibinfo{person}{Basil Mustafa}, \bibinfo{person}{Aakanksha Chowdhery},
  \bibinfo{person}{Yun Liu}, \bibinfo{person}{Simon Kornblith},
  \bibinfo{person}{David Fleet}, \bibinfo{person}{Philip Mansfield},
  \bibinfo{person}{Sushant Prakash}, \bibinfo{person}{Renee Wong},
  \bibinfo{person}{Sunny Virmani}, \bibinfo{person}{Christopher Semturs},
  \bibinfo{person}{S~Sara Mahdavi}, \bibinfo{person}{Bradley Green},
  \bibinfo{person}{Ewa Dominowska}, \bibinfo{person}{Blaise~Aguera y Arcas},
  \bibinfo{person}{Joelle Barral}, \bibinfo{person}{Dale Webster},
  \bibinfo{person}{Greg~S. Corrado}, \bibinfo{person}{Yossi Matias},
  \bibinfo{person}{Karan Singhal}, \bibinfo{person}{Pete Florence},
  \bibinfo{person}{Alan Karthikesalingam}, {and} \bibinfo{person}{Vivek
  Natarajan}.} \bibinfo{year}{2023}\natexlab{b}.
\newblock \bibinfo{title}{Towards Generalist Biomedical AI}.
\newblock
\newblock
\showeprint[arxiv]{2307.14334}~[cs.CL]


\bibitem[Tulk~Jesso et~al\mbox{.}(2022)]%
        {tulk2022inclusion}
\bibfield{author}{\bibinfo{person}{Stephanie Tulk~Jesso},
  \bibinfo{person}{Aisling Kelliher}, \bibinfo{person}{Harsh Sanghavi},
  \bibinfo{person}{Thomas Martin}, {and} \bibinfo{person}{Sarah
  Henrickson~Parker}.} \bibinfo{year}{2022}\natexlab{}.
\newblock \showarticletitle{Inclusion of clinicians in the development and
  evaluation of clinical artificial intelligence tools: a systematic literature
  review}.
\newblock \bibinfo{journal}{\emph{Frontiers in Psychology}}
  \bibinfo{volume}{13} (\bibinfo{year}{2022}), \bibinfo{pages}{830345}.
\newblock


\bibitem[Valencia et~al\mbox{.}(2023)]%
        {valencia2023less}
\bibfield{author}{\bibinfo{person}{Stephanie Valencia},
  \bibinfo{person}{Richard Cave}, \bibinfo{person}{Krystal Kallarackal},
  \bibinfo{person}{Katie Seaver}, \bibinfo{person}{Michael Terry}, {and}
  \bibinfo{person}{Shaun~K Kane}.} \bibinfo{year}{2023}\natexlab{}.
\newblock \showarticletitle{“The less I type, the better”: How AI Language
  Models can Enhance or Impede Communication for AAC Users}. In
  \bibinfo{booktitle}{\emph{Proceedings of the 2023 CHI Conference on Human
  Factors in Computing Systems}}. \bibinfo{pages}{1--14}.
\newblock


\bibitem[Verma et~al\mbox{.}(2021)]%
        {verma2021improving}
\bibfield{author}{\bibinfo{person}{Himanshu Verma}, \bibinfo{person}{Roger
  Schaer}, \bibinfo{person}{Julien Reichenbach}, \bibinfo{person}{Mario
  Jreige}, \bibinfo{person}{John~O Prior}, \bibinfo{person}{Florian
  Ev{\'e}quoz}, {and} \bibinfo{person}{Adrien Depeursinge}.}
  \bibinfo{year}{2021}\natexlab{}.
\newblock \showarticletitle{On improving physicians’ trust in AI: Qualitative
  inquiry with imaging experts in the oncological domain}.
\newblock \bibinfo{journal}{\emph{BMC Medical Imaging, in review}}
  (\bibinfo{year}{2021}).
\newblock


\bibitem[Wang et~al\mbox{.}(2023)]%
        {wang2023designing}
\bibfield{author}{\bibinfo{person}{Qiaosi Wang}, \bibinfo{person}{Michael
  Madaio}, \bibinfo{person}{Shaun Kane}, \bibinfo{person}{Shivani Kapania},
  \bibinfo{person}{Michael Terry}, {and} \bibinfo{person}{Lauren Wilcox}.}
  \bibinfo{year}{2023}\natexlab{}.
\newblock \showarticletitle{Designing Responsible AI: Adaptations of UX
  Practice to Meet Responsible AI Challenges}. In
  \bibinfo{booktitle}{\emph{Proceedings of the 2023 CHI Conference on Human
  Factors in Computing Systems}}. \bibinfo{pages}{1--16}.
\newblock


\bibitem[Wilcox et~al\mbox{.}(2023)]%
        {wilcox2023ai}
\bibfield{author}{\bibinfo{person}{Lauren Wilcox}, \bibinfo{person}{Robin
  Brewer}, {and} \bibinfo{person}{Fernando Diaz}.}
  \bibinfo{year}{2023}\natexlab{}.
\newblock \showarticletitle{AI Consent Futures: A Case Study on Voice Data
  Collection with Clinicians}.
\newblock  (\bibinfo{year}{2023}).
\newblock


\bibitem[W{\'o}jcik(2022)]%
        {wojcik2022foundation}
\bibfield{author}{\bibinfo{person}{Malwina~Anna W{\'o}jcik}.}
  \bibinfo{year}{2022}\natexlab{}.
\newblock \showarticletitle{Foundation Models in Healthcare: Opportunities,
  Biases and Regulatory Prospects in Europe}. In
  \bibinfo{booktitle}{\emph{International Conference on Electronic Government
  and the Information Systems Perspective}}. Springer, \bibinfo{pages}{32--46}.
\newblock


\bibitem[Xie et~al\mbox{.}(2020)]%
        {xie2020chexplain}
\bibfield{author}{\bibinfo{person}{Yao Xie}, \bibinfo{person}{Melody Chen},
  \bibinfo{person}{David Kao}, \bibinfo{person}{Ge Gao}, {and}
  \bibinfo{person}{Xiang'Anthony' Chen}.} \bibinfo{year}{2020}\natexlab{}.
\newblock \showarticletitle{CheXplain: enabling physicians to explore and
  understand data-driven, AI-enabled medical imaging analysis}. In
  \bibinfo{booktitle}{\emph{Proceedings of the 2020 CHI Conference on Human
  Factors in Computing Systems}}. \bibinfo{pages}{1--13}.
\newblock


\bibitem[Xu et~al\mbox{.}(2023)]%
        {xu2023elixr}
\bibfield{author}{\bibinfo{person}{Shawn Xu}, \bibinfo{person}{Lin Yang},
  \bibinfo{person}{Christopher Kelly}, \bibinfo{person}{Marcin Sieniek},
  \bibinfo{person}{Timo Kohlberger}, \bibinfo{person}{Martin Ma},
  \bibinfo{person}{Wei-Hung Weng}, \bibinfo{person}{Attila Kiraly},
  \bibinfo{person}{Sahar Kazemzadeh}, \bibinfo{person}{Zakkai Melamed},
  {et~al\mbox{.}}} \bibinfo{year}{2023}\natexlab{}.
\newblock \showarticletitle{ELIXR: Towards a general purpose X-ray artificial
  intelligence system through alignment of large language models and radiology
  vision encoders}.
\newblock \bibinfo{journal}{\emph{arXiv preprint arXiv:2308.01317}}
  (\bibinfo{year}{2023}).
\newblock


\bibitem[Yang et~al\mbox{.}(2019a)]%
        {yang2019sketching}
\bibfield{author}{\bibinfo{person}{Qian Yang}, \bibinfo{person}{Justin
  Cranshaw}, \bibinfo{person}{Saleema Amershi}, \bibinfo{person}{Shamsi~T
  Iqbal}, {and} \bibinfo{person}{Jaime Teevan}.}
  \bibinfo{year}{2019}\natexlab{a}.
\newblock \showarticletitle{Sketching nlp: A case study of exploring the right
  things to design with language intelligence}. In
  \bibinfo{booktitle}{\emph{Proceedings of the 2019 CHI Conference on Human
  Factors in Computing Systems}}. \bibinfo{pages}{1--12}.
\newblock


\bibitem[Yang et~al\mbox{.}(2023)]%
        {yang2023harnessing}
\bibfield{author}{\bibinfo{person}{Qian Yang}, \bibinfo{person}{Yuexing Hao},
  \bibinfo{person}{Kexin Quan}, \bibinfo{person}{Stephen Yang},
  \bibinfo{person}{Yiran Zhao}, \bibinfo{person}{Volodymyr Kuleshov}, {and}
  \bibinfo{person}{Fei Wang}.} \bibinfo{year}{2023}\natexlab{}.
\newblock \showarticletitle{Harnessing biomedical literature to calibrate
  clinicians’ trust in AI decision support systems}. In
  \bibinfo{booktitle}{\emph{Proceedings of the 2023 CHI Conference on Human
  Factors in Computing Systems}}. \bibinfo{pages}{1--14}.
\newblock


\bibitem[Yang et~al\mbox{.}(2020)]%
        {yang2020re}
\bibfield{author}{\bibinfo{person}{Qian Yang}, \bibinfo{person}{Aaron
  Steinfeld}, \bibinfo{person}{Carolyn Ros{\'e}}, {and} \bibinfo{person}{John
  Zimmerman}.} \bibinfo{year}{2020}\natexlab{}.
\newblock \showarticletitle{Re-examining whether, why, and how human-AI
  interaction is uniquely difficult to design}. In
  \bibinfo{booktitle}{\emph{Proceedings of the 2020 chi conference on human
  factors in computing systems}}. \bibinfo{pages}{1--13}.
\newblock


\bibitem[Yang et~al\mbox{.}(2019b)]%
        {yang2019unremarkable}
\bibfield{author}{\bibinfo{person}{Qian Yang}, \bibinfo{person}{Aaron
  Steinfeld}, {and} \bibinfo{person}{John Zimmerman}.}
  \bibinfo{year}{2019}\natexlab{b}.
\newblock \showarticletitle{Unremarkable AI: Fitting intelligent decision
  support into critical, clinical decision-making processes}. In
  \bibinfo{booktitle}{\emph{Proceedings of the 2019 CHI conference on human
  factors in computing systems}}. \bibinfo{pages}{1--11}.
\newblock


\bibitem[Yildirim et~al\mbox{.}(2022)]%
        {yildirim2022experienced}
\bibfield{author}{\bibinfo{person}{Nur Yildirim}, \bibinfo{person}{Alex Kass},
  \bibinfo{person}{Teresa Tung}, \bibinfo{person}{Connor Upton},
  \bibinfo{person}{Donnacha Costello}, \bibinfo{person}{Robert Giusti},
  \bibinfo{person}{Sinem Lacin}, \bibinfo{person}{Sara Lovic},
  \bibinfo{person}{James~M O'Neill}, \bibinfo{person}{Rudi~O'Reilly Meehan},
  {et~al\mbox{.}}} \bibinfo{year}{2022}\natexlab{}.
\newblock \showarticletitle{How Experienced Designers of Enterprise
  Applications Engage AI as a Design Material}. In
  \bibinfo{booktitle}{\emph{Proceedings of the 2022 CHI Conference on Human
  Factors in Computing Systems}}. \bibinfo{pages}{1--13}.
\newblock


\bibitem[Yildirim et~al\mbox{.}(2023a)]%
        {yildirim2023creating}
\bibfield{author}{\bibinfo{person}{Nur Yildirim}, \bibinfo{person}{Changhoon
  Oh}, \bibinfo{person}{Deniz Sayar}, \bibinfo{person}{Kayla Brand},
  \bibinfo{person}{Supritha Challa}, \bibinfo{person}{Violet Turri},
  \bibinfo{person}{Nina Crosby~Walton}, \bibinfo{person}{Anna~Elise Wong},
  \bibinfo{person}{Jodi Forlizzi}, \bibinfo{person}{James McCann},
  {et~al\mbox{.}}} \bibinfo{year}{2023}\natexlab{a}.
\newblock \showarticletitle{Creating Design Resources to Scaffold the Ideation
  of AI Concepts}. In \bibinfo{booktitle}{\emph{Proceedings of the 2023 ACM
  Designing Interactive Systems Conference}}. \bibinfo{pages}{2326--2346}.
\newblock


\bibitem[Yildirim et~al\mbox{.}(2023b)]%
        {yildirim2023investigating}
\bibfield{author}{\bibinfo{person}{Nur Yildirim}, \bibinfo{person}{Mahima
  Pushkarna}, \bibinfo{person}{Nitesh Goyal}, \bibinfo{person}{Martin
  Wattenberg}, {and} \bibinfo{person}{Fernanda Vi{\'e}gas}.}
  \bibinfo{year}{2023}\natexlab{b}.
\newblock \showarticletitle{Investigating How Practitioners Use Human-AI
  Guidelines: A Case Study on the People+ AI Guidebook}. In
  \bibinfo{booktitle}{\emph{Proceedings of the 2023 CHI Conference on Human
  Factors in Computing Systems}}. \bibinfo{pages}{1--13}.
\newblock


\bibitem[Yildirim et~al\mbox{.}(2021)]%
        {yildirim2021technical}
\bibfield{author}{\bibinfo{person}{Nur Yildirim}, \bibinfo{person}{John
  Zimmerman}, {and} \bibinfo{person}{Sarah Preum}.}
  \bibinfo{year}{2021}\natexlab{}.
\newblock \showarticletitle{Technical Feasibility, Financial Viability, and
  Clinician Acceptance: On the Many Challenges to AI in Clinical Practice.}. In
  \bibinfo{booktitle}{\emph{HUMAN@ AAAI Fall Symposium}}.
\newblock


\bibitem[Yildirim et~al\mbox{.}(2024a)]%
        {yildirim2024sketching}
\bibfield{author}{\bibinfo{person}{Nur Yildirim}, \bibinfo{person}{Susanna
  Zlotnikov}, \bibinfo{person}{Deniz Sayar}, \bibinfo{person}{Jeremy~M. Kahn},
  \bibinfo{person}{Leigh~A. Bukowski}, \bibinfo{person}{Sher~Shah Amin},
  \bibinfo{person}{Kathryn~A. Riman}, \bibinfo{person}{Billie~S. Davis},
  \bibinfo{person}{John~S. Minturn}, \bibinfo{person}{Andrew~J. King},
  \bibinfo{person}{Dan Ricketts}, \bibinfo{person}{Lu Tang},
  \bibinfo{person}{Venkatesh Sivaraman}, \bibinfo{person}{Adam Perer},
  \bibinfo{person}{Sarah~M. Preum}, \bibinfo{person}{James McCann}, {and}
  \bibinfo{person}{John Zimmerman}.} \bibinfo{year}{2024}\natexlab{a}.
\newblock \bibinfo{title}{Sketching AI Concepts with Capabilities and Examples:
  AI Innovation in the Intensive Care Unit}.
\newblock
\newblock
\showeprint[arxiv]{2402.13437}~[cs.HC]


\bibitem[Yildirim et~al\mbox{.}(2024b)]%
        {yildirim2024investigating}
\bibfield{author}{\bibinfo{person}{Nur Yildirim}, \bibinfo{person}{Susanna
  Zlotnikov}, \bibinfo{person}{Aradhana Venkat}, \bibinfo{person}{Gursimran
  Chawla}, \bibinfo{person}{Jennifer Kim}, \bibinfo{person}{Leigh~A. Bukowski},
  \bibinfo{person}{Jeremy~M. Kahn}, \bibinfo{person}{James McCann}, {and}
  \bibinfo{person}{John Zimmerman}.} \bibinfo{year}{2024}\natexlab{b}.
\newblock \bibinfo{title}{Investigating Why Clinicians Deviate from Standards
  of Care: Liberating Patients from Mechanical Ventilation in the ICU}.
\newblock
\newblock
\showeprint[arxiv]{2402.13464}~[cs.HC]


\bibitem[Yu et~al\mbox{.}(2023)]%
        {yu2023evaluating}
\bibfield{author}{\bibinfo{person}{Feiyang Yu}, \bibinfo{person}{Mark Endo},
  \bibinfo{person}{Rayan Krishnan}, \bibinfo{person}{Ian Pan},
  \bibinfo{person}{Andy Tsai}, \bibinfo{person}{Eduardo~Pontes Reis},
  \bibinfo{person}{Eduardo Kaiser Ururahy~Nunes Fonseca},
  \bibinfo{person}{Henrique Min~Ho Lee}, \bibinfo{person}{Zahra Shakeri~Hossein
  Abad}, \bibinfo{person}{Andrew~Y. Ng}, \bibinfo{person}{Curtis~P. Langlotz},
  \bibinfo{person}{Vasantha~Kumar Venugopal}, {and} \bibinfo{person}{Pranav
  Rajpurkar}.} \bibinfo{year}{2023}\natexlab{}.
\newblock \showarticletitle{Evaluating progress in automatic chest {X}-ray
  radiology report generation}.
\newblock \bibinfo{journal}{\emph{Patterns}} \bibinfo{volume}{4},
  \bibinfo{number}{9} (\bibinfo{year}{2023}).
\newblock
\urldef\tempurl%
\url{https://doi.org/10.1016/j.patter.2023.100802}
\showDOI{\tempurl}


\bibitem[Yu et~al\mbox{.}(2018)]%
        {yu2018artificial}
\bibfield{author}{\bibinfo{person}{Kun-Hsing Yu}, \bibinfo{person}{Andrew~L
  Beam}, {and} \bibinfo{person}{Isaac~S Kohane}.}
  \bibinfo{year}{2018}\natexlab{}.
\newblock \showarticletitle{Artificial intelligence in healthcare}.
\newblock \bibinfo{journal}{\emph{Nature biomedical engineering}}
  \bibinfo{volume}{2}, \bibinfo{number}{10} (\bibinfo{year}{2018}),
  \bibinfo{pages}{719--731}.
\newblock


\bibitem[Zaj{\k{a}}c et~al\mbox{.}(2023)]%
        {zajkac2023clinician}
\bibfield{author}{\bibinfo{person}{Hubert~D Zaj{\k{a}}c}, \bibinfo{person}{Dana
  Li}, \bibinfo{person}{Xiang Dai}, \bibinfo{person}{Jonathan~F Carlsen},
  \bibinfo{person}{Finn Kensing}, {and} \bibinfo{person}{Tariq~O Andersen}.}
  \bibinfo{year}{2023}\natexlab{}.
\newblock \showarticletitle{Clinician-facing AI in the Wild: Taking Stock of
  the Sociotechnical Challenges and Opportunities for HCI}.
\newblock \bibinfo{journal}{\emph{ACM Transactions on Computer-Human
  Interaction}} \bibinfo{volume}{30}, \bibinfo{number}{2}
  (\bibinfo{year}{2023}), \bibinfo{pages}{1--39}.
\newblock


\bibitem[Zhang et~al\mbox{.}(2023)]%
        {zhang2023deliberating}
\bibfield{author}{\bibinfo{person}{Angie Zhang}, \bibinfo{person}{Olympia
  Walker}, \bibinfo{person}{Kaci Nguyen}, \bibinfo{person}{Jiajun Dai},
  \bibinfo{person}{Anqing Chen}, {and} \bibinfo{person}{Min~Kyung Lee}.}
  \bibinfo{year}{2023}\natexlab{}.
\newblock \showarticletitle{Deliberating with AI: Improving Decision-Making for
  the Future through Participatory AI Design and Stakeholder Deliberation}.
\newblock \bibinfo{journal}{\emph{Proceedings of the ACM on Human-Computer
  Interaction}} \bibinfo{volume}{7}, \bibinfo{number}{CSCW1}
  (\bibinfo{year}{2023}), \bibinfo{pages}{1--32}.
\newblock


\bibitem[Zimmerman et~al\mbox{.}(2007)]%
        {zimmerman2007research}
\bibfield{author}{\bibinfo{person}{John Zimmerman}, \bibinfo{person}{Jodi
  Forlizzi}, {and} \bibinfo{person}{Shelley Evenson}.}
  \bibinfo{year}{2007}\natexlab{}.
\newblock \showarticletitle{Research through design as a method for interaction
  design research in HCI}. In \bibinfo{booktitle}{\emph{Proceedings of the
  SIGCHI conference on Human factors in computing systems}}.
  \bibinfo{pages}{493--502}.
\newblock


\end{thebibliography}


\section{Appendix A: EXAMPLE RADIOLOGY REPORT AND PROTOTYPE FLOWS} \label{appendix}
\textcolor{black}{This section contains (a) an example radiology report from the MIMIC-CXR dataset \cite{johnson2019mimic}, and (b)} the click-through prototypes used for concepts where we had more than a single frame \textit{(Draft Report Generation, Augmented Report Review, Visual Search and Querying)}.

\begin{figure*}
\centering
  \includegraphics[width=1.2\columnwidth]{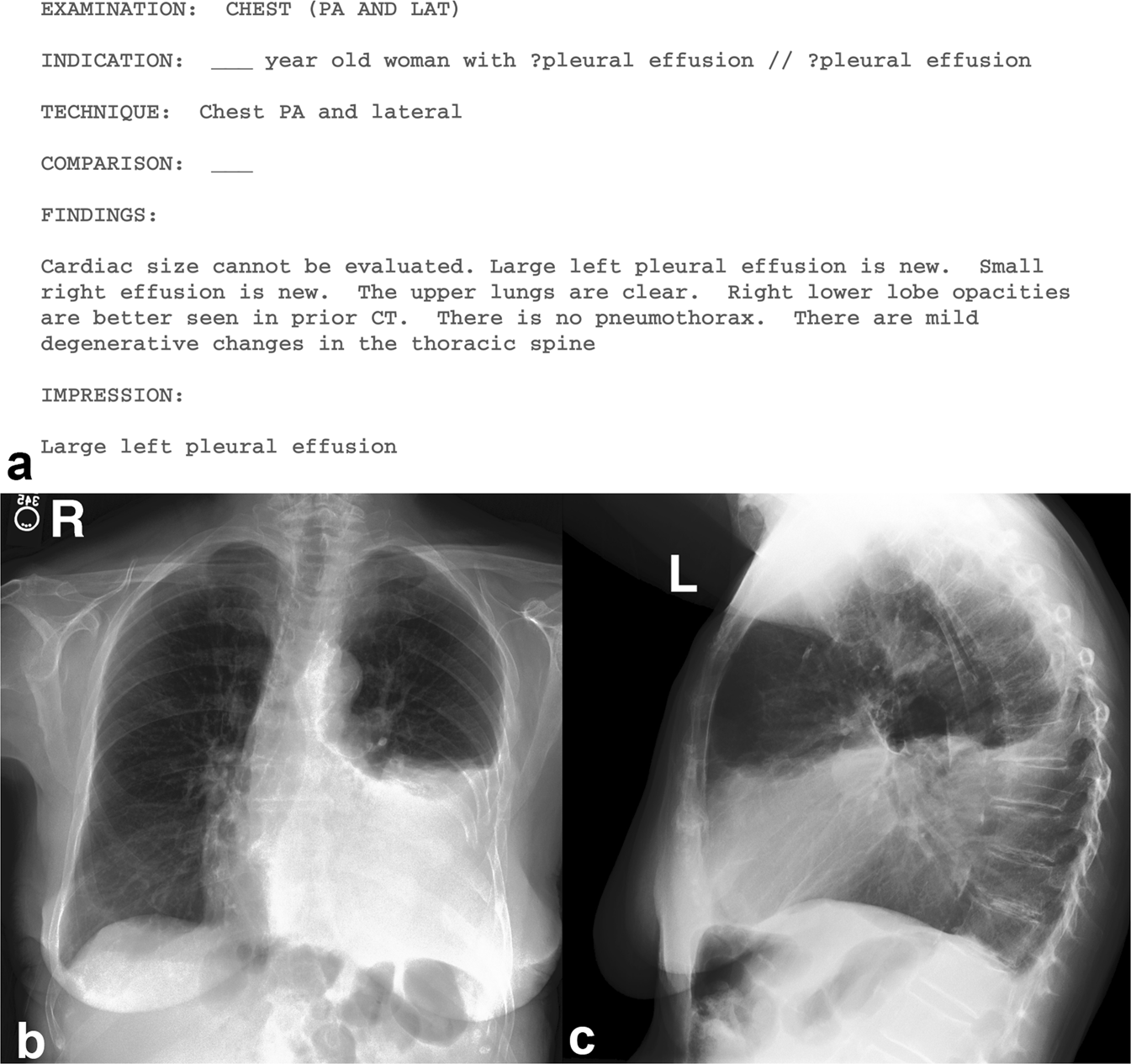}
  \caption{\label{fig:report} An example image-report pair from the MIMIC-CXR dataset \cite{johnson2019mimic} with (a) the radiology report and (b, c) the chest X-rays including the (b) frontal view and (c) the lateral view.
  \Description{An example image-report pair showing a radiology report text written in response to a patient's chest X-ray that has frontal and lateral views.}
}
\end{figure*}

\begin{figure*}
\centering
  \includegraphics[width=2\columnwidth]{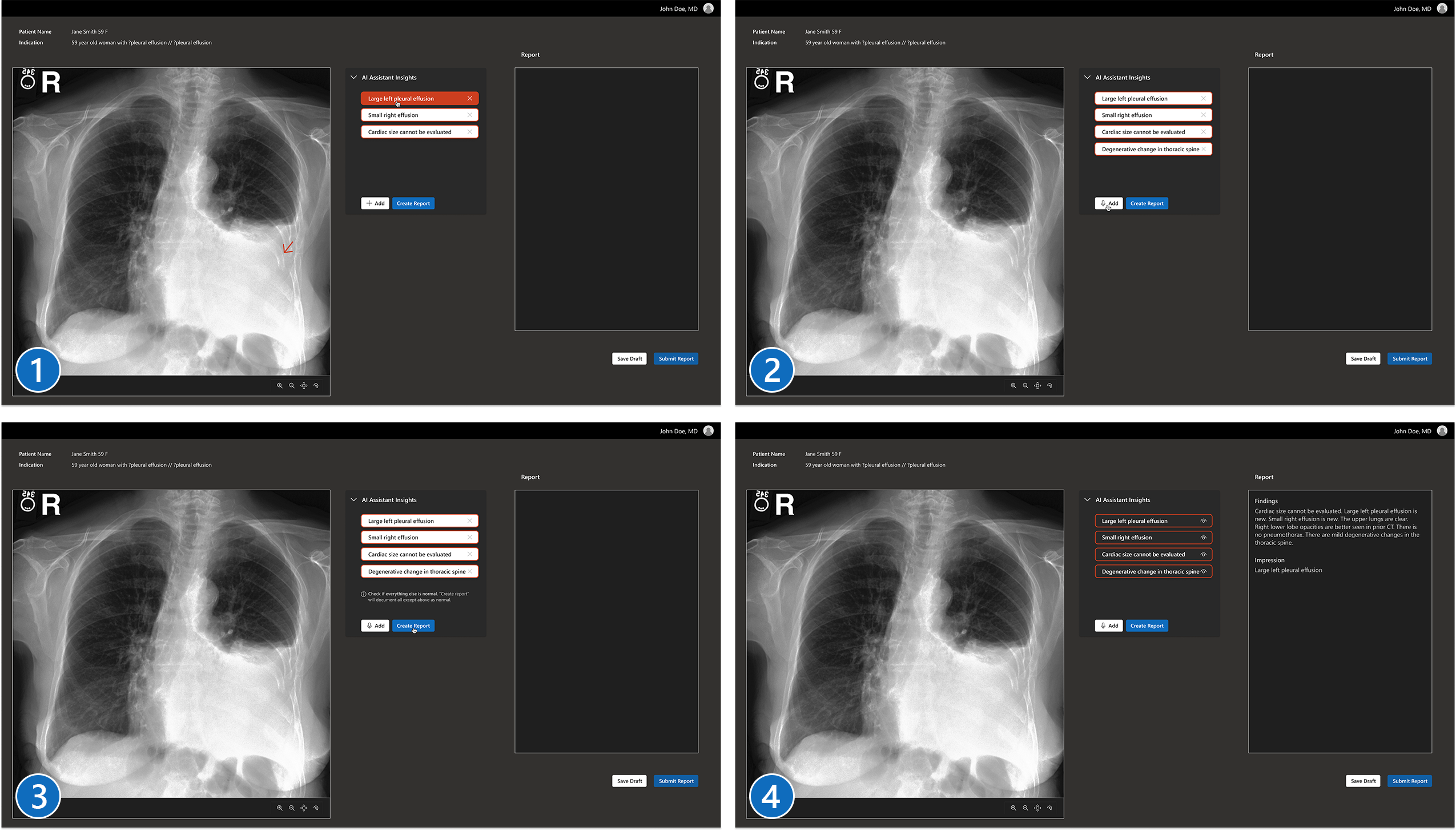}
  \caption{\label{fig:flow1} Click-through prototype flow illustrating the \textit{Draft Report Generation} concept.
  \Description{Click-through prototype flow for the Draft Report Generation concept displaying how bullet point findings can be used to view abnormalities on the image, and how they are edited to create a final prose report.}
}
\end{figure*}

\begin{figure*}
\centering
  \includegraphics[width=2\columnwidth]{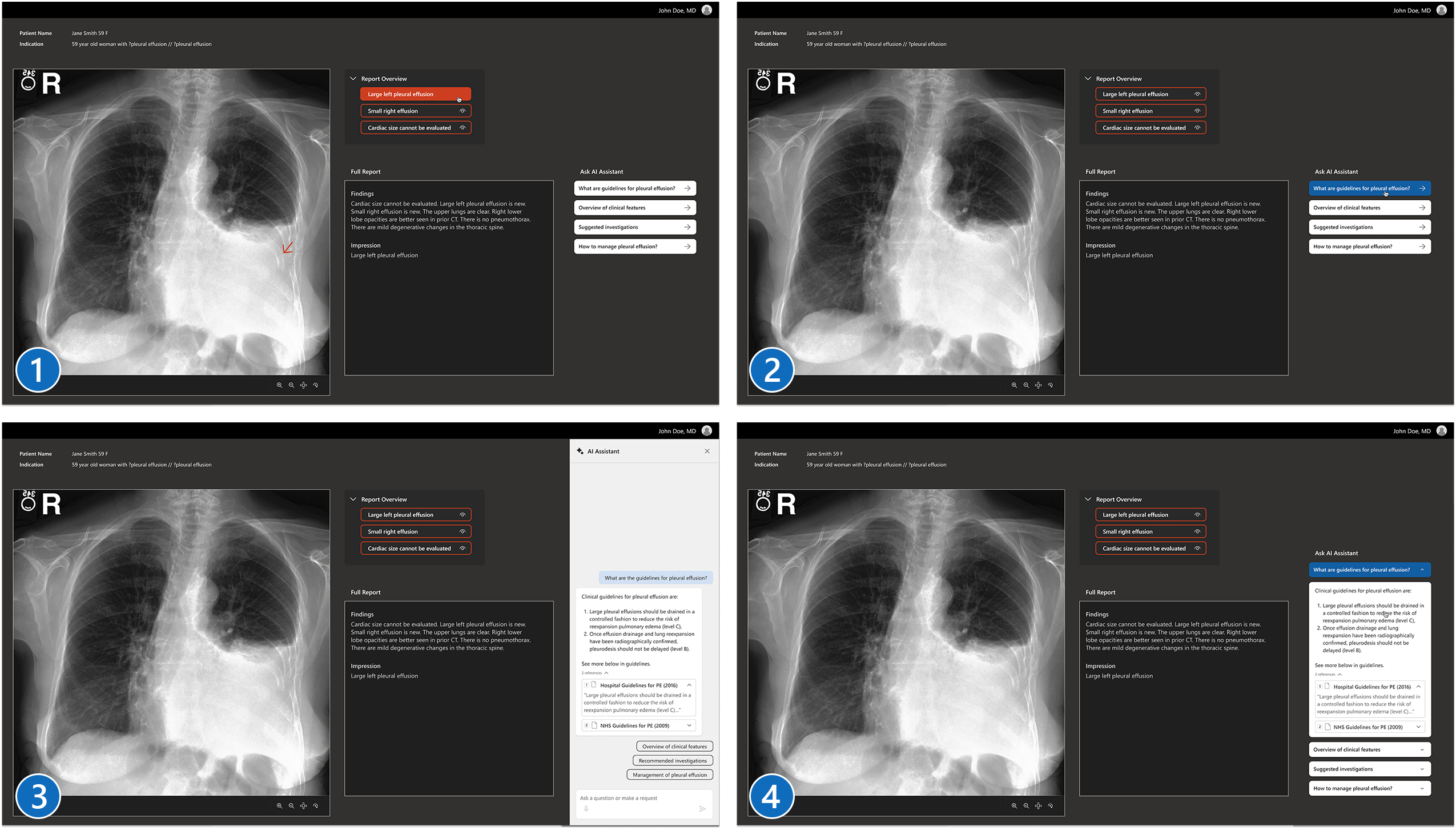}
  \caption{\label{fig:flow1} Click-through prototype flow illustrating the \textit{Augmented Report Review} concept.
  \Description{Click-through prototype flow for the Augmented Report Review concept displaying how the report review section findings can be used to view abnormalities on the image, and how the AI assistant can provide contextual information through queries in form of pre-run prompts or chat.}
}
\end{figure*}

\begin{figure*}
\centering
  \includegraphics[width=2\columnwidth]{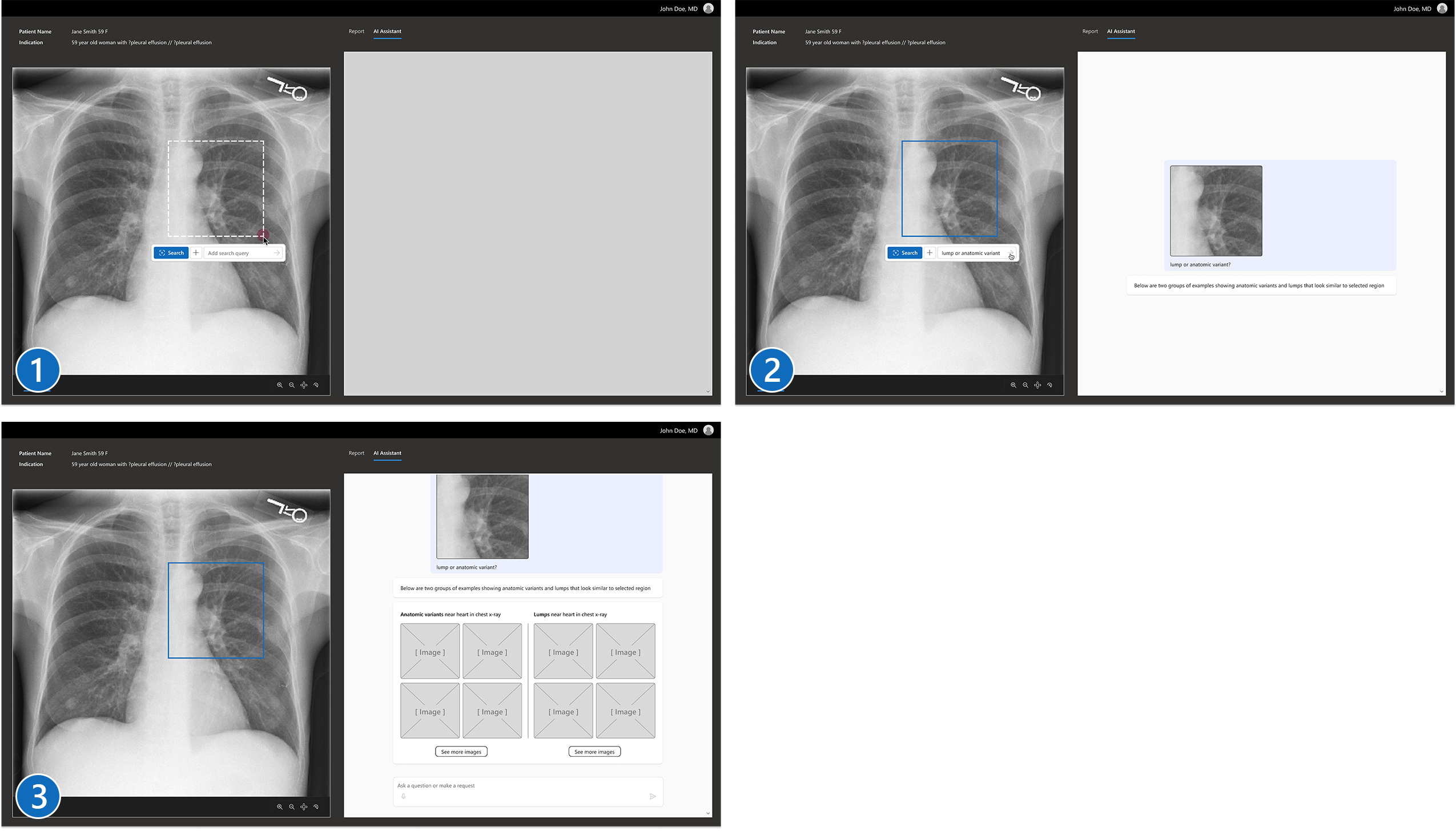}
  \caption{\label{fig:flow1} Click-through prototype flow illustrating the \textit{Visual Search and Querying} concept.
  \Description{Click-through prototype flow for the Visual Search and Querying concept displaying how an image region can be selected and queried for image search or image and text search, and how the AI assistant might display similar image results with accompanying text.}
}
\end{figure*}








\end{document}